\documentclass[prd,twocolumn,showpacs,amsmath,amssymb,nofootinbib,eqsecnum,floatfix]{revtex4}

\usepackage{graphicx}
\usepackage{bm}
\begin{document}
\title{A survey of spinning test particle orbits in Kerr spacetime}
\author{Michael D. Hartl}
\email{mhartl@tapir.caltech.edu}
\affiliation{
Department of Physics, California Institute of Technology, Pasadena CA 91125}
\date{February 17, 2003}
\begin{abstract}

We investigate the dynamics of the Papapetrou equations in Kerr spacetime.
These equations provide a model for the motion of a relativistic spinning test
particle orbiting a rotating (Kerr) black hole.  We perform a thorough
parameter space search for signs of chaotic dynamics by calculating the
Lyapunov exponents for a large variety of initial conditions.  We find that the
Papapetrou equations admit many chaotic solutions, with the strongest chaos
occurring in the case of eccentric orbits with pericenters close to the limit
of stability against plunge into a maximally spinning Kerr black hole. Despite
the presence of these chaotic solutions, we show that physically realistic
solutions to the Papapetrou equations are not chaotic; in all cases, the
chaotic solutions either do not correspond to realistic astrophysical systems,
or involve a breakdown of the test-particle approximation leading to the
Papapetrou equations (or both).   As a result, the gravitational radiation from
bodies spiraling into much more massive black holes (as detectable, for
example, by LISA, the Laser Interferometer Space Antenna) should not exhibit
any signs of chaos.

\end{abstract}
\pacs{04.70.Bw, 04.80.Nn, 95.10.Fh}
\maketitle

\section{INTRODUCTION}
\label{sec:introduction}

The present paper continues the investigation initiated in~\cite{Hartl_2002_1},
which considered the dynamics of spinning test particles (i.e., compact
astrophysical objects) orbiting a spinning black hole (Kerr spacetime). 
The primary objective is to determine whether the orbits of spinning compact
objects spiraling into much more massive black holes are chaotic or not. 
Evidence for chaotic orbits in relativistic binaries has been presented
in~\cite{SuzukiMaeda1997, LevinPRL2000, Levin2000, Cornish2001,
CornishLevinSR2002, CornishLevin2002} (although only~\cite{SuzukiMaeda1997} was
directly concerned with the extreme mass ratio systems we consider here).  An
astrophysical example of the systems we consider is a solar-mass black hole or
neutron star orbiting a supermassive black hole in a galactic nucleus. Such
systems are potentially important sources of gravitational radiation in
frequency bands detectable by space-based gravitational wave detectors such as
the proposed Laser Interferometer Space Antenna (LISA) mission.   The methods
of data analysis for signals from such detectors typically rely on matched
filters, which use a discrete mesh in parameter space to achieve a given
signal-to-noise ratio.  The presence of chaos would cause the number of
templates needed to fit a given signal to grow exponentially with decreasing
mesh size, potentially overwhelming the computers performing the analysis.

Many studies have used the Papapetrou equations to investigate the dynamics of
spinning test particles in the background spacetime of a central black hole
(including most recently~\cite{TMSS1996, SuzukiMaeda1997, Hartl_2002_1,
Semerak1999, SuzukiMaeda1998}; see~\cite{Semerak1999}
and~\cite{SuzukiMaeda1998} for a more thorough literature review). We found
in~\cite{Hartl_2002_1} that the Papapetrou equations formally admit chaotic
solutions in Kerr spacetime, but the chaos seemed to disappear for physically
realistic spins.  This conclusion was tentative, since we investigated only a
few values of the many parameters appearing in the equations.  In the present
study, we undertake a thorough search of parameter space in order to determine
the prevalence of chaotic orbits, both for dynamically interesting (but
physically unrealistic) orbits with high spin parameter~$S$ and for
astrophysically relevant systems with smaller spin.

As in~\cite{Hartl_2002_1}, we use Lyapunov exponents to measure the divergence
of nearby trajectories and to provide an estimate for the timescale of the
divergence.  For chaotic systems, initial conditions separated by a small
distance~$\epsilon_0$ diverge exponentially, with the separation growing as
$\epsilon(\tau) = \epsilon_0\,e^{\lambda\tau}$, where~$\lambda$ is the Lyapunov
exponent.  For nonchaotic orbits, $\lambda=0$, but when~$\lambda$ is nonzero
it provides the $e$-folding timescale $\tau_\lambda = 1/\lambda$ for chaotic
behavior to manifest itself.  In this paper we use two rigorous methods for
determining the maximum Lyapunov exponent~$\lambda_\mathrm{max}$ described
in~\cite{Hartl_2002_1}.  Further details appear below in Sec.~\ref{sec:lyap}
and in Sec.~III of~\cite{Hartl_2002_1}.

Our parameter space search makes use of a convenient orbital parameterization
technique, which allows us to specify desired values of  pericenter~$r_p$ and
orbital inclination~$\iota$ (Sec.~\ref{sec:param}).  We vary these orbital
parameters along with spin magnitude, eccentricity, Kerr spin angular
momentum~$a$, and spin inclination, and calculate the largest Lyapunov exponent
for each set of initial conditions. A complete description of our methods for
searching parameter space appears in Sec.~\ref{sec:results}.  Though not
exhaustive, the resulting survey of parameter space gives a thorough picture of
chaos in the Papapetrou equations with a Kerr background.

We set $G=c=1$, use sign conventions as in MTW~\citep{MTW}, and use 
Boyer-Lindquist coordinates $(r, \theta, \phi)$ for Kerr spacetime.  We use
vector arrows for 4-vectors and boldface for Euclidean vectors. The symbol
$\log$ denotes the natural logarithm.

\section{Spinning test particles}
\label{sec:spinning_test}

We use the Papapetrou equations of motion~\citep{Papapetrou} to model a
spinning test particle.  As reformulated by Dixon~\citep{Dixon}, these
equations describe the motion of a ``pole-dipole particle,'' which neglects
effects smaller than those due to the mass monopole and spin dipole (thus
neglecting the tidal coupling, which is a mass quadrupole effect).  The
Papapetrou equations describe a particle with negligible mass compared to the
masses responsible for generating the background spacetime. In our case, we
write~$\mu$ for the test particle's mass and~$M$ for the mass of the central
Kerr black hole; the test particle limit then requires that $\mu\ll M$.

\subsection{Equations of motion}
\label{sec:eom}

Dixon's formulation for the equations of motion uses an antisymmetric spin
tensor $S^{\mu\nu}$ to represent the spin angular momentum of the particle. 
The covariant derivative of the 4-momentum $p^\mu$ deviates from the
nonspinning (geodesic) case by a term representing a coupling of the spin to
the background spacetime curvature.  If we write $v^\mu$ for the 4-velocity,
the full equations of motion appear as follows: 
\begin{eqnarray}
\label{eq:Dixon}
\frac{dx^\mu}{d\tau} & = & v^\mu\nonumber\\
\nabla_{\vec v}\,p^\mu&=&
	-\textstyle{\frac{1}{2}}R^{\mu}_{\ \nu\alpha\beta}\,
	v^\nu S^{\alpha\beta}\\
\nabla_{\vec v}\,S^{\mu\nu}&=&2p^{[\mu}v^{\nu]}\nonumber.
\end{eqnarray}
Here $R^{\mu}_{\ \nu\alpha\beta}$ is the Riemann curvature tensor of the
spacetime, which in our case corresponds to the Riemann tensor for the Kerr
metric.  Note that in the case of vanishing spin (all components of
$S^{\mu\nu}$ equal to zero) or flat spacetime (all components of $R^{\mu}_{\
\nu\alpha\beta}$ equal to zero), we recover the geodesic equation $\nabla_{\vec
v}\,p^\mu=0$.

For numerical and conceptual purposes, we use a reformulation of the equations
of motion in terms of the momentum 1-form $p_\mu$ and spin 1-form $S_\mu$.  The
spin 1-form can be derived from the 4-momentum and spin tensor using
\begin{equation}
\label{eq:S_vector}
S_{\mu}=\textstyle{\frac{1}{2}}\epsilon_{\mu\nu\alpha\beta}\,
	u^\nu S^{\alpha\beta},
\end{equation}
where $u^\nu = p^\nu / \mu$.
The spin 1-form is easier to visualize than the spin tensor, and the Papapetrou
equations are better-behaved numerically in the $S\rightarrow0$ limit when
expressed in terms of the spin 1-form.
Details of this formulation appear in~\cite{Hartl_2002_1} and the appendix.

\subsection{General constraints}
\label{sec:constraints}

The Papapetrou system of equations~(\ref{eq:Dixon}) is highly constrained, even
in arbitrary spacetimes.  The 4-momentum and spin satisfy normalization
conditions that are conserved by the equations of motion:
\begin{equation}
\label{eq:p_squared}
p^\mu p_\mu = -\mu^2
\end{equation}
and
\begin{equation}
\label{eq:S_squared}
\textstyle{\frac{1}{2}}S^{\mu\nu} S_{\mu\nu} = S^2.
\end{equation}
A further condition is required to identify a unique worldline for the center of
mass:
\begin{equation}
\label{eq:pSortho}
p_\mu S^{\mu\nu} = 0.
\end{equation}
A more detailed discussion of this ``spin supplementary condition'' appears
in~\cite{Hartl_2002_1}.

\subsection{Kerr constraints}
\label{sec:Kerr_constraints}

The Papapetrou equations share an important property with geodesic motion,
namely, to each symmetry of the background spacetime (typically represented by
a Killing vector) there corresponds a constant of the motion.  Kerr spacetime
has two such symmetries, which provide two constraints in addition to those
described in the previous section.  (The Killing tensor that leads to the Carter
constant in the geodesic does not correspond to a conserved quantity when the
spin is nonzero; see Sec.~\ref{sec:empirical_orbit} below.) 

For each Killing vector~$\vec \xi$ there corresponds a constant given by the
standard expression for geodesics plus a contribution due to a coupling with 
the spin tensor:
\begin{equation}
\label{eq:C}
C_{\xi}=\xi^\mu p_\mu-\textstyle{1\over2}\xi_{\mu;\nu} S^{\mu\nu}.
\end{equation}
Kerr spacetime's two Killing vectors
$\vec\xi^{t}=\partial/\partial t$ and $\vec\xi^{\phi}=\partial/\partial\phi$
then give energy and 
$z$~angular momentum conservation:
\begin{equation}
\label{eq:E}
-E=p_t-\textstyle{1\over2}g_{t\mu,\nu}S^{\mu\nu}
\end{equation}
and
\begin{equation}
\label{eq:J}
J_z=p_\phi-\textstyle{1\over2}g_{\phi\mu,\nu}S^{\mu\nu}.
\end{equation}

\subsection{The spin parameter $S$}
\label{sec:spin_param}

The spin magnitude~$S$ that appears in Eq.~(\ref{eq:S_squared}) quantifies the
size of the spin and thus plays a crucial role in determining the behavior of
spinning test particle systems. As in~\cite{Hartl_2002_1}, we measure distances
and times in terms of~$M$ and momenta in terms of~$\mu$.  In these units, $S$
is measured in units of $\mu M$. When measured in traditional geometrized
units, the spin of a black hole can be as large as its mass squared, i.e.,
$S_\mathrm{geom,\,max} = \mu^2$.  In this case the spin parameter goes like
$S\sim \mu^2/(\mu M) = \mu/M$, which is necessarily small for a test particle
system.  This result applies to all astrophysically relevant systems, as shown
in~\cite{Hartl_2002_1}.  Thus, we have the important physical constraint that
\emph{the spin parameter $S$ must satisfy $S\ll1$ for all physically realistic
systems}.  A thorough discussion of various possibilities (including neutron
stars and white dwarfs) in~\cite{Hartl_2002_1} showed that realistic values
of~$S$ for LISA sources fall in the range $10^{-4}$--$10^{-7}$.

The smallness of the spin's effect is not dependent on
our choice to measure $S$ in terms of $\mu M$.  If instead
we measure $S$ in terms of $\mu^2$, then the equation of motion for
$S^{\mu\nu}$ becomes (rewriting $\nabla_{\vec v}$ as $D/d\tau$)
\begin{equation} 
\frac{D(S^{\mu\nu}/\mu^2)}{d(\tau/M)} =
2\,\frac{1}{\mu}p^{[\mu}v^{\nu]}, 
\end{equation} 
where we have factored out from each
variable its corresponding scale factor.  This gives 
\begin{equation}
\frac{DS^{\mu\nu}}{d\tau} = \frac{\mu}{M}\left(2p^{[\mu}v^{\nu]}\right),
\end{equation} 
so that in these units the spin's effective magnitude is
suppressed by a factor of $\mu/M$.  Measuring $S$ in terms of $\mu M$
effectively absorbs the unavoidable smallness of the spin effect into the spin
parameter itself.

\section{Parameterizing initial conditions}
\label{sec:param}

The many constraints on the equations of motion lead to considerable
complexity in parameterizing the initial conditions. We summarize here the
primary parameterization method described in~\cite{Hartl_2002_1}, and then
discuss in detail a method for parameterizing initial conditions using the
geometrical properties of the orbit.

\subsection{Energy and angular momentum parameterization}
\label{sec:ELparam}

It is easy to parameterize the Papapetrou equations directly in terms of the
momentum and spin components, but more physically relevant parameterizations
are more difficult. The parameterization method discussed
in~\cite{Hartl_2002_1} solves for the initial conditions using the integrals of
the motion~$E$ and~$J_z$.  This method also serves as the foundation for  a
more sophisticated parameterization in terms of orbital parameters
(Sec.~\ref{sec:orbital_param}).  

The energy and angular momentum parameterization method proceeds as follows. 
If we think of the system in terms of the spin 1-form~$S_\mu$, we have twelve
variables, four each for position, 4-momentum, and spin.  Since Kerr spacetime
is static and axially symmetric, without loss of generality we may set the
initial time and axial angle to zero: $\tau = \phi = 0$.  (We use the proper
time~$\tau$ as our time parameter, as discussed in Sec.~\ref{sec:lyap} below.) 
We then provide the initial values for $r$, $\theta$, and $p_r$, together with
the constants $E$, $J_z$, and $S$.  Finally, we provide two components of the
spin vector in an orthonormal basis.  It is easiest to specify the radial and
$\theta$ components $S^{\hat r}$ and $S^{\hat \theta}$; since the $r$-$\theta$
part of the Kerr metric is diagonal, we may then easily derive the 1-form
components $S_r$ and~$S_\theta$.

Having specified values for seven of the variables, we must now use the five
equations that relate them.  Since we measure the particle's momentum in terms
of its rest mass~$\mu$, the momentum normalization relation is
\begin{equation}
p^\mu p_\mu = -1,
\end{equation}
which we use to eliminate~$p_\theta$.  We then solve the spin
normalization condition 
\begin{equation}
S^\mu S_\mu = S^2
\end{equation} to eliminate~$S_\phi$.  The
spin-momentum orthogonality constraint and the two integrals of the motion 
then
give three equations in the three remaining unknowns $p_t$, $p_\phi$, 
and~$S_t$:
\begin{eqnarray}
0&=&p_\mu S_\nu g^{\mu\nu}\label{eq:1}\\
E&=&-p_t+\textstyle{1\over2}g_{t\mu,\nu}S^{\mu\nu}\\
J_z&=&p_\phi-\textstyle{1\over2}g_{\phi\mu,\nu}S^{\mu\nu}\label{eq:3}
\end{eqnarray}
(The first of these equations is equivalent to the 
condition~$p_\mu S^{\mu\nu}=0$, as discussed in the appendix.)

We solve Eqs.~(\ref{eq:1})--(\ref{eq:3}) with a Newton-Raphson root finder,
which works robustly given good initial guesses. The terms
$\textstyle{1\over2}g_{t\mu,\nu}S^{\mu\nu}$ and 
$\textstyle{1\over2}g_{\phi\mu,\nu}S^{\mu\nu}$ are typically small, even in the
physically unrealistic case of $S\sim1$, so $E$ and $J_z$ are good initial
guesses for their corresponding momenta.  We typically use $S_\theta$ as the
initial guess for $S_t$. In all cases, we verify \emph{a posteriori} that all
constraints are satisfied to within machine precision.

\subsection{Orbital geometry parameterization}
\label{sec:orbital_param}

While the method described above is sufficient for determining a set of valid
initial conditions, parameterizing orbits by energy and angular momentum is not
particularly intuitive.  It is much more natural to think in terms of the
orbital geometry, so we prefer to parameterize by the pericenter~$r_p$, the
eccentricity~$e$, and the orbital inclination angle~$\iota$.  Such parameters
only have precise meaning for geodesic orbits, but are 
nevertheless still useful
for the case of spinning test particles.

\subsubsection{Kerr geodesics}
\label{sec:Kerr_geod}

The first step in parameterizing solutions of the Papapetrou-Dixon equations
using orbital parameters is to solve the geodesic case.  The traditional method
for specifying a geodesic in terms of conserved quantities uses the energy~$E$,
the $z$~angular momentum~$L_z$, and the Carter constant~$Q$~\cite{MTW}.  In
order to use the orbital parameters, we adopt a mapping from $(r_p, e, \iota)$
to $(E, L_z, Q)$ based on unpublished notes supplied by Teviet Creighton and
Scott Hughes (and implemented in Hughes's Kerr geodesic
integrator~\cite{HughesIntegrator}).

In order to use the more intuitive orbital parameters, we must determine the
set $(E, L_z, Q)$ given the set $(r_p, e, \iota)$.  We obtain two of the
necessary equations by noting that the radial velocity~$dr/d\tau$ vanishes at
pericenter and
apocenter, since these radii correspond to turning points in the radial motion.
The equation for the
time-evolution of the Boyer-Lindquist radius~$r$ is~\cite{MTW}:
\begin{equation}
\Sigma^2\left(\frac{dr}{d\tau}\right)^2 = R(r),
\end{equation}
where
\begin{equation}
\label{eq:R}
R(r) = [E(r^2+a^2) - a L_z]^2 -\Delta[r^2+(L_z-aE)^2 + Q],
\end{equation}
and we use the standard auxiliary
variables
\begin{equation}
\label{eq:Sigma}
\Sigma = r^2+a^2\cos^2\theta
\end{equation}
and
\begin{equation}
\label{eq:Delta}
\Delta=r^2-2Mr+a^2.
\end{equation}
The quantity~$a$ is the Kerr spin parameter~$J/M$, i.e., the central black 
hole's spin angular momentum per unit mass, which is dimensionless in our
normalized units.
From Eq.~(\ref{eq:R}) we see that $dr/d\tau=0$ implies that $R(r)=0$, 
so we obtain one equation at each turning point:\footnote{In 
this paper we never consider exactly circular orbits,
but we note that our prescription fails in this case:  the
conditions~(\ref{eq:peri}) and~(\ref{eq:apo}) are identical when $e=0$, since
$r_a=r_p$.  For exactly circular orbits one must use the additional condition
$R'(r_p)=0$, where $R' = dR/dr$.}
\begin{equation}
\label{eq:peri}
R(r_p) = 0
\end{equation}
and
\begin{equation}
\label{eq:apo}
R(r_a) = 0,
\end{equation}
where the apocenter is defined by
\begin{equation}
r_a = \left(\frac{1+e}{1-e}\right)\,r_p.
\end{equation}

The final equation required to complete the mapping is~\cite{Ryan1996}
\begin{equation}
\label{eq:Qiota}
Q = L_z^2\,\tan^2\iota.
\end{equation}
The value of~$\iota$ resulting from this definition agrees closely with the
maximum value of $|\pi/2-\theta|$ for a numerically determined solution
to the equations of motion, i.e., it faithfully captures a geometric
property of the orbit.

Eqs.~(\ref{eq:peri})--(\ref{eq:Qiota}) give three equations in three
unknowns, which are easy to solve using a nonlinear root finder as long as good
initial guesses for the energy, angular momentum, and Carter constant 
can be found. The approach
we adopt uses the degenerate cases of circular equatorial orbits to provide
the raw material for analytical guesses.  The energies for prograde
and retrograde circular orbits in the equatorial plane are 
\begin{equation}
\label{eq:pro}
E^\mathrm{pro}(r) = \frac{1-2v^2+\tilde av^3}{\sqrt{1-3v^2+2\tilde av^3}}
\end{equation}
and
\begin{equation}
\label{eq:ret}
E^\mathrm{ret}(r) = \frac{1-2v^2-\tilde av^3}{\sqrt{1-3v^2-2\tilde av^3}},
\end{equation}
where we write $v\equiv\sqrt{M/r}$ and $\tilde a = a/M$ for
notational simplicity.
The initial guess for the energy is then an
average of these energies,
weighted using the inclination angle,\footnote{We adopt the convention that
$a\geq0$, so that $\iota$ indicates whether the orbit is prograde
($0\leq\iota<\frac{\pi}{2}$) or retrograde ($\frac{\pi}{2}<\iota\leq\pi$).} with
``radius'' given by the semimajor axis of an ellipse with pericenter~$r_p$ and
eccentricity~$e$: 
\begin{equation}
\label{eq:E_guess}
E^\mathrm{guess} =\textstyle{\frac{1}{2}}\left[\alpha_+\,
E^\mathrm{pro}(r_\mathrm{semi})+
\alpha_- \,E^\mathrm{ret}(r_\mathrm{semi})\right],
\end{equation}
where
\begin{equation}
\alpha_\pm = 1\pm\cos\iota,
\end{equation}
and
\begin{equation}
r_\mathrm{semi} = \frac{r_p}{1-e}.
\end{equation}
It is possible (though rare) for Eq.~(\ref{eq:E_guess}) to yield an energy guess
greater than 1; in this case, we simply set $E^\mathrm{guess} = 1$.

Once we have a guess for the energy, we can find the corresponding guess for
the angular momentum.  Using the value from
Eq.~(\ref{eq:E_guess}) and the expression for the angular momentum for
a circular equatorial orbit gives
\begin{equation}
\label{eq:L_guess}
L_z^\mathrm{guess} = \cos\iota\,\sqrt{\frac{1-e^2}{2(1-E^\mathrm{guess})}}
\end{equation}
as an initial guess for the angular momentum.  Finally, the guess for
the Carter constant is 
\begin{equation}
\label{eq:Q_guess}
Q^\mathrm{guess} = \left(L_z^\mathrm{guess}\right)^2\,\tan^2\iota.
\end{equation}
Plugging Eqs.~(\ref{eq:E_guess}), (\ref{eq:L_guess}), and (\ref{eq:Q_guess})
into the nonlinear root finder yields the actual values of $E$, $L_z$, and $Q$
to within machine precision in fewer than 10 iterations.

One caveat about our parameterization method is worth mentioning: some values
of $(r_p, e, \iota)$ correspond to unstable Kerr orbits, and in this case the
method returns a set $(E, L_z, Q)$ corresponding to an orbit with a different
pericenter from the one requested.  We can illustrate this behavior by
factoring Eq.~(\ref{eq:R}), which is a quartic function in~$r$:
\begin{equation} 
\label{eq:R_factored} 
R(r) = (1-E^2)(r_1-r)(r-r_2)(r-r_3)(r-r_4).
\end{equation} 
The roots are ordered so that $r_1\geq r_2\geq r_3\geq r_4$.  Bound motion
occurs for $r_1\geq r\geq r_2$, which implies that $r_1=r_a$ and $r_2 = r_p$,
but this only works for stable orbits. In the case that the orbit
\emph{requested} is unstable, the set $(r_p, e, \iota)$ returned by the
algorithm instead corresponds to a nearby, stable orbit. In this case, the 
numerically calculated roots of $R(r)$ satisfy
$r_3=r_{p,\,\mathrm{requested}}$, but the parameterization method returns $r_2$
as the pericenter, i.e., it returns a nearby stable orbit with pericenter $r_p
= r_2 > r_{p,\,\mathrm{requested}}$.  As a result, the \emph{actual}
pericenter is larger than the value requested.

We use routines from~\cite{HughesIntegrator} to identify the boundary between
stable and unstable orbits (so that the latter may be excluded), but the code
has a few minor bugs, and the identification procedure is not infallible.  As
a result, some such orbits appear in the results below (Sec.~\ref{sec:results})
and can be identified by having pericenters different from those requested.  It
is essential to understand, however, that the orbits returned by the
parameterization algorithm are never unstable, and represent perfectly valid
(stable) solutions to the equations of motion.

\subsubsection{From Kerr geodesics to Papapetrou initial conditions}
\label{sec:papa_param}

Once we have the set $(E, L_z, Q)$ for the geodesic case, we can use the
well-known properties of the Kerr metric to solve for all the parameters
necessary for the method described in Sec.~\ref{sec:ELparam} above.    The
first step is to force the constants of the motion to agree by setting
$E_\mathrm{Papapetrou} = E_\mathrm{Kerr}$ and $J_z = L_z$. Next, 
we set $\theta_0=\pi/2$ (corresponding to the equatorial plane), 
since this is the only angle guaranteed to be shared by all
Kerr geodesic orbits.\footnote{This
restricts our sample to Papapetrou orbits that cross the equatorial plane. 
This almost certainly represents the vast majority of valid Papapetrou
solutions, but there remains the intriguing possibility of spinning particle
solutions that orbit permanently above or below the equatorial plane.  We leave
an examination of this possibility to future investigators.} 
Finally, we must choose an initial value~$r_0$ for the
Boyer-Lindquist radius, which in the case of $\theta_0=\pi/2$ coincides with
the crossing of the equatorial plane.  One possibility is simply to use the
average value of the pericenter and apocenter,
\begin{equation}
\label{eq:r0}
r_0 = \textstyle{\frac{1}{2}}(r_p+r_a).
\end{equation}
This radius is guaranteed to lie between pericenter and apocenter, and
the prescription in Eq.~(\ref{eq:r0}) works fine for the mildly eccentric orbits
considered in this paper, but a more robust method must take into account that
highly eccentric orbits should in general cross the equatorial plane near
pericenter.  (Requiring a plane-crossing far from pericenter would force the
inclination angle to be small, which is a constraint we do not wish to
impose.)  This suggests choosing an initial value of~$r$ close to pericenter. 
One flexible method is to choose
\begin{equation}
\label{eq:r0_alt}
r_0 = r_p (1+ \alpha e),
\end{equation}
where $\alpha$ is a number of order unity.  This reduces correctly to $r_p$ in
the $e=0$ (circular) limit, and selects an equatorial plane crossing near
pericenter in the $e\sim1$ limit.  If we set $\alpha = 2$ in
Eq.~(\ref{eq:r0_alt}), then this method agrees exactly with Eq.~(\ref{eq:r0}) 
in the cases $e=0$ and $e=0.5$, and differs from Eq.~(\ref{eq:r0}) by less than
15\% when $e=0.6$.  Most of the results in this paper use Eq.~(\ref{eq:r0}),
but Eq.~(\ref{eq:r0_alt}) is preferable in general.

Once the initial~$\theta$ and~$r$ are known, we can determine all components of
the Kerr 4-momentum~$p_K^\mu$, but three of the Papapetrou momenta are
determined by the constraints (Sec.~\ref{sec:ELparam}).  We are therefore free
to force only one of the four components of the Papapetrou momentum~$p_P^\mu$
to be the same as its Kerr counterpart.  For the bulk of this paper, we set the
radial components equal ($p_K^r = p_P^r$), since it is the radial momentum that
is most closely tied to the stability and boundedness of the orbits.  This
choice results in Papapetrou orbits with pericenters and eccentricities fairly
close to the Kerr geodesic values, but with much higher orbital inclination
angles (Fig.~\ref{fig:color_map}).  The alternate choice of $p_K^\theta =
p_P^\theta$ results in Papapetrou orbits with inclinations similar to their
Kerr counterparts, but with very different pericenters
(Fig.~\ref{fig:color_map_alt}).  See Sec.~\ref{sec:results} for further
discussion.

We determine a value for the radial Papapetrou momentum $p_P^r$ using the
Kerr geodesic parameters $E_K$, $L_z$, and
$Q$ by applying Eq.~(\ref{eq:R}) and the equation
\begin{equation}
\rho^2 p^r = \mu\sqrt{R(r)},
\end{equation}
where $\rho = r^2 + a^2\cos^2\theta$~\cite{MTW}. We then convert to $p_{r}$
using $p_{r} = p^r g_{rr}$. Proceeding exactly as in Sec.~\ref{sec:ELparam}, we
specify two of the spin components and eliminate two variables using 4-momentum
and spin normalization, and then solve numerically for $p_t$, $p_\phi$, and
$S_t$.  The result is a set of initial conditions for the Papapetrou equations
with the same energy and angular momentum as a Kerr geodesic with the desired
values of~$r_p$, $e$, and~$\iota$.

It sometimes happens that the Papapetrou initial conditions derived in this
manner specify an orbit that is unstable against plunge into the black hole. 
Since there is no ``effective potential'' for a generic Papapetrou orbit as
there is for Kerr geodesics, there is no way \emph{a priori} to detect this
instability.  Plunge orbits are detected at runtime by testing for radial
coordinates less than the horizon radius.\footnote{In practice, the most common
runtime error is actually a numerical underflow in the integration stepsize as
the particle approaches the horizon.}  Orbits that plunge are removed from
consideration since by definition they cannot be chaotic.

\subsubsection{Empirical orbital parameters}
\label{sec:empirical_orbit}

In making the transition from geodesics to solutions of the Papapetrou
equations, we are able to enforce the conditions $E_\mathrm{Papapetrou} =
E_\mathrm{Kerr}$ and $J_z = L_z$, but this is no guarantee that the
Papapetrou orbit has the corresponding orbital parameters $r_p$ and~$\iota$:
the spin-coupling term $-\frac{1}{2}\xi_{\mu;\nu}S^{\mu\nu}$ [cf.
Eq.~(\ref{eq:C})] has a potentially large effect on the empirical values of the
pericenter and orbital inclination.  This effect is most pronounced when we
consider high spin parameter values, i.e., $S\sim1$.  In these dynamically
interesting cases, the \emph{empirical} pericenter and inclination will differ
in general from the values requested in the parameterization.

The empirical Papapetrou pericenter~$r_{p,\,P}$ is easy to find: we simply
integrate the orbit with a small stepsize for a large number of periods, and
then record the minimum radius achieved.  In practice, this works robustly,
reproducing almost exactly the requested Kerr value of~$r_{p,\,K}$ in the
limit~$S\ll 1$.  The only exception involves values of $r_{p,\,K}$ that
correspond to requested  unstable orbits, as discussed in
Sec.~\ref{sec:Kerr_geod} above.  Each of these orbits has an empirical
pericenter larger than the pericenter requested: the requested pericenter
corresponds to the root~$r_3$ in Eq.~(\ref{eq:R_factored}), but the empirical
pericenter returned by the algorithm corresponds to the larger root~$r_2$.

Having found the empirical pericenter for an orbit, we must next find its
empirical orbital inclination angle~$\iota_P$.  In order to
reproduce the definition from Eq.~(\ref{eq:Qiota}), we need to find an
analogue of the Carter constant~$Q$ for spinning test particles.  Kerr
spacetime has a Killing tensor $K_{\mu\nu}$~\cite{Wald} that satisfies
\begin{equation}
K_{\mu(\nu;\alpha)} = 0,
\end{equation}
which gives rise to an extra conserved quantity in the case of geodesic motion:
\begin{equation}
K = K_{\mu\nu} p^\mu p^\nu.
\end{equation}
This quantity is related to the traditional Carter constant~$Q$ by~\cite{MTW}
\begin{equation}
\label{eq:KerrK}
K = Q + (L_z - aE)^2.
\end{equation}

When the test particle has nonzero spin, the quantity defined by
Eq.~(\ref{eq:KerrK}) is no longer constant,
but there is an analogous expression that is conserved to linear order. 
Adapting a result from~\cite{TMSS1996}, we can write this approximately
conserved quantity as
\begin{equation}
\label{eq:approx_C}
C = K_{\mu\nu} p^\mu p^\nu - 2p^\mu S^{\rho\sigma}(f^\nu_{\
\sigma}\,f_{\mu\rho\nu} - f_\mu^{\ \nu}\,f_{\rho\sigma\nu}),
\end{equation}
where
\begin{equation}
\label{eq:f_tensor1}
f_{\mu\nu} = 2a\cos\theta e^1_{[\mu}e^0_{\nu]} + 2r e^2_{[\mu}e^3_{\nu]},
\end{equation}
\begin{equation}
\label{eq:f_tensor2}
f_{\mu\nu\sigma} = 
6\left(
\frac{a\sin\theta}{\sqrt{\Sigma}}e^0_{[\mu}e^1_\nu e^2_{\sigma]}
+\sqrt{\frac{\Delta}{\Sigma}}e^1_{[\mu}e^2_\nu e^3_{\sigma]}
\right),
\end{equation}
and the $\{e^a_{\ \mu}\}$ are the standard orthonormal tetrad for the Kerr
metric.  The effective Carter ``constant'' for spinning particles is then
\begin{equation}
\label{eq:Qeff}
Q_\mathrm{eff} = C - (J_z - aE)^2,
\end{equation}
where we use the full angular momentum~$J_z$ (which includes the contribution
from the spin) in place of~$L_z$.
The quantity~$Q_\mathrm{eff}$ is nearly but not exactly not constant, 
so in order to define an empirical inclination
angle we find the maximum effective~$Q$ over an orbit, and then
define~$\iota_P$ by 
\begin{equation}
\label{eq:iemp}
Q_\mathrm{eff,\,max} = J_z^2\,\tan^2\iota_P.
\end{equation}
As in the geodesic case, the value of~$\iota$ obtained from Eq.~(\ref{eq:iemp})
agrees well with the maximum value of $|\pi/2-\theta|$ over an
orbit.\footnote{This is true only when we force the Kerr and Papapetrou values
of $p^r$ to agree (Sec.~\ref{sec:papa_param}), which is the case for all orbits
considered except for the initial conditions 
shown in Fig.~\ref{fig:color_map_alt}.  In that case we
revert to the simpler method of finding the maximum value  of~$|\pi/2-\theta|$
over several orbits.} When $S=0$,  Eq.~(\ref{eq:iemp}) reduces to the
definition of the orbital inclination for geodesics, Eq.~(\ref{eq:Qiota}).

\section{Lyapunov exponents}
\label{sec:lyap}

\subsection{The principal exponent}
\label{sec:principal}

The Lyapunov exponents for a chaotic dynamical system quantify the chaos and
give insight into its dynamics (revealing, for example, whether it is
Hamiltonian or dissipative).  For a dynamical system of~$n$ degrees of freedom,
in general there are~$n$ Lyapunov exponents, which describe the time-evolution
of an infinitesimal ball centered on an initial condition.  This initial ball
evolves into an ellipsoid under the action of the Jacobian matrix of the
system, and the Lyapunov exponents are related to the average stretching of the
ellipsoid's principal axes.  We described in~\cite{Hartl_2002_1} a general
method for calculating all~$n$ of the system's exponents, but in the present
study we are interested only in the presence or absence of chaos in the
Papapetrou system, so we need only calculate the principal Lyapunov exponent,
i.e., the exponent corresponding to the direction of greatest stretching.  A
nonzero principal exponent indicates the presence of chaos. 

When studying a differentiable dynamical system, we typically introduce a set
of  variables ${\bf y}=\{y_i\}$ to represent the system's phase space, together
with the autonomous set of differential equations
\begin{equation}
\label{eq:f}
\frac{d\mathbf{y}}{d\tau}=\mathbf{f}(\mathbf{y})
\end{equation}
which determine the dynamics.  Associated with each initial condition is a
solution (or \emph{flow}).  The principal Lyapunov exponent quantifies the
\emph{local} divergence of nearby initial conditions, so any fully
rigorous method necessarily involves the local behavior of the system, i.e., its
derivative.  
For a multidimensional system, this derivative map is given by the
Jacobian matrix, 
\begin{equation}
\label{eq:jacobian_def}
({\bf Df})_{ij}=\frac{\partial f_i}{\partial x^j}.
\end{equation}

It is the Jacobian map that determines the time-evolution of infinitesimally
separated trajectories.
If we consider a point~$\mathbf{y}$ on the flow and a nearby point
$\mathbf{y} + \delta\mathbf{y}$, then we have
\begin{equation}
\mathbf{f}(\mathbf{y} + \delta\mathbf{y}) = 
\frac{d}{d\tau}(\mathbf{y} + \delta\mathbf{y}) = \frac{d\mathbf{y}}{d\tau}
+ \frac{d(\delta\mathbf{y})}{d\tau},
\end{equation}
so that the separation~$\delta\mathbf{y}$ satisfies
\begin{equation}
\frac{d(\delta\mathbf{y})}{d\tau} = \mathbf{f}(\mathbf{y} + \delta\mathbf{y})
- \frac{d\mathbf{y}}{d\tau} = \mathbf{f}(\mathbf{y} + \delta\mathbf{y})
- \mathbf{f}(\mathbf{y}).
\end{equation}
Since
\begin{equation}
{\bf f}({\bf y}+\delta{\bf y})-{\bf f}({\bf y})={\bf Df}\cdot
\delta{\bf y}+{\cal O}(\|\delta{\bf y}\|^2),
\end{equation}
we can write the time-evolution of the deviation vector as
\begin{equation}
\frac{d(\delta\mathbf{y})}{d\tau} = {\bf Df}\cdot
\delta{\bf y}+{\cal O}(\|\delta{\bf y}\|^2).
\end{equation}
If we identify the ``infinitesimal'' deviation $\delta\mathbf{y}$ with 
an element~$\bm{\xi}$ of the tangent space at~$\mathbf{y}$, we effectively take
the limit as $\delta\mathbf{y}\rightarrow 0$; the equation of motion
for~$\bm{\xi}$ is then
\begin{equation}
\label{eq:dxidt}
\frac{d\bm{\xi}}{d\tau}={\bf Df}\cdot\bm{\xi}.
\end{equation}
This equation describes the time-evolution of the separation between nearby
initial conditions in a rigorous way.  
We will refer to $\bm{\xi}$ as a tangent vector, since
formally is an element of the tangent space to the phase space.

If the Jacobian~${\bf Df}$ 
in Eq.~(\ref{eq:dxidt}) were some constant matrix~$\mathbf{A}$,
the solution would be the matrix exponential
\begin{equation}
\bm{\xi}(\tau) = \exp(\mathbf{A}\tau)\cdot\bm{\xi}_0.
\end{equation}
For long times, the solution would be dominated by the largest eigenvector
of~$\mathbf{A}$, and would grow like
\begin{equation}
\|\bm{\xi}(\tau)\| \approx e^{\lambda_\mathrm{max}\tau},
\end{equation}
where $\lambda_\mathrm{max}$ is the largest eigenvalue.\footnote{The most
common choice for the norm $\|\cdot\|$ is the Euclidean norm, but we use a
slightly different norm in the case of the Papapetrou equations
(Sec.~\ref{sec:papa_lyap} below).}  (Here we have used $\|\bm{\xi}_0\|=1$.)
Turning things around, if we measured the time-evolution of $\bm{\xi}$, we
could find an approximation for the exponent using
\begin{equation}
\lambda_\mathrm{max} = \frac{\log{\|\bm{\xi}(\tau)\|}}{\tau}.
\end{equation}  

The general case follows by considering Jacobian matrices that are
time-dependent.  In this case, we are unable to define a unique principal
exponent valid for all times, but there still is a unique \emph{average}
exponent that describes the average stretching of the principal eigenvector.
Our method is to track the evolution of a tangent vector as it
evolves into the principal axis of the
ellipsoid.  If we use $r_e(\tau) = \|\bm{\xi}(\tau)\|$ to denote the length of 
the longest principal ellipsoid axis, the principal
Lyapunov exponent is then defined by
\begin{equation}
\label{eq:lambda}
\lambda_\mathrm{max} = \lim_{\tau\rightarrow\infty} 
\frac{\log{[r_e(\tau)]}}{\tau}.
\end{equation}
We use an infinite time limit in this formal definition, but 
of course in practice a numerical approach relies on a finite cutoff to
obtain a numerical approximation.

The \emph{Jacobian method} for determining the largest Lyapunov exponent
involves solving Eqs.~(\ref{eq:f}) and~(\ref{eq:dxidt}) as a coupled system of
differential equations in order to follow the time-evolution of a ball of
initial conditions.  One possibility is then to use Eq.~(\ref{eq:lambda}) to
estimate the system's largest exponent.  A related technique, which provides
more accurate exponents, is to sample $\log{[r_e(\tau)]}$ at regular time
intervals, and then perform a least-squares fit on the simulation data.  Since
$\log{[r_e(\tau)]}=\lambda\tau$, the slope of the resulting line then gives an
estimate for the Lyapunov exponent.  This is the method we implement in
practice.

A less rigorous but still useful technique, which
we call the \emph{deviation vector} approach, involves solving only 
Eq.~(\ref{eq:f}), but for two initial conditions: $\mathbf{y}_0$ and
$\mathbf{y}_0+\delta\mathbf{y}_0$.  If the solutions to these initial
conditions a time~$\tau$ later are $\mathbf{y}$ and $\mathbf{y}'$, respectively,
then the approximate principal Lyapunov exponent is
\begin{equation}
\lambda_\mathrm{max} = 
\frac{\log{\|\delta\mathbf{y}\|/\|\delta\mathbf{y}_0}\|}{\tau},
\end{equation}
where $\delta\mathbf{y} \equiv \mathbf{y}' - \mathbf{y}$.  This approach has a
serious drawback: no matter how small the initial size of the deviation,
eventually the method saturates as the difference between $\mathbf{y}$ and 
$\mathbf{y}'$ grows so large that it no longer samples the
\emph{local} difference between two trajectories.\footnote{It is possible to
re-scale the deviation once it reaches a certain size, but this method is
error-prone since it can depend sensitively on the precise method of rescaling. 
The constrained nature of the Papapetrou equations also presents difficulties
for rescaling, as discussed in Sec.~\ref{sec:papa_lyap}.}  On the other hand,
because we need only solve Eq.~(\ref{eq:f}) and not Eq.~(\ref{eq:dxidt}), the
deviation vector method is significantly faster than the Jacobian method (by a
factor of approximately~5 for the system considered here).  
We therefore adopt the deviation vector method as
our principal tool for broad surveys of parameter space.  The method for
handling the saturation problem is discussed in Sec.~\ref{sec:chaos_detector}.

\begin{figure*}
\includegraphics[width=3in]{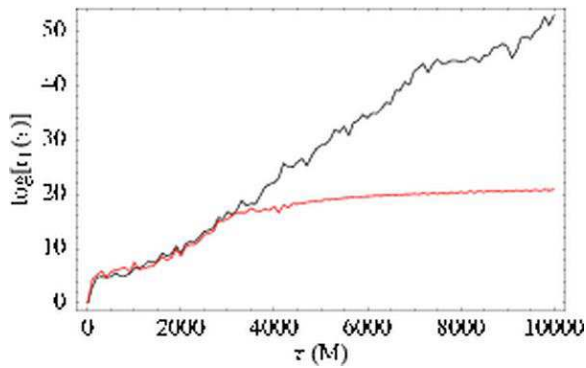}\\
\caption{
\label{fig:kerr_lyap_compare}
Rigorous Jacobian method compared to unrescaled deviation vector method for an
$S=1$ particle in maximal ($a=1$) Kerr spacetime.  The
vertical axis is the natural logarithm of the largest
principal axis~$r_1$ of the phase
space ellipsoid; the slope is the principal Lyapunov
exponent, $\lambda_\mathrm{max}\approx5\times10^{-3}\,M^{-1}$.  
The
unrescaled deviation vector method started  with a deviation of size~$10^{-7}$,
and it saturates at~$\sim\!\!16$. This corresponds to a growth of
$e^{16}\approx9\times10^{6}$, which means that the separation has grown to a
size of order unity.  In conventional units, this indicates
radial separations of
order~$GM/c^2$ and velocity separations of order~$c$.  The norm is calculated
using the projected norm described in Sec.~\ref{sec:papa_lyap}.}
\end{figure*}

A comparison of the Jacobian and deviation vector methods appears in
Fig.~\ref{fig:kerr_lyap_compare}.  It is apparent that the two methods agree
closely until the deviation vector approach reaches the saturation limit.

\subsection{The Papapetrou case}
\label{sec:papa_lyap}

The discussion in the previous section was of a general nature, applicable to
virtually any dynamical system described by differential equations.  Here we
describe some of the details needed to apply the general methods to the
Papapetrou equations.  In particular, we must discuss further the key ideas of
constrained deviation vectors and phase space metric.

For the Papapetrou system, the phase space vector~$\mathbf{y}$ has~12
components:
\begin{equation}
{\mathbf{y}} = (t, r,
\mu, \phi, p_t, p_r, p_\theta, p_\phi, S_t, S_r, S_\mu, S_\phi).
\end{equation}
The tangent vector~$\bm{\xi}$ has one component for each variable.
The set of equations~Eq.~(\ref{eq:f}) is simply the Papapetrou equations written
out in full:
\begin{eqnarray*}
	\dot x^\mu & = & v^\mu\\
	\dot p_\mu & = & -R^{*\ \ \alpha\beta}_{\mu\nu}v^\nu p_\alpha S_\beta+
		\Gamma^\alpha_{\ \beta\mu}p_\alpha v^\beta\\
	\dot S_\mu & = & -p_\mu \left(R^{*\alpha\ \gamma\delta}_{\ \ \beta} 
        S_\alpha v^\beta p_\gamma S_\delta\right)+
		\Gamma^\alpha_{\ \beta\mu}S_\alpha v^\beta,
\end{eqnarray*}
where we have used the formulation in terms of the spin 1-form described in the
appendix.  The second necessary equation, Eq.~(\ref{eq:dxidt}), requires the
Jacobian matrix, 
\begin{equation}
\label{eq:jacobian}
\left(
\begin{array}{rcl}
	\displaystyle{\frac{\partial\dot x^\mu}{\partial x^\nu}} & 
		\displaystyle{\frac{\partial\dot x^\mu}{\partial p_\nu}} &
		\displaystyle{\frac{\partial\dot x^\mu}{\partial S_\nu}}\medskip\\
	\displaystyle{\frac{\partial\dot p_\mu}{\partial x^\nu}} & 
		\displaystyle{\frac{\partial\dot p_\mu}{\partial p_\nu}} &
		\displaystyle{\frac{\partial\dot p_\mu}{\partial S_\nu}}\medskip\\
	\displaystyle{\frac{\partial\dot S_\mu}{\partial x^\nu}} & 
		\displaystyle{\frac{\partial\dot S_\mu}{\partial p_\nu}} &
		\displaystyle{\frac{\partial\dot S_\mu}{\partial S_\nu}}\\
\end{array}
\right),
\end{equation}
whose explicit form appears in~\cite{Hartl_2002_1}.

It is important to mention that the tangent vector~$\bm{\xi}$---or,
equivalently, the deviation vector~$\delta\mathbf{y}$---cannot have completely
arbitrary components.  On the contrary, the deviation must be chosen carefully,
in order to ensure that, given a point~$\mathbf{y}$ that satisfies the
constraints from Sec.~\ref{sec:constraints}, the deviated vector
$\mathbf{y}+\delta\mathbf{y}$ also satisfies the constraints.  Otherwise, the
relation ${\bf f}({\bf y}+\delta{\bf y})-{\bf f}({\bf y})={\bf Df}\cdot
\delta{\bf y}+{\cal O}(\|\delta{\bf y}\|^2)$ is not satisfied.  (This is the
principal complication in implementing a rescaled version of the deviation
vector method: the rescaled vector would violate the constraint.) In practice,
we are able to find a valid deviation vector by  applying the same techniques
used to satisfy the constraints in the first place (Sec.~\ref{sec:param}); for
details, see~\cite{Hartl_2002_1}.  

One final detail is the notion of metric: implicit in the definition of the
Lyapunov exponent, Eq.~(\ref{eq:lambda}), is a metric on phase space used to
calculate the norm of the tangent vector~$\bm{\xi}$.  We adopt a metric
introduced in~\cite{KarasVokrouhlicky1992}, which involves projecting the
deviation vector onto the spacelike hypersurface defined by the zero angular
momentum observers (ZAMOs).  The projection is effected using the projection
tensor $P^\mu_{\ \nu} = \delta^\mu_{\ \nu} + U^\mu U_\nu$, where $U^\mu$ is the
ZAMO 4-velocity. The spatial and momentum variables are then projected
according to  $x^\mu \rightarrow \tilde x^i = P^i_{\ \mu}\,x^\mu$, $p_\mu
\rightarrow \tilde p_i = P^\mu_{\ i}\,p_\mu$, and $S_\mu \rightarrow \tilde S_i
= P^\mu_{\ i}\,S_\mu$.  After the projection, we calculate the Euclidean norm
in the 3-dimensional hypersurface.  We note that while this prescription is
convenient, and reduces correctly in the nonrelativistic limit, the magnitudes
of the Lyapunov exponents are similar for several other possible choices of
metric~\cite{Hartl_2002_1}.

\subsection{Chaos detector}
\label{sec:chaos_detector}

Since we are concerned with calculating Lyapunov exponents for a large number
of parameter values, we use the (unrescaled) deviation vector method because of
its speed.  As mentioned in Sec.~\ref{sec:principal}, this method has the
property of saturation (as illustrated in Fig.~\ref{fig:kerr_lyap_compare}),
which is ordinarily a problem, but here we use it to our advantage as a
sensitive detector of chaos.

Our method for determining whether a particular initial condition is chaotic is
to consider a nearby initial condition separated by a small 
vector~$\delta\mathbf{y}_0$ (with norm $\|\delta\mathbf{y}_0\|$  typically of
order $10^{-7}$ or $10^{-8}$) and then integrate until the system reaches 90\%
of the saturation limit, defined as a separation $\delta\mathbf{y}$ with unit
 norm.  If we write $r_e=\|\delta\mathbf{y}\|/\|\delta\mathbf{y}_0\|$, the
approximate Lyapunov exponent satisfies 
\begin{equation} 
\log{[r_e(\tau)]} = \lambda\tau,
\end{equation} 
so that~$\lambda$ is the slope of the line $\log{[r_e(\tau)]}$ vs.\ $\tau$. 
Saturation occurs when $r_e = 1/\|\delta\mathbf{y}_0\|$, so that the integration
ends when $\log{[r_e(\tau)]} = -\log{(0.9\,\|\delta\mathbf{y}_0\|)}$.
We record the value of $\log{[r_e(\tau)]}$ at regular time intervals
(typically every time~$T=100\,M$ in our case),
and upon reaching 90\% saturation perform a least-squares fit on the simulation
data to determine the exponent.

We note that the cutoff value of~$0.9$ is somewhat arbitrary and is the result
of numerical experimentation.  When using the (unrescaled) deviation vector
approach, most of the chaotic systems saturate---that is, plots of
$\log{[r_e(\tau)]}$ vs.\ $\tau$ flatten out
(Fig.~\ref{fig:kerr_lyap_compare})---when the separation is of order unity,
corresponding to radial separations of order~$M$, velocity separations of
order~$1$, and angular separations of order~$1~\mathrm{radian}$.  The 90\%
prescription ends the integration before the growth flattens out, so that the
numerical estimate for the exponent is still reasonably accurate.

\begin{figure}
\includegraphics[width=3in]{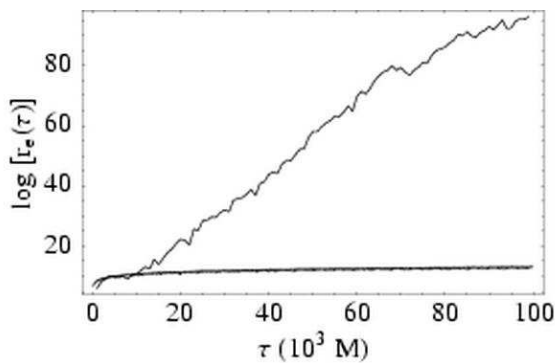}
\caption{\label{fig:lyap_compare_e=0.6}
A comparison of chaotic and nonchaotic initial conditions. The slopes of the
lines are the approximate Lyapunov exponents.  Each initial
condition shares the same values of $S=0.1$, $e=0.6$, and $r_p=1.21$. They 
differ only in inclination angle: $\iota=31^\circ$ (chaotic) and
$\iota=28.5^\circ$ (not chaotic).  Their respective Lyapunov exponents are
$\lambda=1.0\times10^{-3}\,M^{-1}$ and $\lambda=3.0\times10^{-3}\,M^{-1}$.
See Sec.~\ref{sec:e} for details.}
\end{figure}

For the large maps of parameter space, we typically integrate up to
$\tau_\mathrm{final} = 10^5\,M$ or saturation, whichever comes first.  We
choose this maximum time mainly for practical reasons: it is the longest
integration possible in a reasonable amount of time.  [We integrate as deeply
as $10^7\,M$ (Sec.~\ref{sec:deep}) for individual orbits, but such long
integrations are impractical for more than a handful of parameter
combinations.]  Dramatically longer integrations are also not particularly
useful, since the timescale for gravitational radiation reaction is on the
order of $\tau_\mathrm{GW}=M^2/\mu$~\cite{Hughes2001_2}, which for the most
relevant LISA sources is $10^4$--$10^6\,M$ (i.e.,
$\mu\sim10^{-6}$--$10^{-4}\,M$).  Searching for chaos in such systems on a
timescale longer than $\sim\!\!10^7\,M$ is probably pointless, since the
radiation reaction would dominate the dynamics in this case. Finally, it
appears that chaos, when present in the Papapetrou system, manifests itself on
relatively short timescales ($10^2$--$10^4\,M$), or else not at all.  The onset
of chaos is marked by a qualitative change from power law growth (which appears
as logarithmic growth on our plots of $\log{[r_e(\tau)]}$ vs.\ $\tau$) to
exponential growth (which is linear on the same plots).  An example of two
similar initial conditions giving rise to qualitatively different dynamical
behavior appears in Fig.~\ref{fig:lyap_compare_e=0.6}.

\begin{figure}
\includegraphics[width=3in]{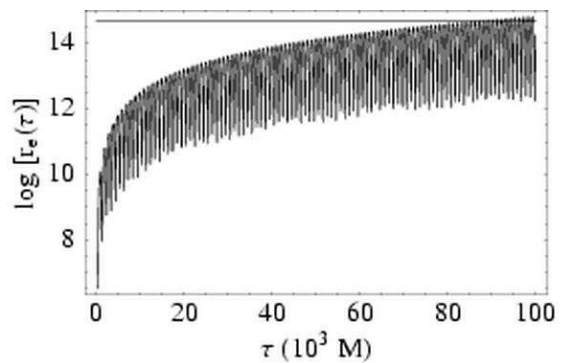}
\caption{\label{fig:mimic} A chaos mimic: $\log{[r_e(\tau)]}$ vs.\ $\tau$ for an
$S=0$ orbit. 
The size of the initial deviation vector is $\epsilon_0=3.3\times10^{-7}$.
The value of $\log{[r_e(\tau)]}$
periodically rises up to the saturation level (shown as a horizontal line at
$14.82$, since $e^{14.82}\,\epsilon_0 = 0.9$).  The system's spin satisfies
$S=0$, and is
hence fully integrable, which implies no chaos.  We detect such spurious chaos
by demanding several saturation points in a row for a positive detection.
The corresponding orbit appears in Fig.~\ref{fig:mimic_orbit}.}
\end{figure}

\begin{figure}
\includegraphics[width=3in]{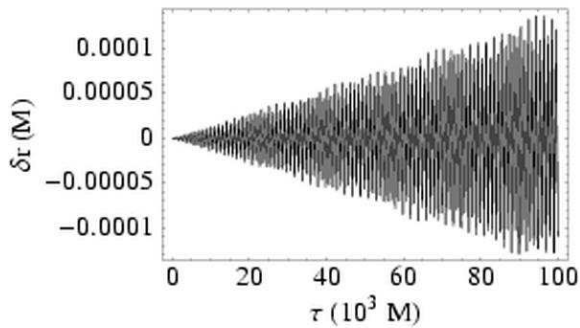}
\caption{\label{fig:longdrplot} 
The difference~$\delta r$ between the Boyer-Lindquist radii of two nearby
trajectories for a chaos mimic.  The growth in the separation is substantial,
but not exponential.  The initial conditions are the same as 
in Fig.~\ref{fig:mimic}.}
\end{figure}

\begin{figure}
\includegraphics[width=3in]{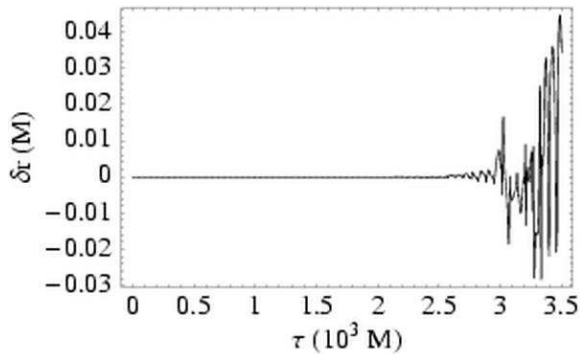}
\caption{\label{fig:drplot} The difference~$\delta r$ between the 
Boyer-Lindquist radii of two nearby
trajectories for a chaotic orbit.  The initial condition is taken from the inner
region of Fig.~\ref{fig:e=0.5_S=1} ($r_{p,\,K} = 2.0$, $e_K= 0.5$, $\iota_K =
10^\circ$, $S=1$).
On this linear scale, the separation seems to
burst unexpectedly, in contrast to the
relatively smooth
linear growth for the nonchaotic orbit shown in Fig~\ref{fig:longdrplot}.}
\end{figure}

One important refinement to the technique described above is to require several
90\% saturation points \emph{in a row} before declaring the orbit to be
chaotic.  This is necessary because some nonchaotic orbits have very high
amplitudes on plots of $\log{[r_e(\tau)]}$ vs.\ $\tau$ (Fig.~\ref{fig:mimic}).
This phenomenon occurs mainly for orbits with many periods in the deeply
relativistic zone near the horizon.  Such orbits may reach ``90\% saturation''
briefly as part of their oscillation, but quickly return to separations far
below our chaotic cutoff.  We therefore adopt the criterion of \emph{three 90\%
saturation points in a row} (with a time~$T=100\,M$ sampling interval) as a
robust practical test for chaos. See Sec.~\ref{sec:chaos_mimics} for further
discussion.

Our confidence of this method's robustness derives from comparing the method
above to the Jacobian method for the same initial conditions.  Since the
Jacobian method does not saturate, the agreement of the two methods indicates
that our procedure provides an accurate detector for chaos (as in
Fig.~\ref{fig:lyap_long_e=0.6} below).  

\subsection{Implementation notes}

We integrate the Papapetrou equations on a computer using Bulirsch-Stoer and
Runge-Kutta integrators implemented in the~C programming language, as described
in~\cite{Hartl_2002_1}.  The derivatives and Jacobian matrix are extensively
hand-optimized for speed.  We monitor errors using constraints and conserved
quantities, with a global error goal of $10^{-13}$.   The errors are at the
$10^{-11}$ level or better for highly chaotic orbits after $10^5\,M$.  Orbits
with low spin or high pericenter are even more accurate, often achieving the
error goal of $10^{-13}$.

The many plots in Sec.~\ref{sec:results} are typically generated using driver
programs written in the Perl programming language, which in turns calls the~C
integrator repeatedly.  This general paradigm---using an interpreted language
such as Perl to call optimized routines in a compiled language such as C---is
one we warmly recommend.

\section{Results}
\label{sec:results}

\begin{figure*}
\begin{tabular}{ccccc}
\includegraphics[height=2in]{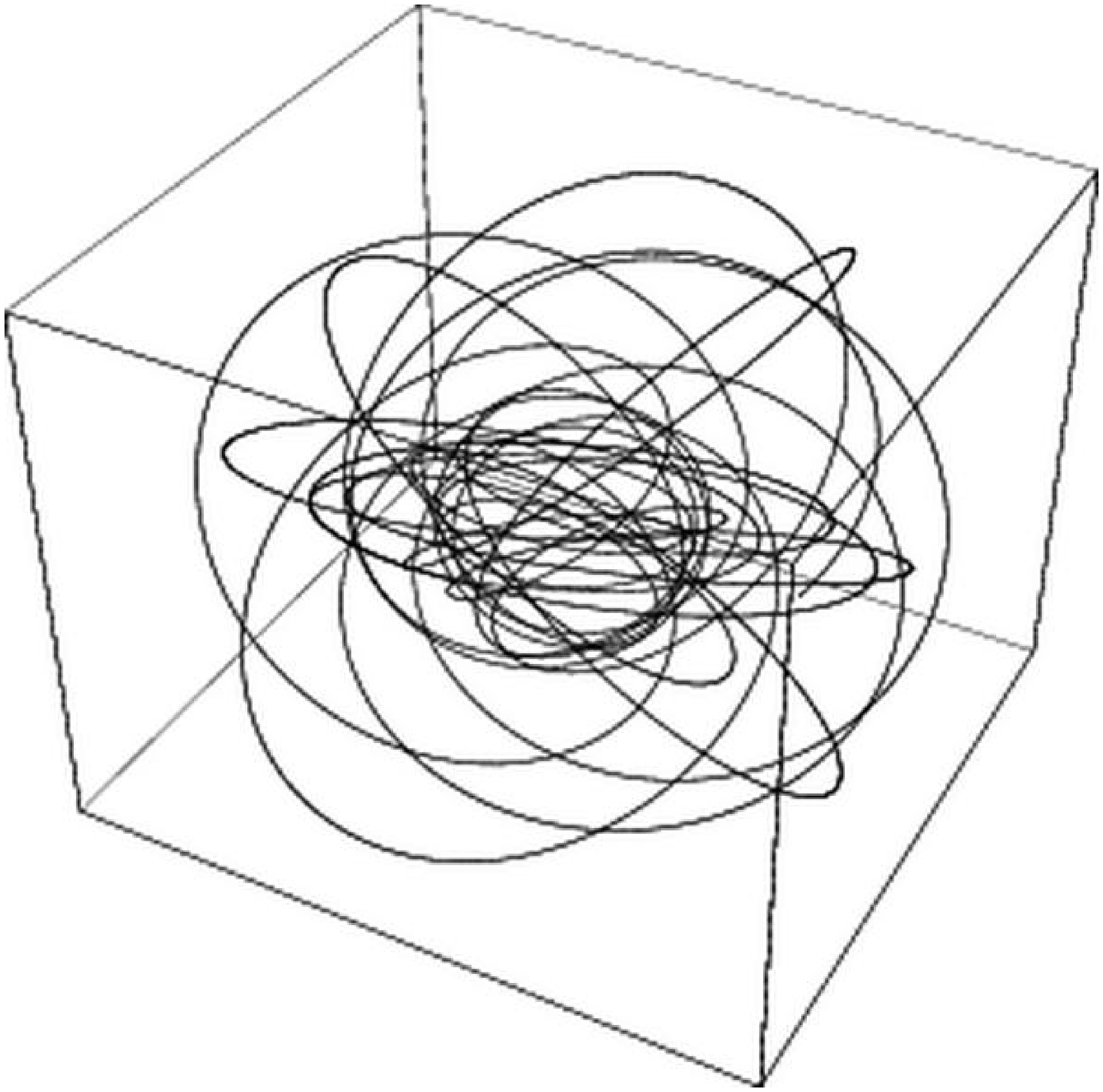}
    & \hspace{0.25in}
	& \includegraphics[height=2in]{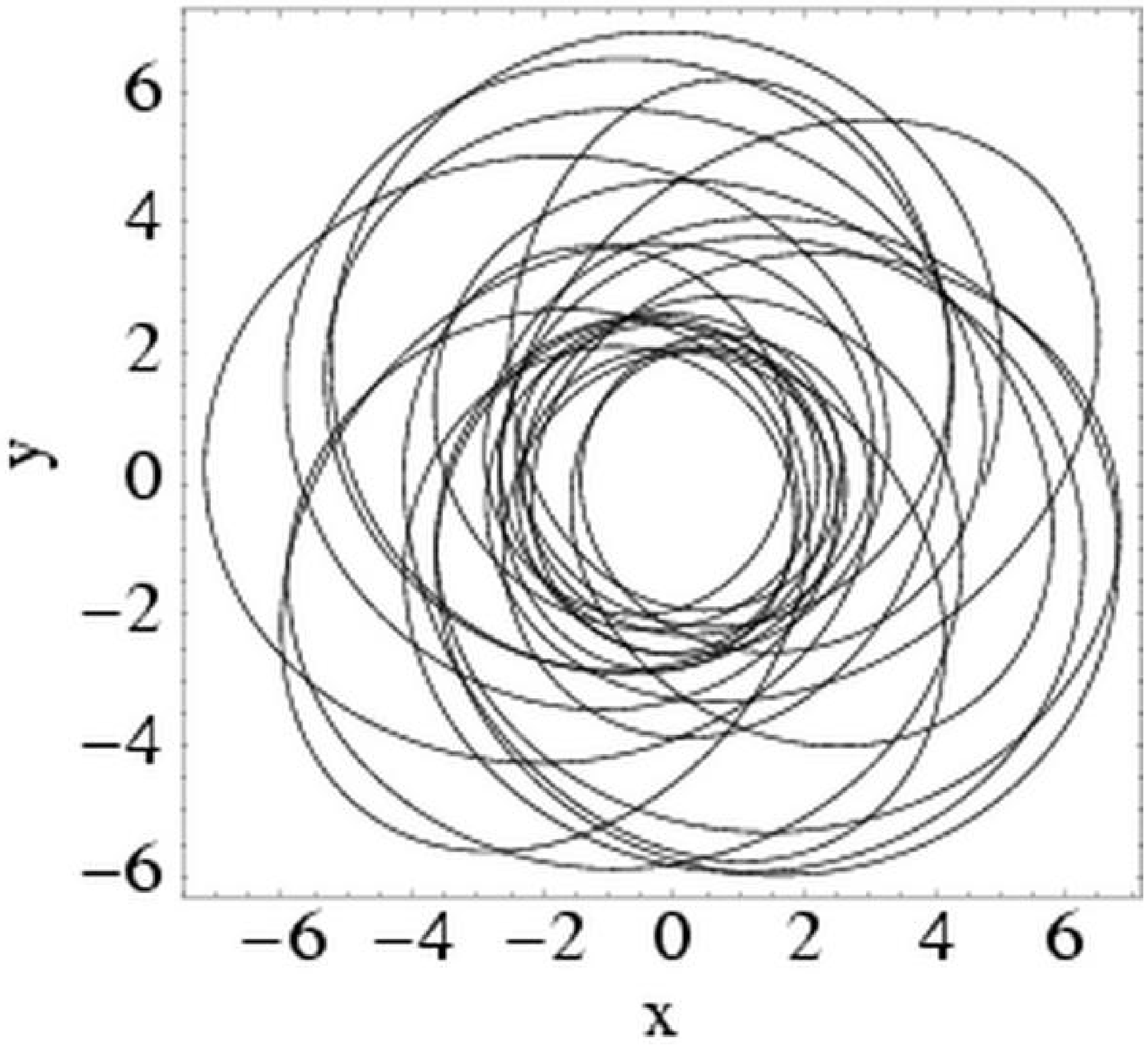}
    & \hspace{0.25in}
    & \includegraphics[height=2in]{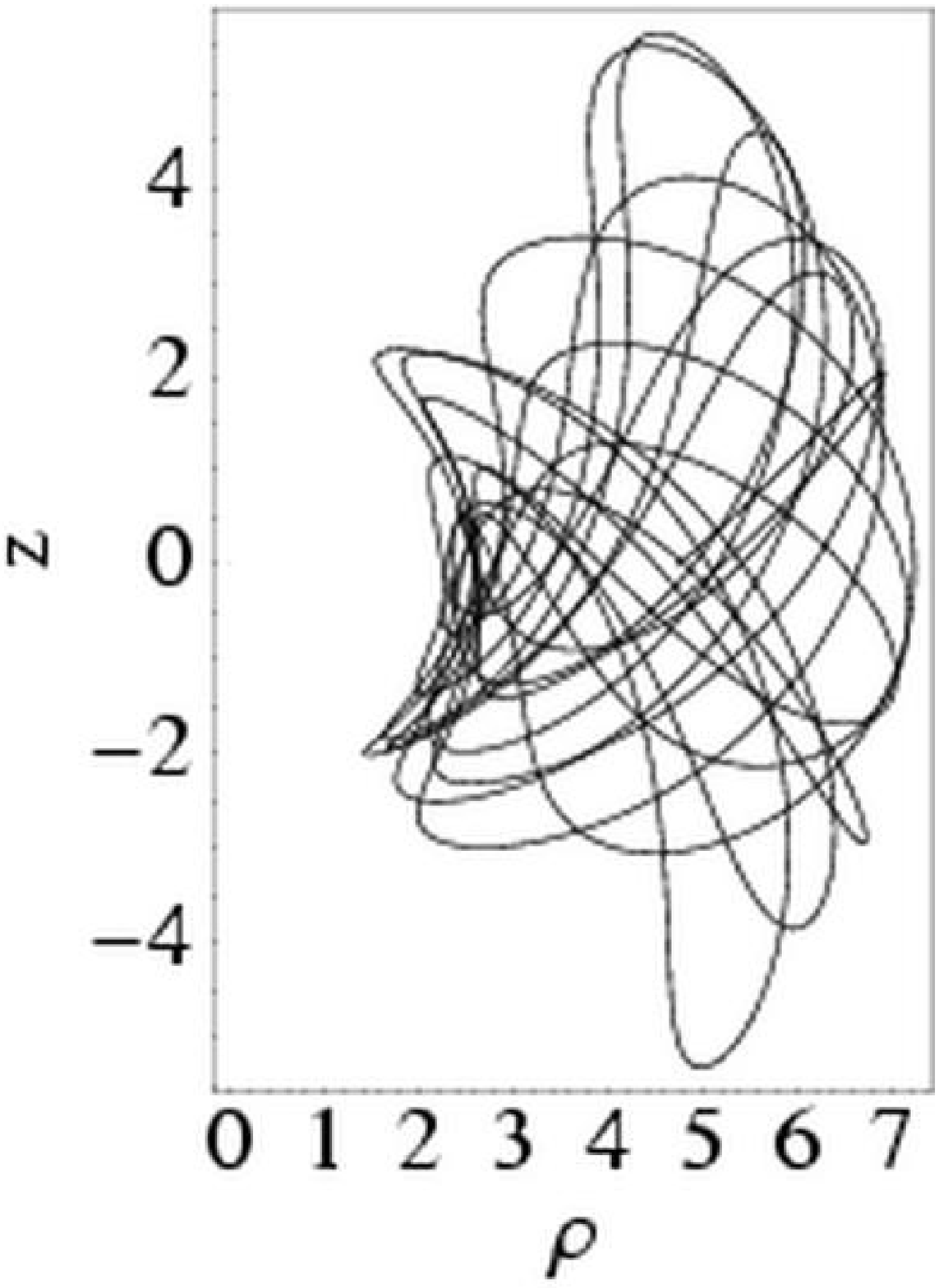}\\
(a) & & (b) & & (c)\medskip\\
\end{tabular}
\caption{\label{fig:spin_orbit}
The orbit of a maximally spinning~($S=1$)
test particle in maximal ($a=1$) Kerr spacetime, 
plotted in Boyer-Lindquist coordinates.
(a) The orbit embedded in three-dimensional space, treating the Boyer-Lindquist
coordinates as ordinary spherical polar coordinates;
(b) $y=r\sin\theta\sin\phi$ vs.\ $x=r\sin\theta\cos\phi$; 
(c) $z$ vs.\ $\rho=\sqrt{x^2+y^2}$.  The requested orbital inclination angle
is $\iota=6^\circ$, but the strong spin coupling gives rise to an empirical
value closer to $\iota_P=46^\circ$. The empirical pericenter is
$r_p=2.219\,M$, which is fairly close to the requested value of $2.367\,M$.}
\end{figure*}

\begin{figure*}
\begin{tabular}{ccl}
\includegraphics[width=3in]{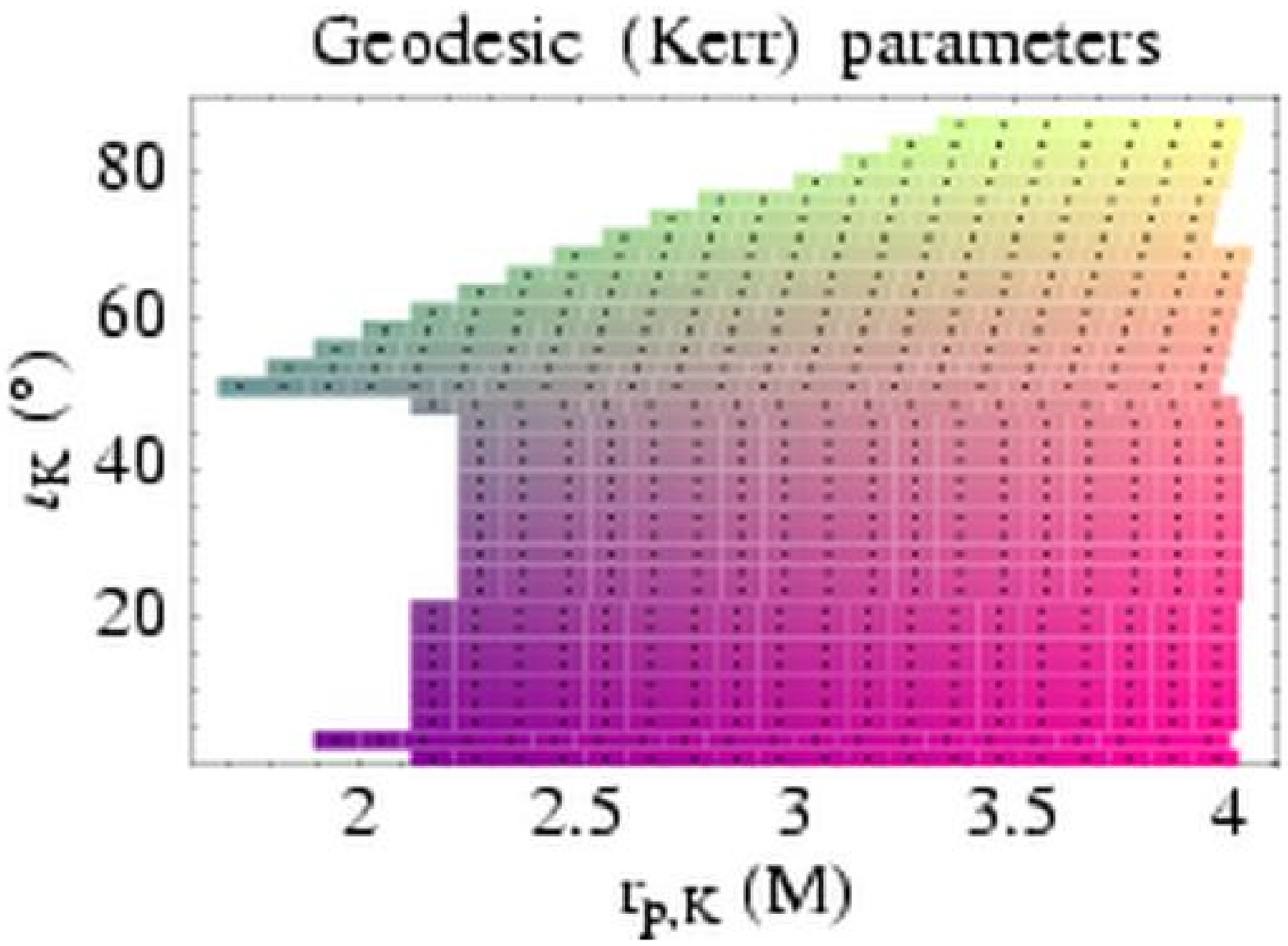}
	& \includegraphics[width=3in]{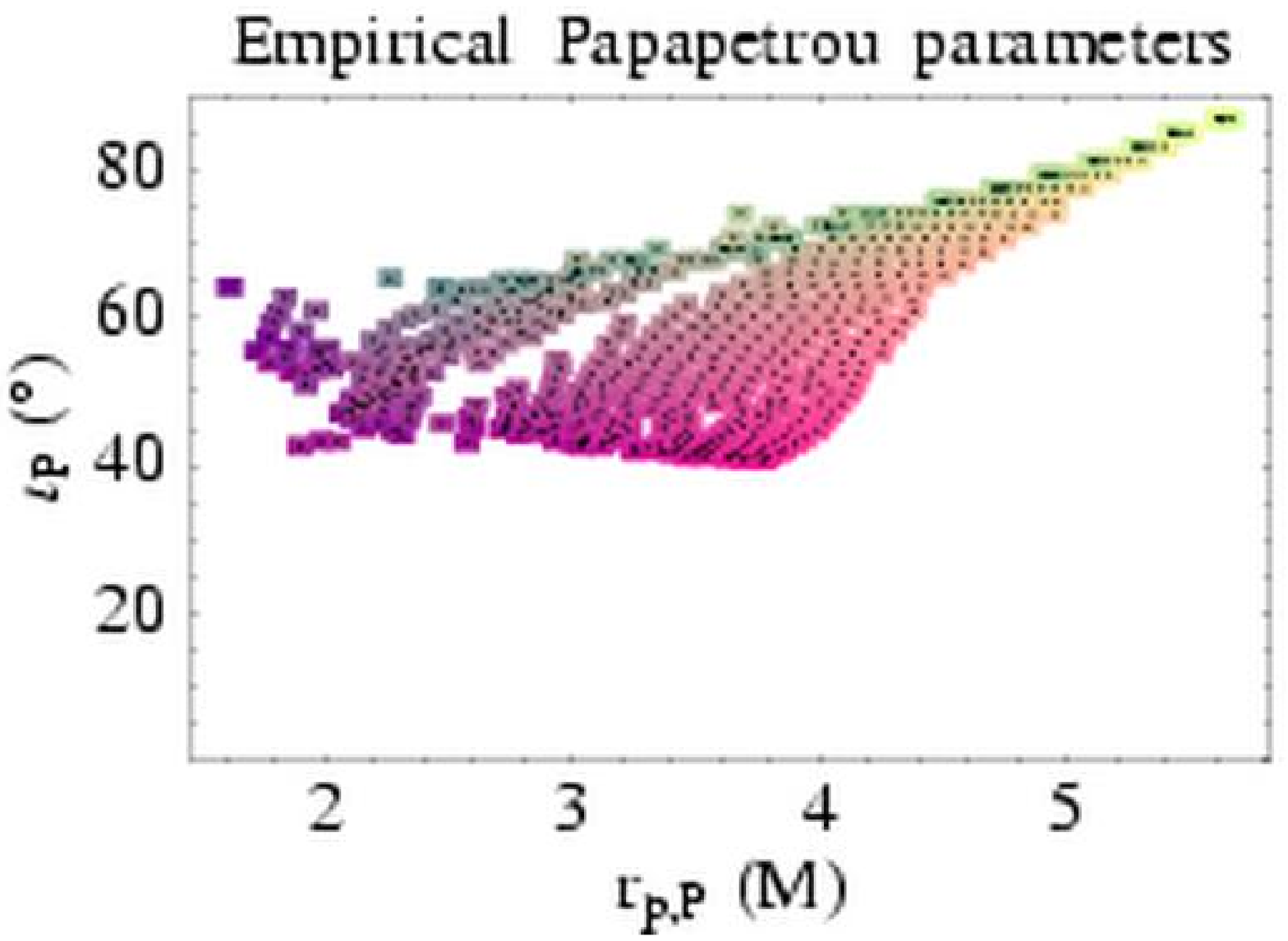}\\
(a) & (b)\medskip\\
\end{tabular}
\caption{\label{fig:color_map} Shading/coloring 
of parameter space for an $S=1$ orbit.  The shading/color on the left is
repeated on the right, so that we can visually determine the mapping of
$(r_{p,\,K},\iota_K)$ to
$(r_{p,\,P},\iota_P)$.
It is evident that orbits with low
requested pericenters and orbital inclination angles are mapped to
low-pericenter orbits at high inclinations, and the entire parameter space is
compressed.  Not that the gap in stable initial conditions visible in~(b) 
is a true gap, not a fold.}
\end{figure*}

\begin{figure*}
\begin{tabular}{ccl}
\includegraphics[width=3in]{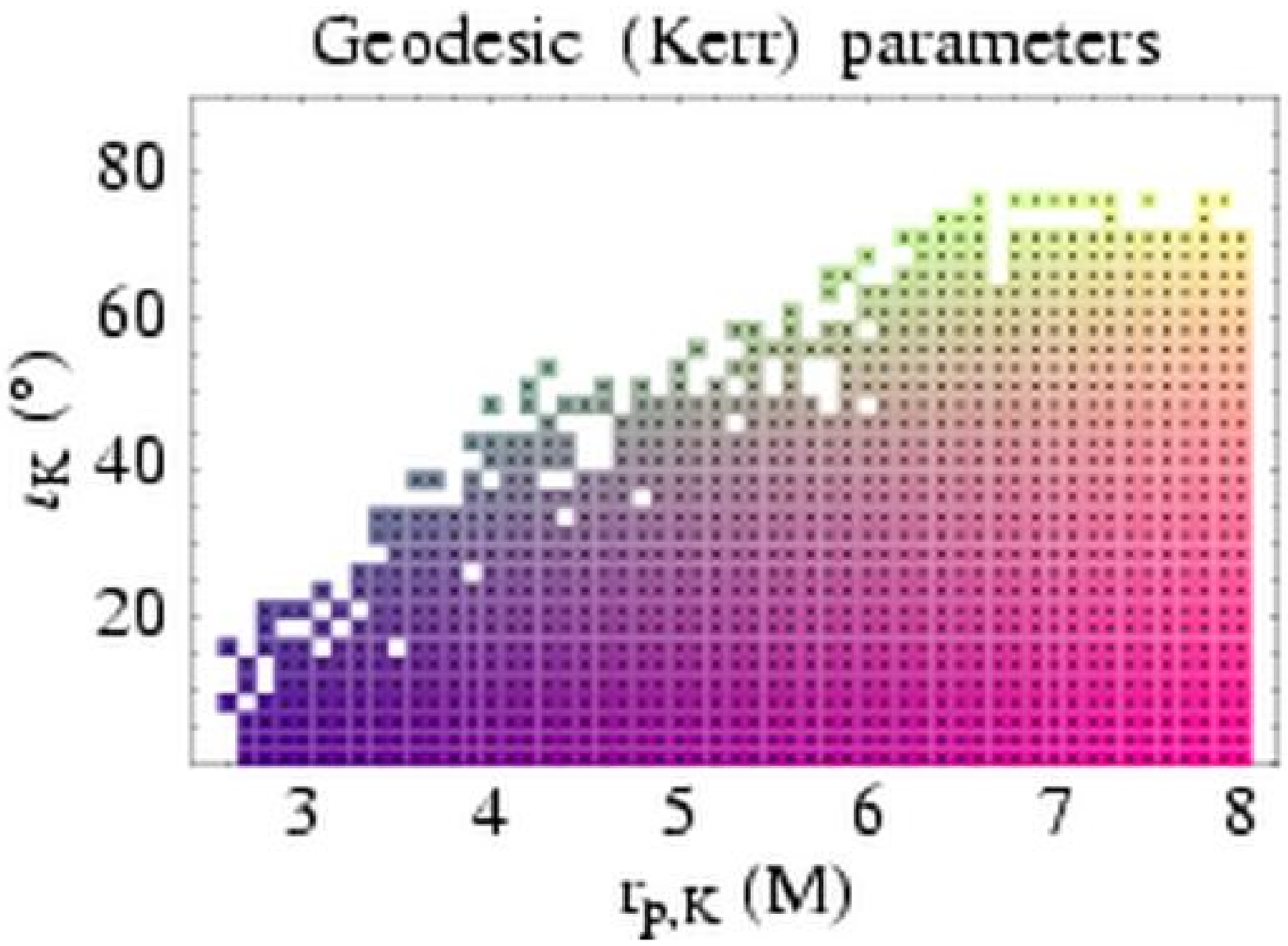}
	& \includegraphics[width=3in]{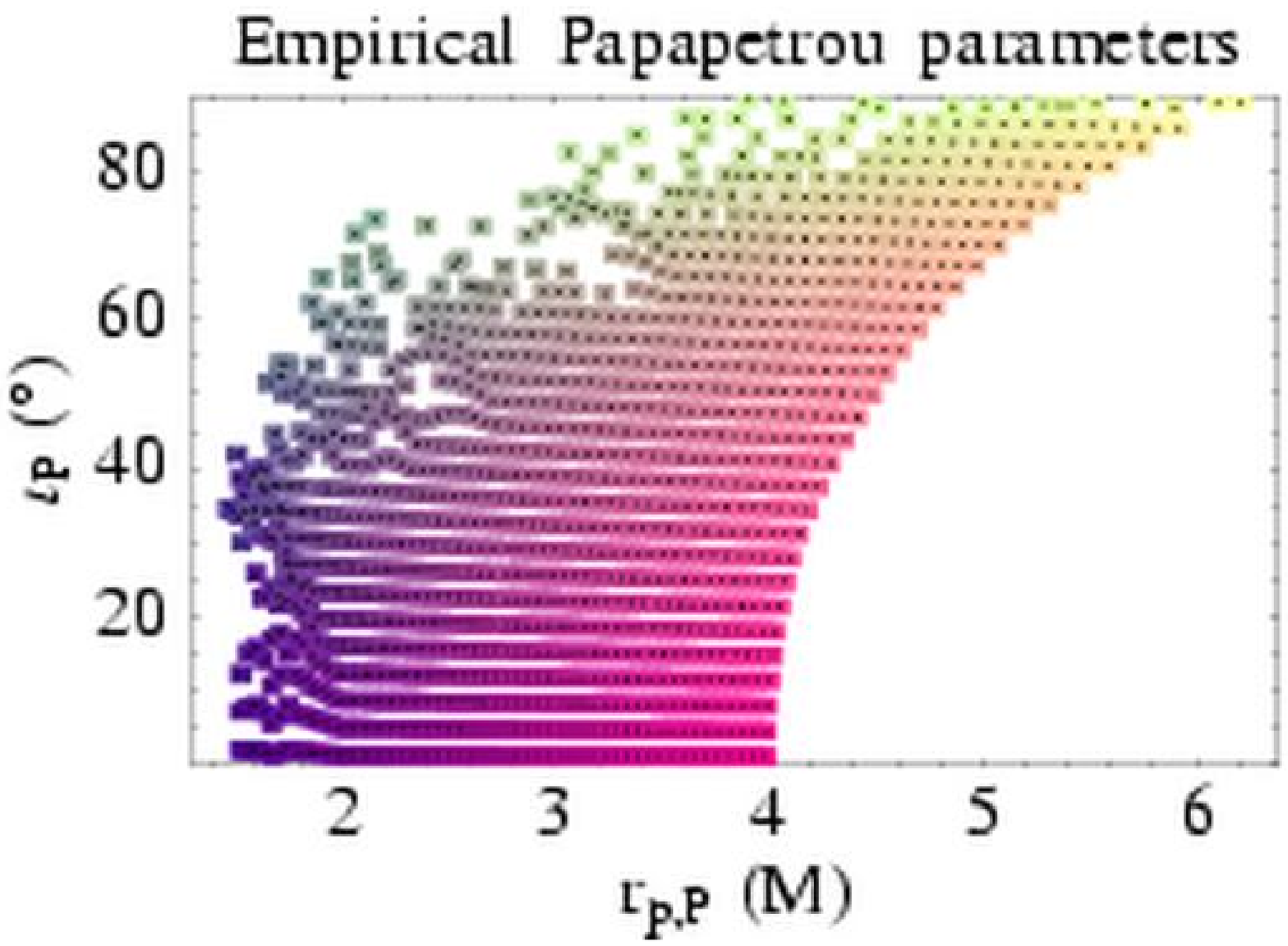}\\
(a) & (b)\medskip\\
\end{tabular}
\caption{\label{fig:color_map_alt} Shading/coloring 
of parameter space for an $S=1$ orbit
illustrating the alternate parameterization method from 
Sec.~\ref{sec:papa_param}.  
As in Fig.~\ref{fig:color_map}, the shading/color on the left is
repeated on the right.  The spatial part of the initial spin is purely in
the~$z$ direction.
The Kerr $p^\theta_K$ and Papapetrou $p^\theta_P$ values are forced to 
agree, which leads to
similar inclination angles in parts (a) and (b), at the cost of dramatically
different pericenters.}
\end{figure*}

We present here the results of parameter variation in the Papapetrou system of
equations.  We represent the effects compactly using several different kinds of
plots, most of which involve plotting inclination vs.\ pericenter, with other
parameters held fixed.  We refer to these as $r_p$-$\iota$ maps.   As discussed
in Sec.~\ref{sec:param}, starting with the Kerr values $r_{p,\,K}$, $\iota_K$,
and $e_K$, we find the corresponding energy~$E_K$, angular momentum $L_{z,K}$,
and Carter constant~$Q_K$, and then force the Papapetrou values to satisfy $E_P
= E_K$, $J_{z,P} = L_{z,K}$ and $p^r_P = p^r_K$. Each $r_p$-$\iota$ map has two
components: part~(a), shown on the left, uses the Kerr parameters $\iota_K$ and
$r_{p,\,K}$ requested by the parameterization (Sec.~\ref{sec:papa_param}), while
part~(b), shown on the right, always shows the empirical Papapetrou values
$\iota_P$ and $r_{p,\,P}$ in the sense of Sec.~\ref{sec:empirical_orbit}.   

One important feature of $r_p$-$\iota$ parameter space apparent in the
empirical plots is the prevalence of large empirical inclination angles for all
values of the Kerr inclination~$\iota_K$.  Fig.~\ref{fig:color_map} shows the
mapping for $S=1$ between the requested Kerr parameters
[Fig.~\ref{fig:color_map}(a)] and the empirical Papapetrou parameters
[Fig.~\ref{fig:color_map}(b)] using a shading scheme (which appears as a more
informative color scale in electronic versions of this paper).  We see that
even the orbits at the bottom of  Fig.~\ref{fig:color_map}(a) get mapped to
high empirical inclination angles; $\iota_K=1^\circ$ orbits are mapped to
inclinations of order~$\iota_P=40^\circ$.  

This compression of parameter space is the result of our choice to force the
Kerr and Papapetrou value of the radial momentum to agree
(Sec.~\ref{sec:papa_param}).  The price we pay for this choice is that the
$\theta$ component of the Papapetrou momentum---which is constrained to satisfy
the equations in Sec.~\ref{sec:constraints}---is relatively large.  This high
$p^\theta_P$ flings even orbits with low requested values of $\iota_K$ to high
inclinations (Fig.~\ref{fig:color_map}).  It is possible to find
low-inclination Papapetrou orbits by requiring that the Kerr and Papapetrou
values of $p^\theta$ agree, but at the cost of forcing $p^r$ to be very
different---again a result of the constraints.  The resulting parameter space
(Fig.~\ref{fig:color_map_alt}) is not nearly as compressed in inclination
angle, but the Papapetrou pericenters are compressed and shifted down, and many
requested values of $r_p,\,K$ are lost as plunge orbits.  In addition, the
empirical values of the eccentricity are typically not close to the value
requested (reaching, e.g., $e_P = 0.75$ for $e_K = 0.5$).  Because of the
deficiencies of this alternate parameterization method, we choose the
fixed~$p^r$ plots are our primary investigative tool in this paper.

\subsection{Varying pericenter and orbital inclination}
\label{sec:rp_i}

\begin{figure*}
\begin{tabular}{ccl}
\includegraphics[width=3in]{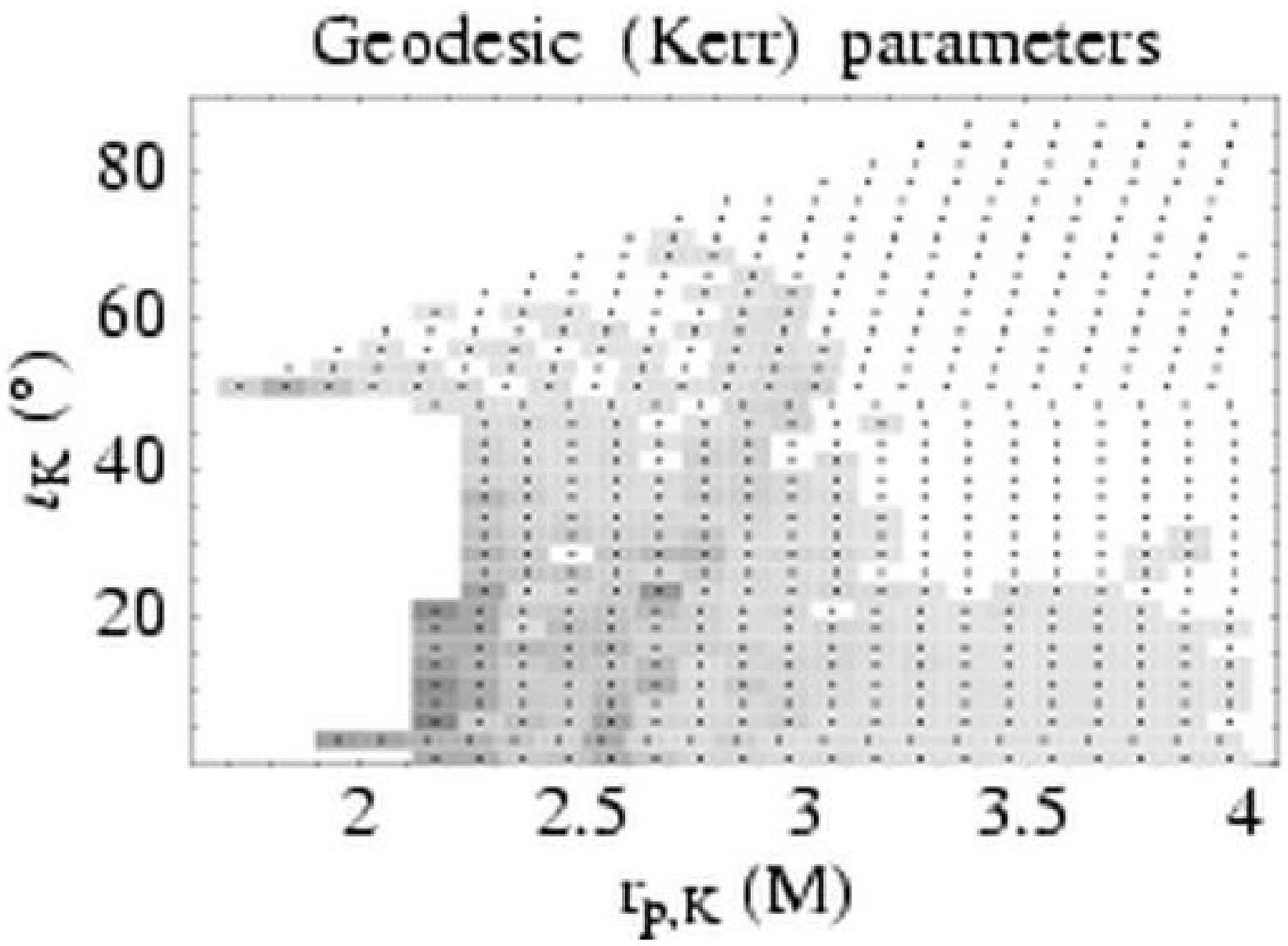}
	& \includegraphics[width=3in]{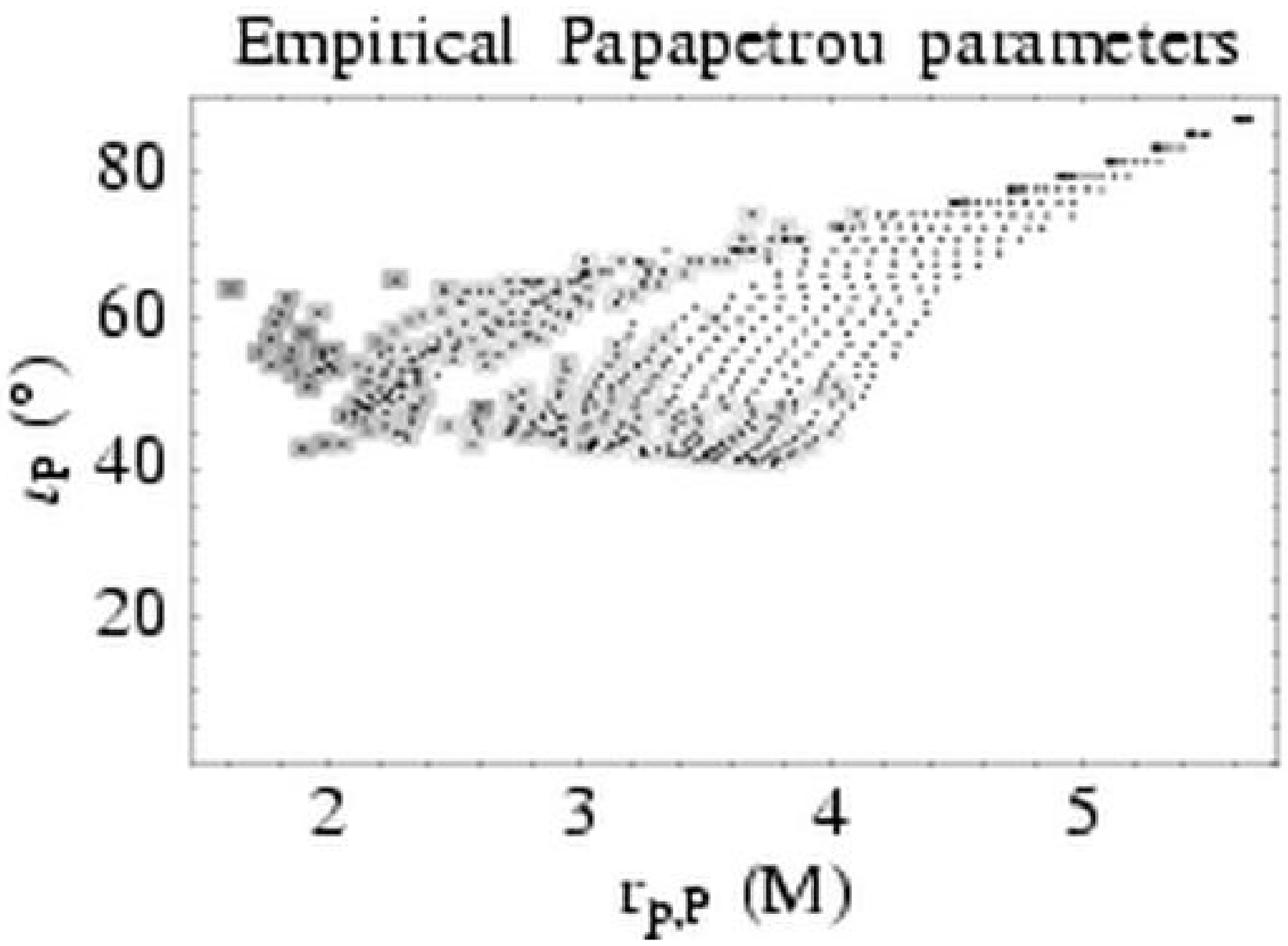}
    & \includegraphics[height=2in]{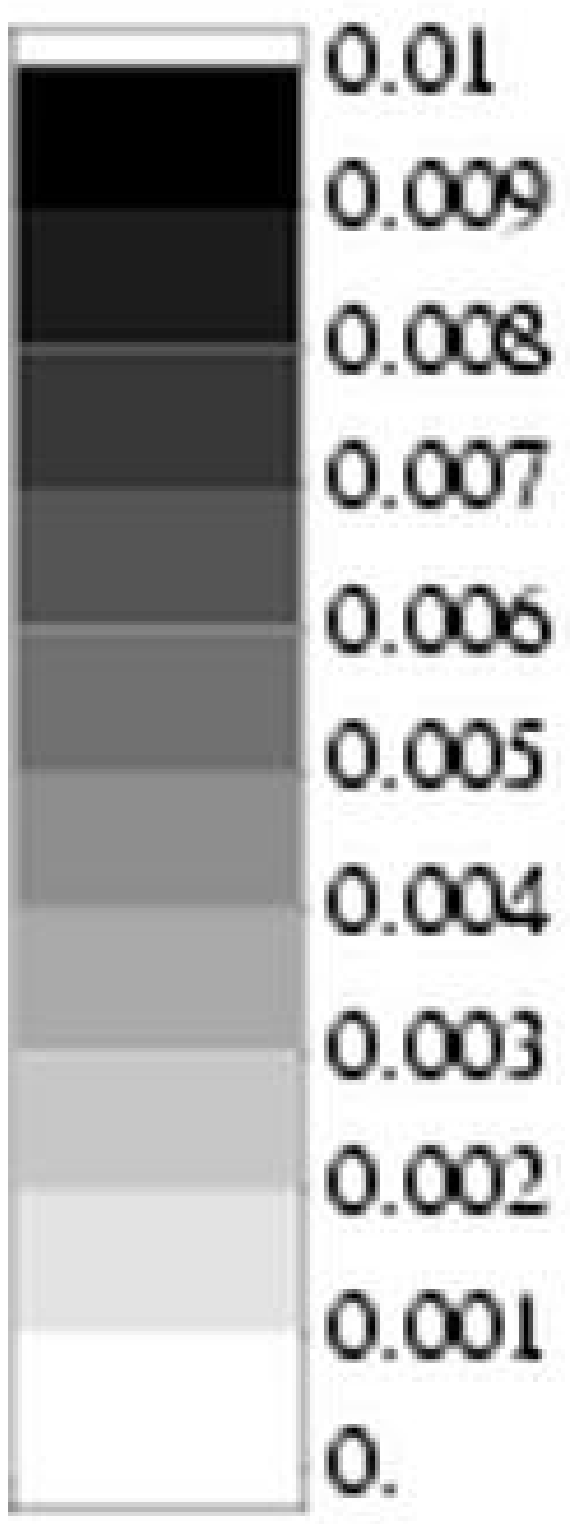}\\
(a) & (b) & \medskip\\
\end{tabular}
\caption{\label{fig:e=0.5_S=1}
$r_p$-$\iota$ map for $a=1$, $e=0.5$, $S=1$, $S^{\hat r}=S^{\hat z} = 0.2\,S$:
chaos strength as a function of pericenter and orbital inclination angle.
(a)~Requested parameters; (b)~empirical parameters.
The initial
conditions are the same as in Fig.~\ref{fig:color_map}.
The rectangles are shaded according to the Lyapunov exponent for the
initial condition represented by each point, with darker shades of gray
representing larger exponents and hence stronger chaos.  
The point at $r_{p,\,K}=2.3$ and $\iota_K=20^\circ$ is one of the points in
Fig.~\ref{fig:Sincl}, which shows the effects of varying $S^{\hat r}$ and 
$S^{\hat z}$.
The largest exponent in
this plot is $\lambda=4.1\times10^{-3}\,M^{-1}$, corresponding to a timescale of
$1/\lambda =2.4\times10^2\,M$ for a factor of $e$~divergence in nearby initial
conditions.}
\end{figure*}

\begin{figure*}
\begin{tabular}{ccl}
\includegraphics[width=3in]{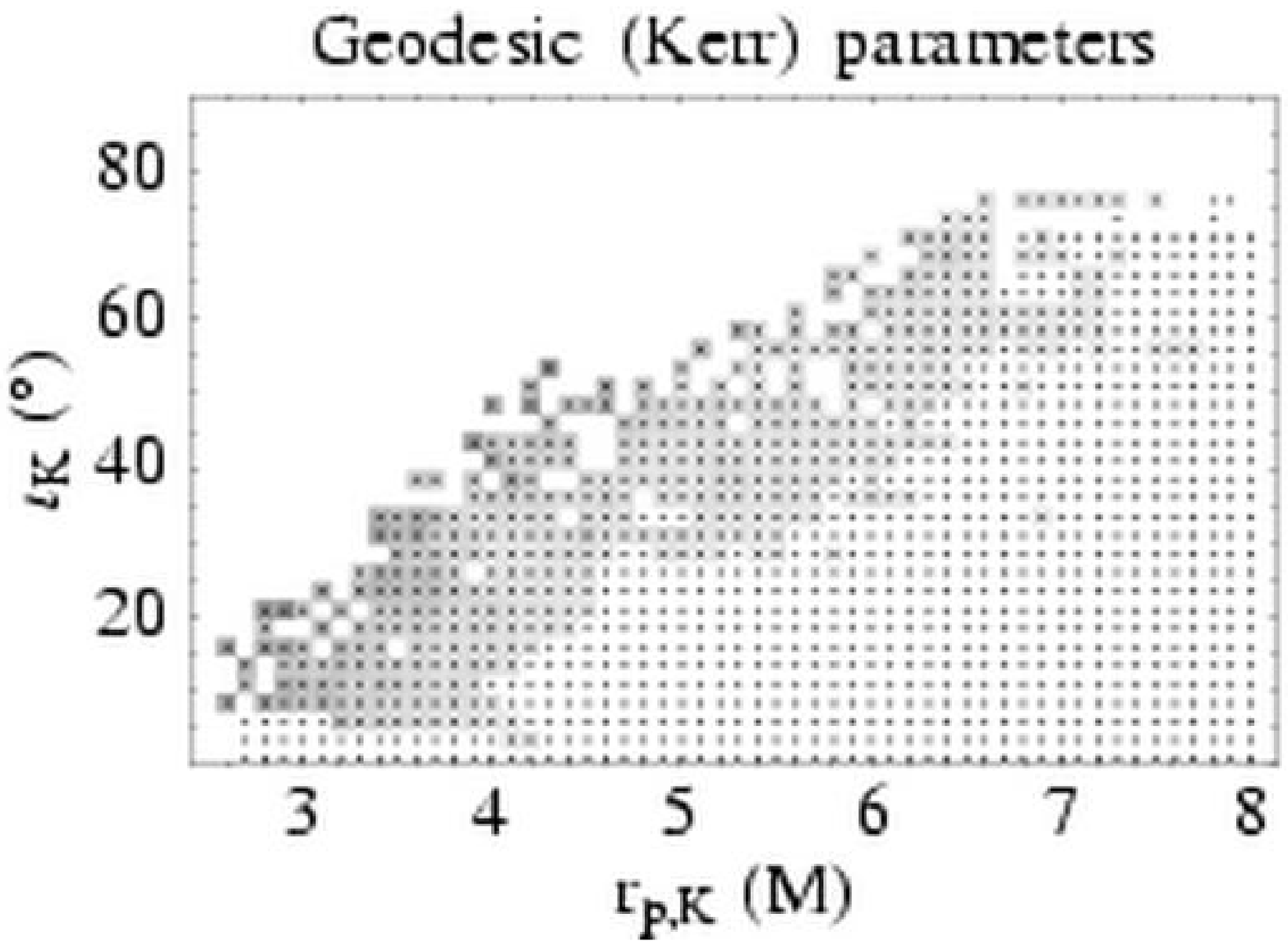}
	& \includegraphics[width=3in]{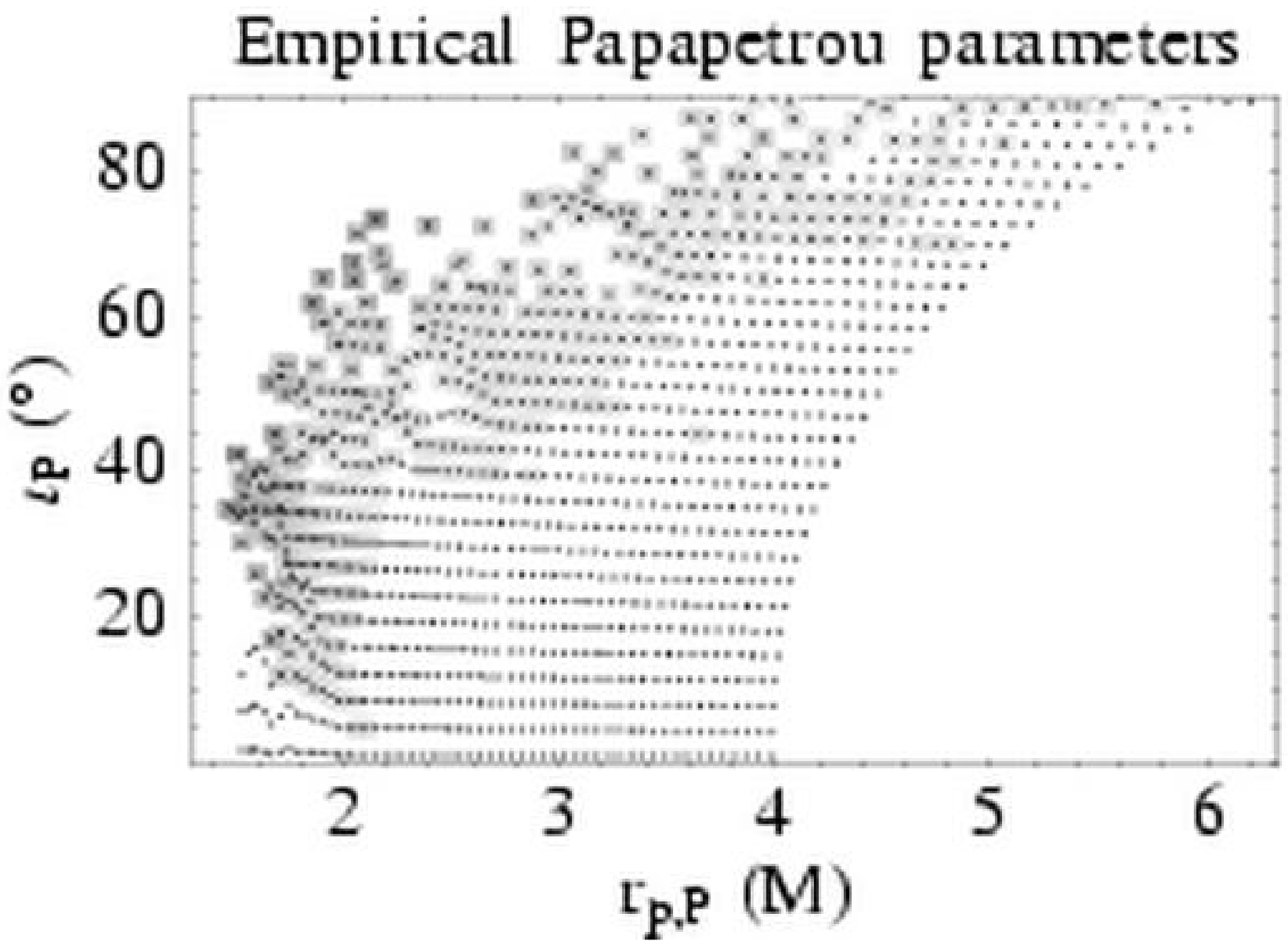}
    & \includegraphics[height=2in]{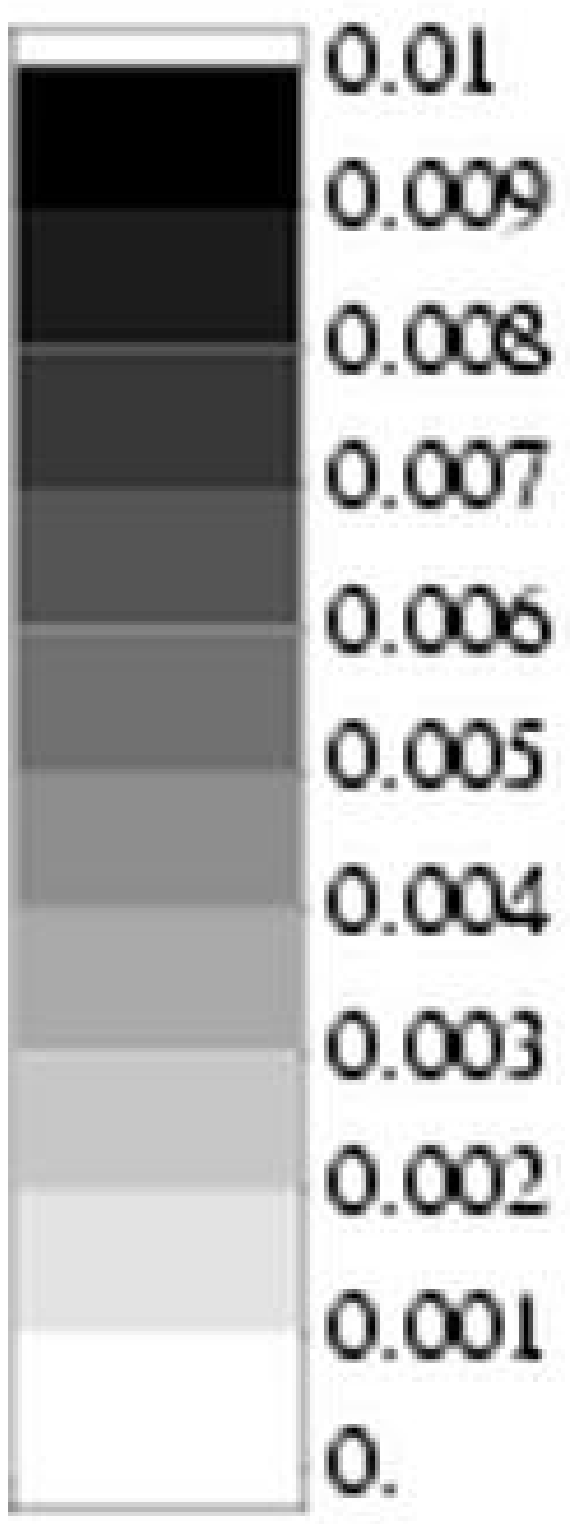}\\
(a) & (b) & \medskip\\
\end{tabular}
\caption{\label{fig:e=0.5_S=1_alt}
$r_p$-$\iota$ map for $a=1$, $e=0.5$, $S=1$, using the
alternate parameterization method that forces $p_K^\theta =
p_P^\theta$.
The initial spatial component of the
spin is purely in the $z$ direction.
(a)~Requested parameters; (b)~empirical parameters.
The initial
conditions are the same as in Fig.~\ref{fig:color_map_alt}.
The rectangles are shaded according to the Lyapunov exponent for the
initial condition represented by each point, with darker shades of gray
representing larger exponents and hence stronger chaos.  
The largest exponent in
this plot is $\lambda=4.0\times10^{-3}\,M^{-1}$, corresponding to a timescale of
$1/\lambda =2.5\times10^2\,M$ for a factor of $e$~divergence in nearby initial
conditions.  These values are virtually identical to the values in 
Fig.~\ref{fig:e=0.5_S=1}.}
\end{figure*}

In this section we show $r_p$-$\iota$ maps for orbits with fixed
eccentricity~$e=0.5$ and spin parameters $S^{\hat r} = S^{\hat z} = 0.2\,S$,
where $S$ is the total spin.\footnote{These ``fixed'' spin parameters give rise
to a variety of spin inclination angles in the fiducial (ZAMO) rest frame; see
Sec.~\ref{sec:Sincl} below.}  We indicate the strength of the chaos at each
point with a grayscale rectangle, with the darkest colors representing the
strongest chaos (and with white indicating no chaos). An example of such a plot
appears in Fig.~\ref{fig:e=0.5_S=1}. (Fig.~\ref{fig:e=0.5_S=1_alt} shows a
similar plot for the alternate parameterization method that forces  $p_K^\theta
= p_P^\theta$.  The maximum exponents in the two cases are virtually
identical.) The most important general results evident from the plots are
two-fold.  First, the largest exponents occur for orbits with pericenters deep
in the relativistic zone near the horizon.  Second, the prevalence of chaotic
orbits is a decreasing function of spin parameter~$S$, with virtually all chaos
gone by the time $S=0.1$.

An example of the value of the empirical $r_{p,\,P}$-$\iota_P$ plots appears in
Fig.~\ref{fig:e=0.5_S=0.5}, which shows orbits of particles with spin~$S=0.5$. 
The appearance of a strongly chaotic point in Fig.~\ref{fig:e=0.5_S=0.5} seems
perplexing, given that it is surrounded but many points with much smaller
exponents.   As is evident from the empirical plot, this point of strong chaos
(which is, in fact, the largest Lyapunov exponent for any of the plots) maps to
a compressed area of initial conditions with small empirical pericenters
[Fig.~\ref{fig:e=0.5_S=0.5}(b)].

From an astrophysical standpoint, the most interesting parameter to vary is the
spin~$S$, since only $S\ll1$ orbits are physically realistic
(Sec.~\ref{sec:spin_param}).  From 
Figs.~\ref{fig:e=0.5_S=0.9}--\ref{fig:e=0.5_S=1e-4}, which involve varying $S$
from 0.9 down to $10^{-4}$, we see that both the prevalence and strength of
chaos decrease significantly as $S$ is decreased.  The Lyapunov exponent ranges
as high as $10^{-2}\,M^{-1}$ when $S=0.5$ (Fig.~\ref{fig:e=0.5_S=0.5}), but the
chaos is weak when  $S=0.2$ (Fig.~\ref{fig:e=0.5_S=0.2}) and is gone below
$S=0.1$ (Figs.~\ref{fig:e=0.5_S=0.1} and~\ref{fig:e=0.5_S=1e-4}).

\subsection{Varying eccentricity}
\label{sec:e}

The choice of $e=0.5$ in the previous section is partially motivated by likely
gravitational wave sources for LISA~\cite{LISAweb}, e.g., a neutron star or
small black hole in an eccentric orbit around a supermassive black hole.  In
this section, we consider a second series of eccentric orbits at fixed $e=0.6$
and varying spin parameter.  We also investigate the case of a near-circular
orbit ($e=0.01$) more appropriate for the circularized gravitational wave
sources important for ground-based detectors such as LIGO.

\begin{figure}
\includegraphics[width=3in]{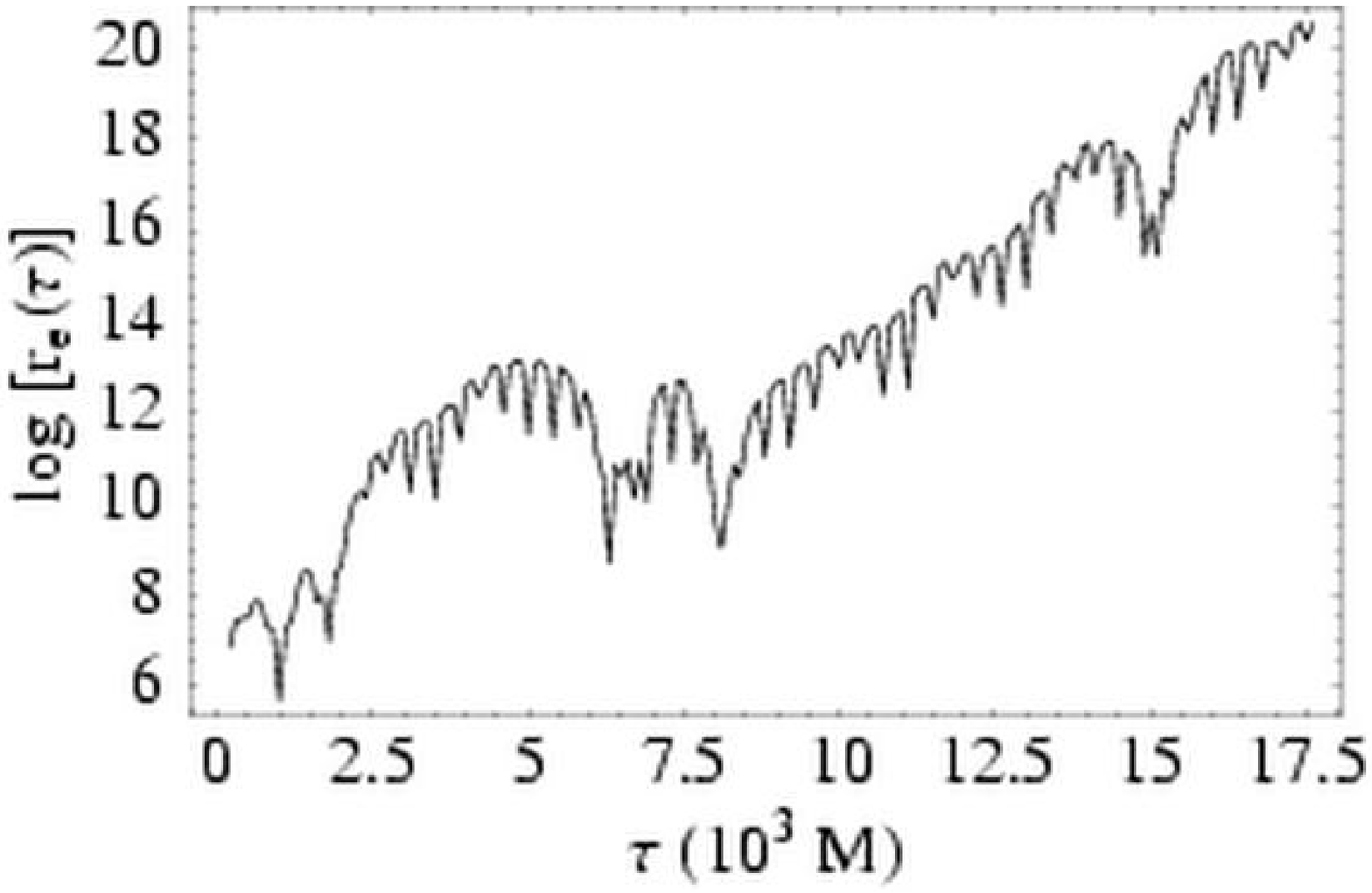}
\caption{\label{fig:lyap_e=0.6}
Natural logarithm of the largest ellipsoid axis vs.\ time.  We use the
unrescaled deviation vector method to investigate the single chaotic initial
condition from Fig.~\ref{fig:e=0.6_S=0.1}, which has $r_{p,\,K}=1.21$, 
$e_k=0.6$, $\iota_K=31^\circ$, and
$S=0.1$. The principal ellipsoid axis grows until it saturates at 
$\tau=17600\,M$.}
\end{figure}

\begin{figure}
\includegraphics[width=3in]{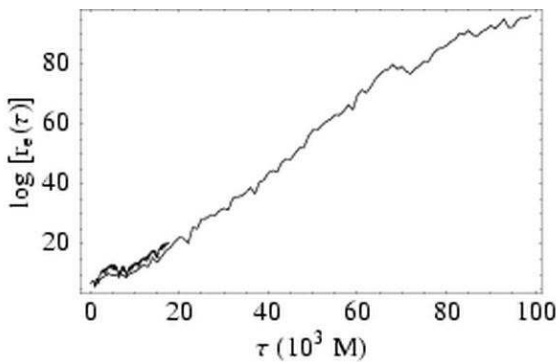}
\caption{\label{fig:lyap_long_e=0.6}
Natural logarithm of the largest ellipsoid axis vs.\ time using the unrescaled
deviation vector method and the rigorous Jacobian method.  The unrescaled
integration is identical to that shown in Fig.~\ref{fig:lyap_e=0.6}. 
The two methods agree
until the deviation method saturates, at which point we stop the deviated vector
integration.  The
continued growth of the Jacobian method confirms that the orbit determined by
the initial condition is
indeed chaotic.}
\end{figure}

The $r_p$-$\iota$ plots for $e=0.6$ follow the same pattern as those with
$e=0.5$.  Chaos is strongest for orbits with small pericenters and values 
of~$S$ of order unity (Figs.~\ref{fig:e=0.6_S=1}--\ref{fig:e=0.6_S=1e-4}). 
There is a single orbit at $S=0.1$ that appears to be chaotic
(Fig.~\ref{fig:e=0.6_S=0.1}), but other than this one exception there is
apparently no chaos below $S=0.1$.  A close examination of the single $S=0.1$
chaotic orbit appears in Figs.~\ref{fig:lyap_e=0.6}--\ref{fig:lyap_long_e=0.6},
which shows that the chaos is real.

\begin{figure*}
\begin{tabular}{ccl}
\includegraphics[width=3in]{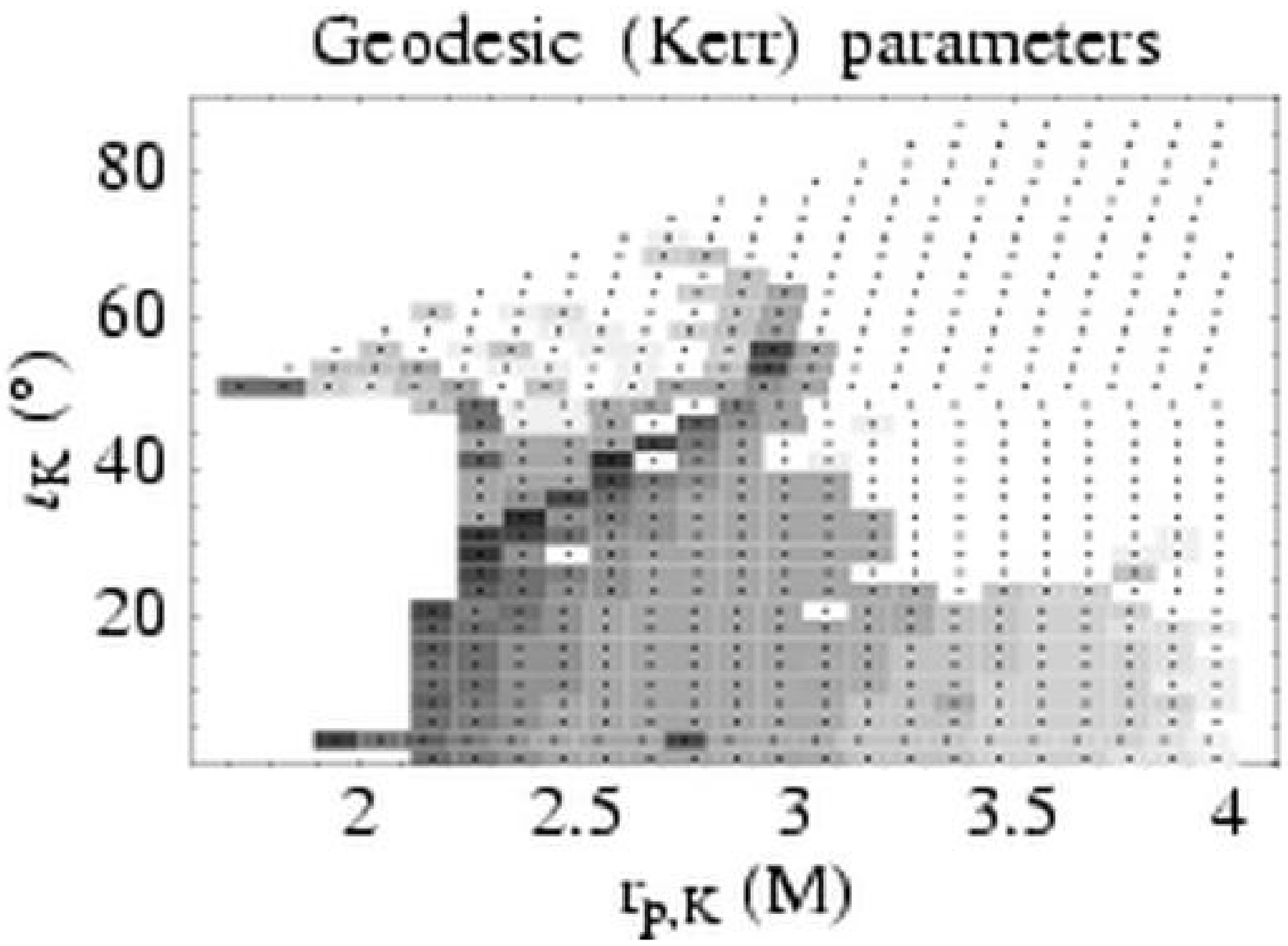}
	& \includegraphics[width=3in]{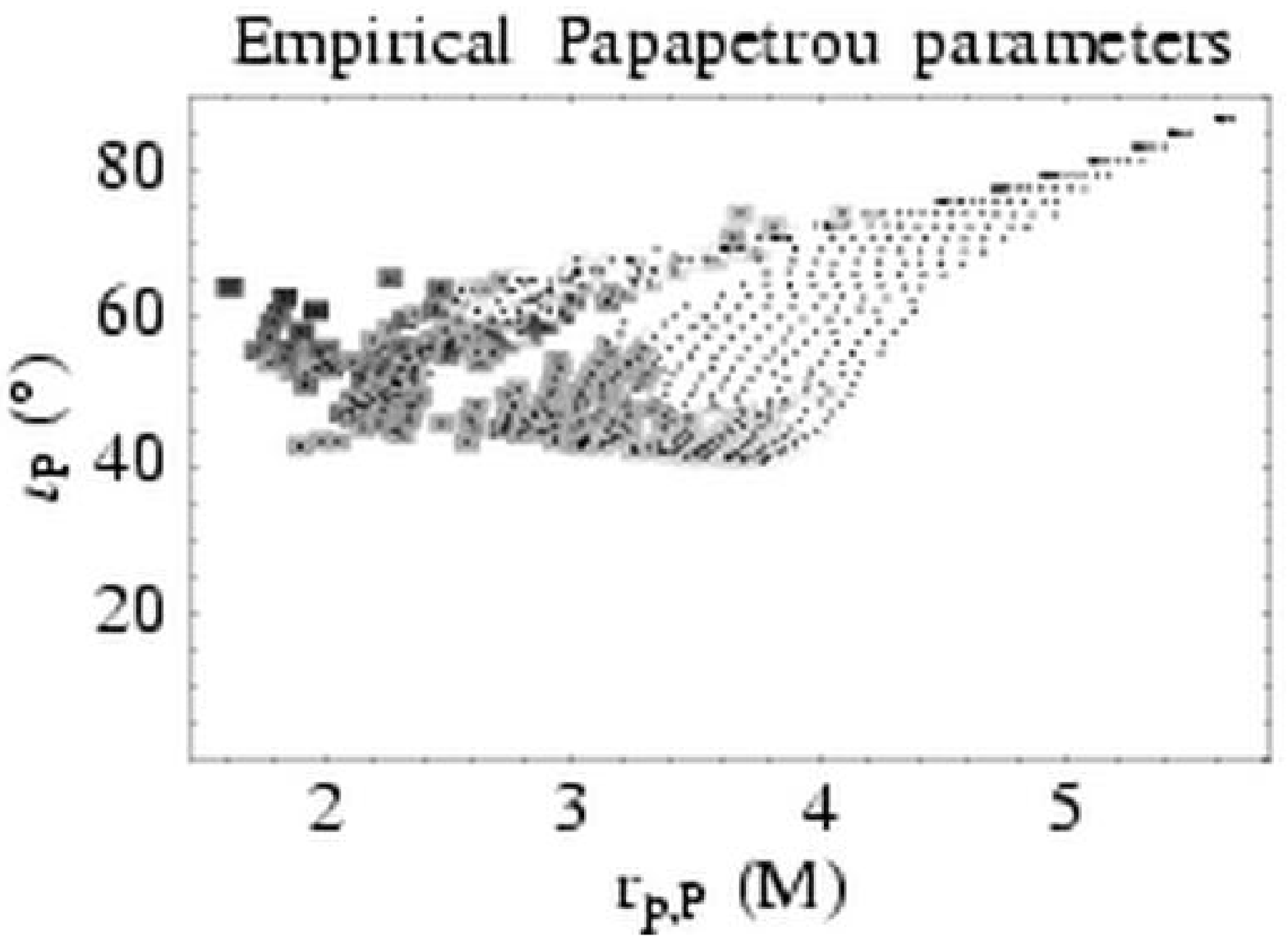}
    & \includegraphics[height=2in]{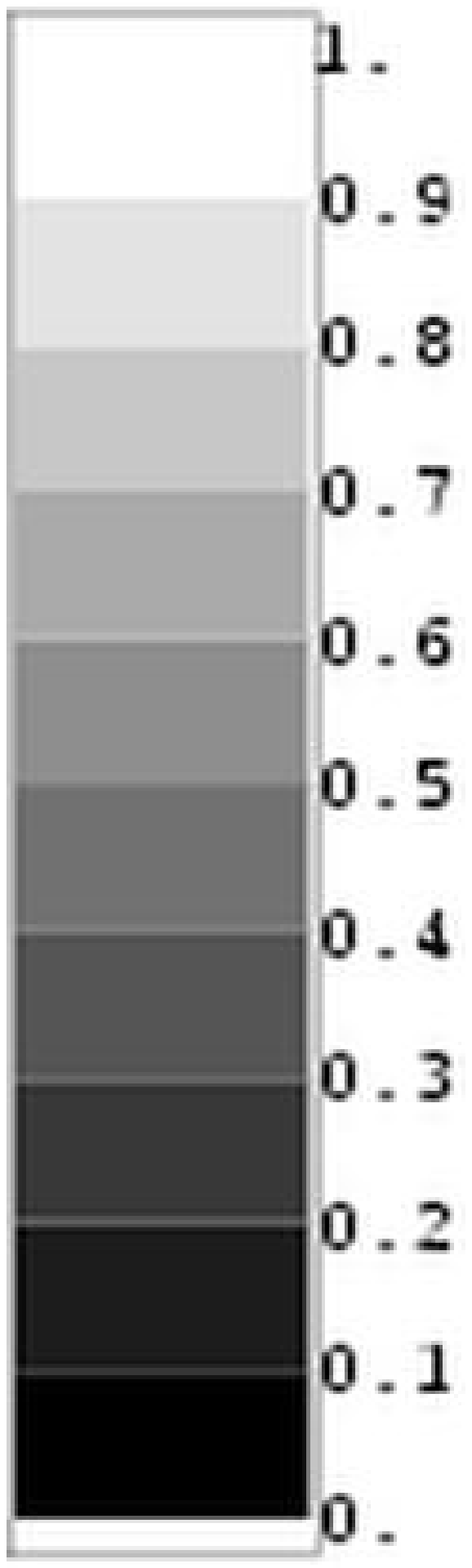}\\
(a) & (b) & \medskip\\
\end{tabular}
\caption{\label{fig:e=0.5_cutoff} Spin cutoff map for $a=1$ and $e=0.5$.  The
rectangles are shaded according to the minimum value of $S$ for which the
corresponding initial condition is still chaotic.  White rectangles indicate
points that are not chaotic even when $S=1$.  The darkest points correspond to a
cutoff value of $S=0.18457$; below this critical value, 
none of the initial conditions are chaotic.}
\end{figure*}

The effect of chaos is smaller for the near-circular ($e=0.01$) orbits
considered (Figs.~\ref{fig:e=0.01_S=1}--\ref{fig:e=0.01_S=1e-4}), with typical
Lyapunov exponents of order $2\times10^{-3}\,M^{-1}$ when $S=1$
(Fig~\ref{fig:e=0.01_S=1}).   Moreover, we find only four points with nonzero
exponents for $S=0.5$ (Fig.~\ref{fig:e=0.01_S=0.5}), in contrast to the more
eccentric orbits, which have strong chaos for $S=0.5$.  By the time $S=0.1$
(Fig.~\ref{fig:e=0.01_S=0.1}), all chaos has completely disappeared for the
near-circular orbits.

\subsection{Varying Kerr parameter $a$}
\label{sec:a}

Here we investigate the effect of the Kerr parameter~$a$ on the strength and
prevalence of chaos.  Although Suzuki \& Maeda found in~\cite{SuzukiMaeda1997}
that there is chaos even in Schwarzschild spacetime, to which Kerr spacetime
reduces when $a=0$, the chaotic orbits found in~\cite{SuzukiMaeda1997} are
exceptional orbits lying on the edge of a generalized effective potential. We
found evidence in~\cite{Hartl_2002_1} that such chaotic orbits are rare.  

The conclusion that chaotic orbits become less prevalent as $a\rightarrow0$ is
supported by a more thorough examination, as illustrated in
Figs.~\ref{fig:a=0.9}--\ref{fig:a=0}. All of these orbits have
eccentricity~$e=0.5$ and spin $S=1$, and $a$ varies from~$0.9$ down to~0. Even
for low values of~$a$, the parameter space is strongly affected by the spin,
with plots of the empirical pericenter and orbital inclination showing
significant distortion.  Nevertheless, we see unambiguously that the chaotic
orbits prevalent when $a=1$ are greatly suppressed as~$a$ decreases, with no
chaotic orbits at all below $a=0.2$.  This appears to be a result of the
increase of the marginally stable radius~$r_\mathrm{ms}$ as $a\rightarrow0$.
When $a=0$, the minimum stable radius is at $r_\mathrm{ms} = 4$ in units of the
central black hole's mass, fully $2\,M$ away from the horizon at $r_H = 2\,M$. 
As discussed in~\cite{Hartl_2002_1}, the extra (spherical) symmetry of the
Schwarzschild metric leads to an additional integral of the motion, which also
has a suppressive effect on chaos.

\begin{figure*}
\begin{tabular}{ccl}
\includegraphics[width=3in]{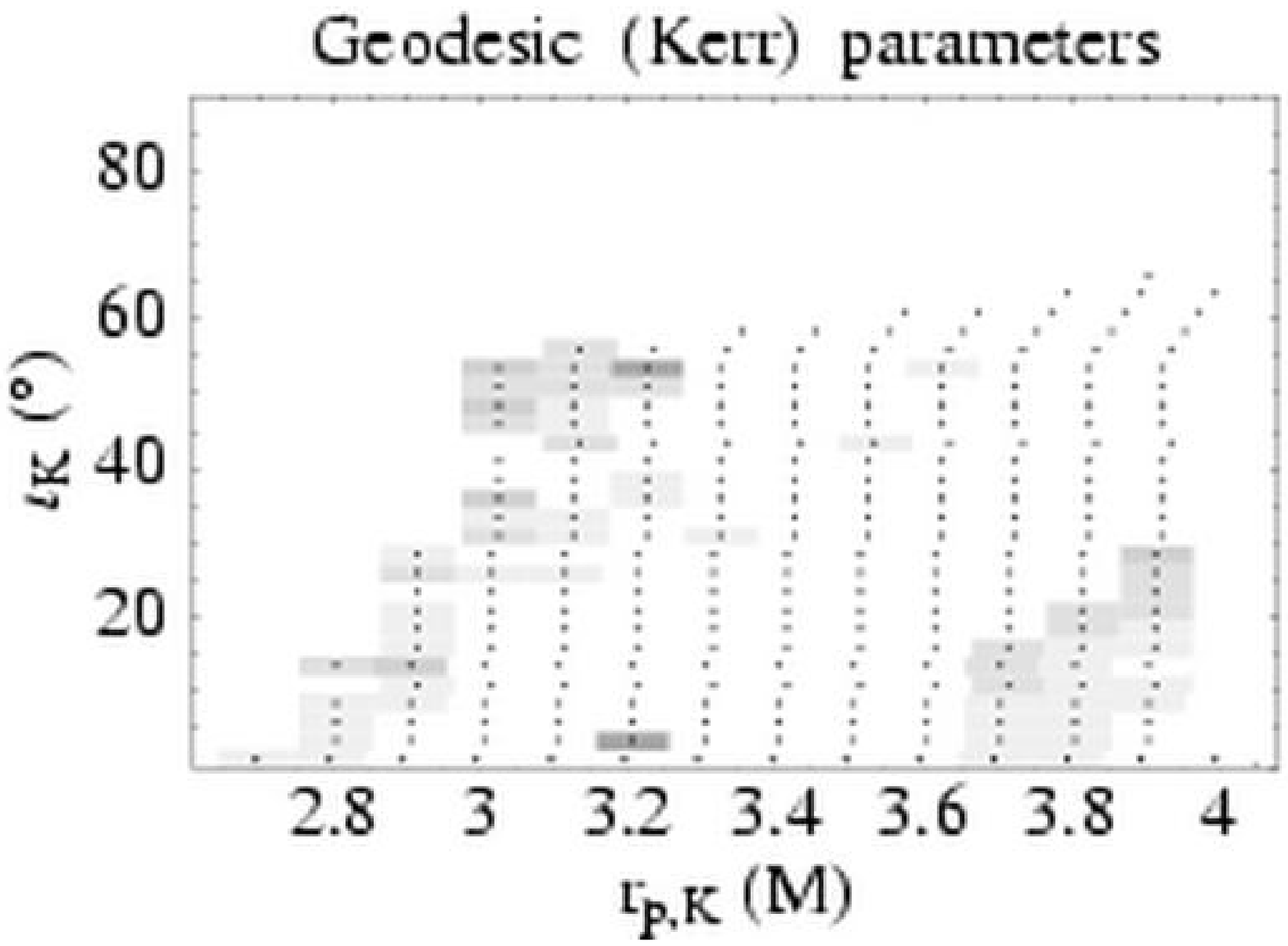}
	& \includegraphics[width=3in]{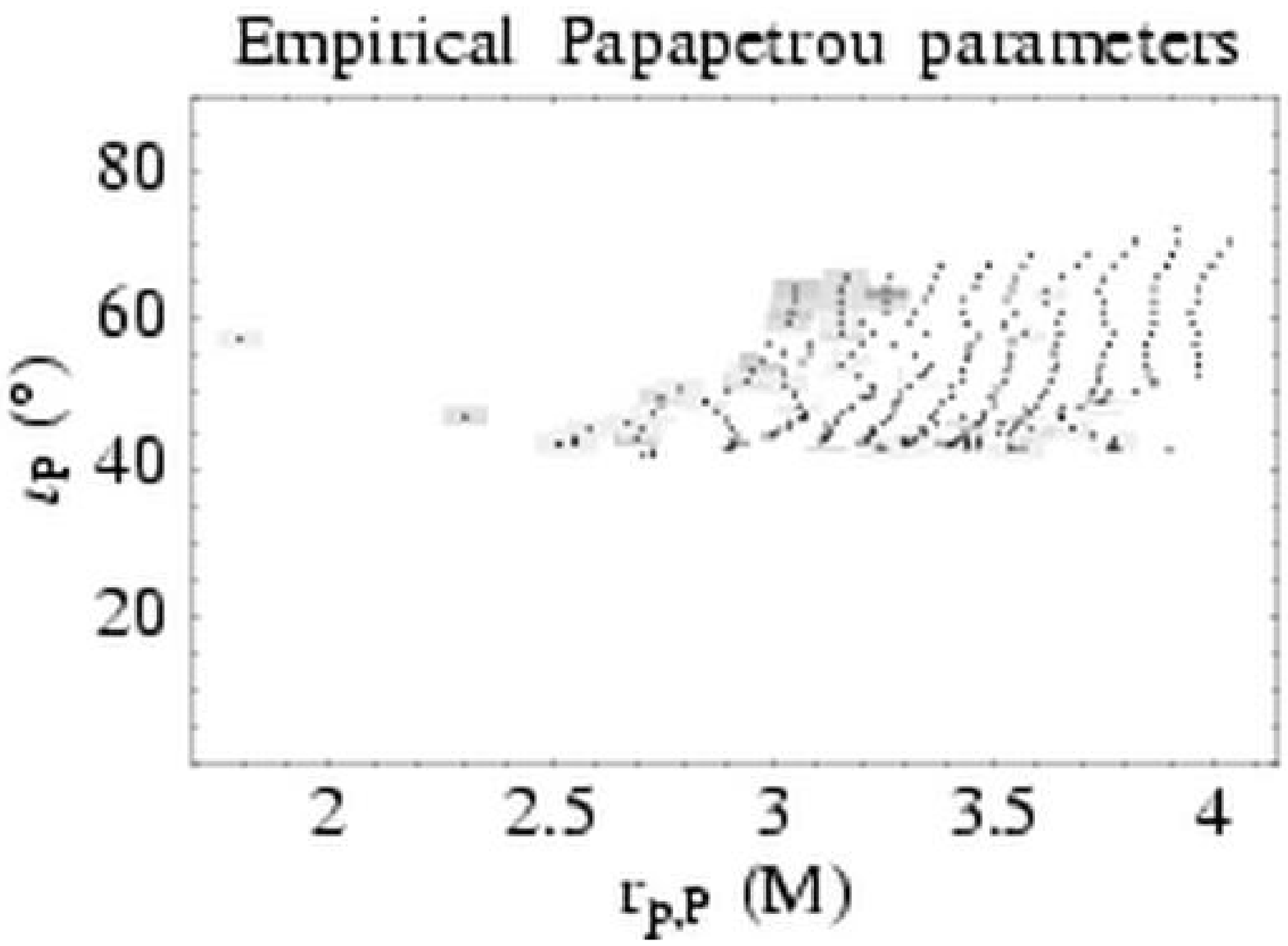}
    & \includegraphics[height=2in]{cutoff_sb.eps}\\
(a) & (b) & \medskip\\
\end{tabular}
\caption{\label{fig:e=0.01_cutoff} 
Spin cutoff map for $a=1$ and $e=0.01$.  The
rectangles are shaded according to the minimum value of $S$ for which the
corresponding initial condition is still chaotic.  White rectangles indicate
points that are not chaotic even when $S=1$.  The darkest points correspond to a
cutoff value of $S=0.65625$; below this critical value, 
none of the initial conditions are chaotic.}
\end{figure*}

\begin{figure*}
\begin{tabular}{ccl}
\includegraphics[width=3in]{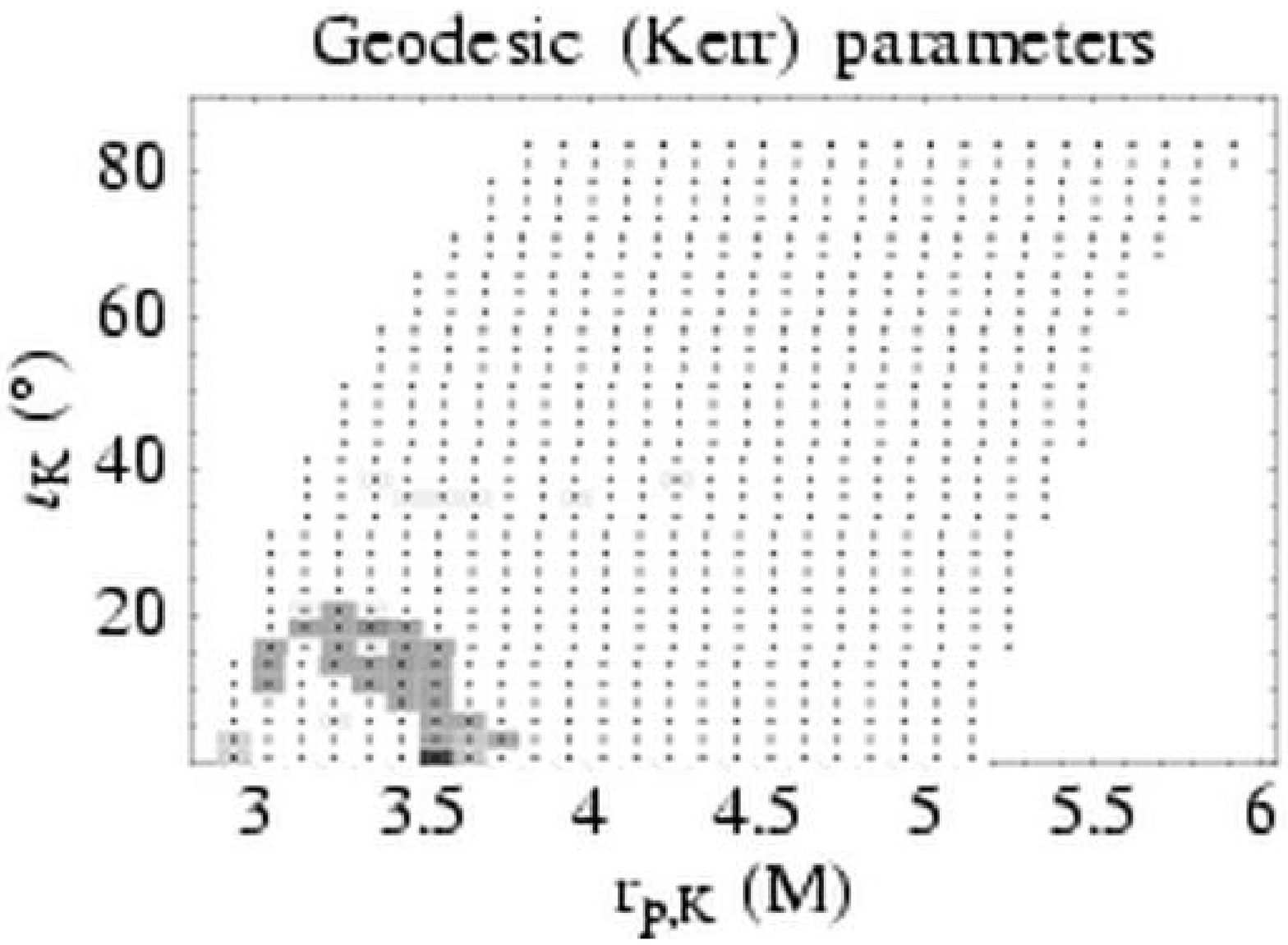}
	& 
    \includegraphics[width=3in]{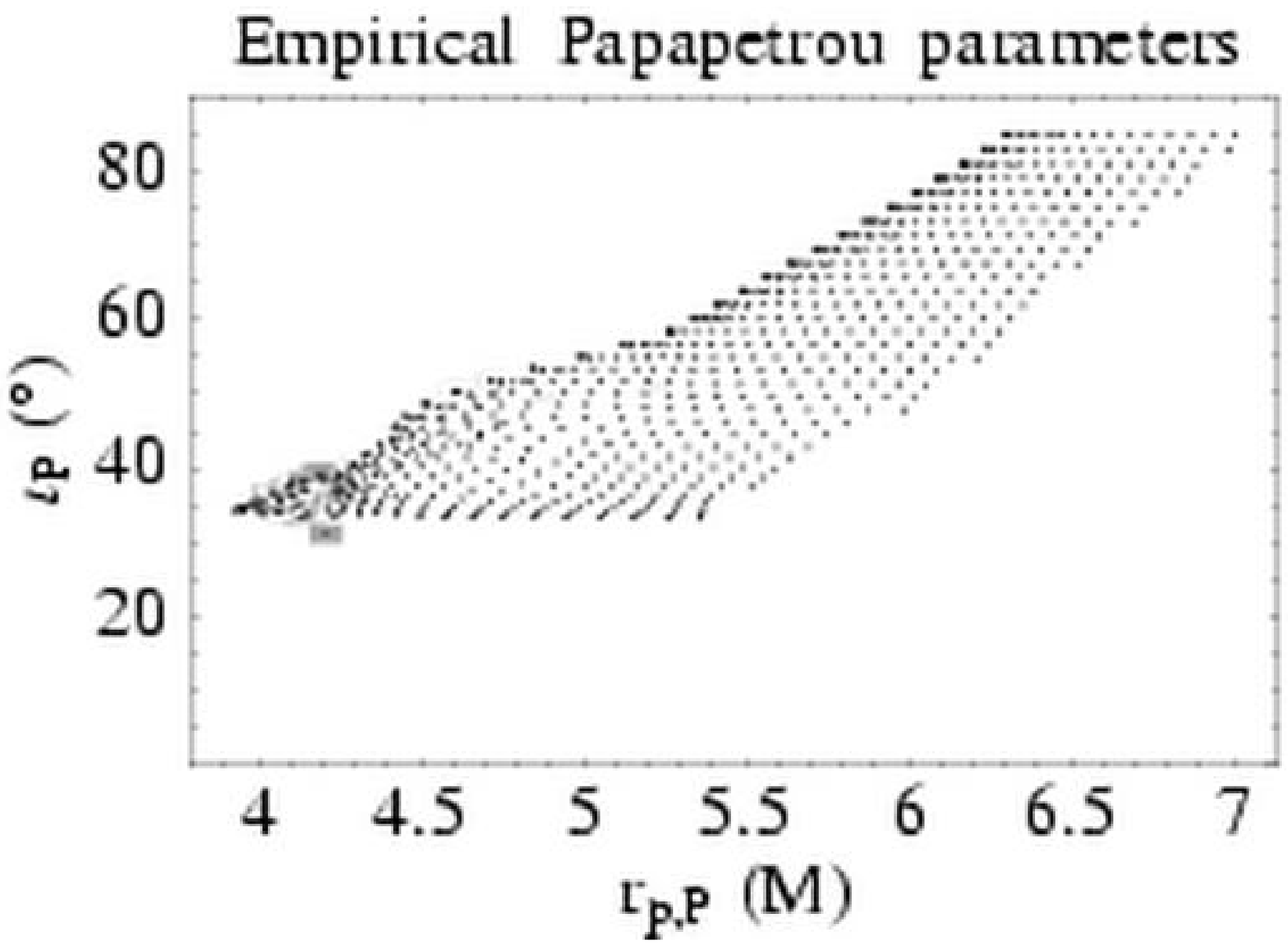}
    & \includegraphics[height=2in]{cutoff_sb.eps}\\
(a) & (b) & \medskip\\
\end{tabular}
\caption{\label{fig:a=0.5_cutoff} Spin cutoff map for $a=0.5$ and $e=0.5$.  The
rectangles are shaded according to the minimum value of $S$ for which the
corresponding initial condition is still chaotic.  White rectangles indicate
points that are not chaotic even when $S=1$.  The darkest points correspond to a
cutoff value of $S=0.28125$; below this critical value, 
none of the initial conditions are chaotic.}
\end{figure*}

\begin{figure*}
\begin{tabular}{ccl}
\includegraphics[width=3in]{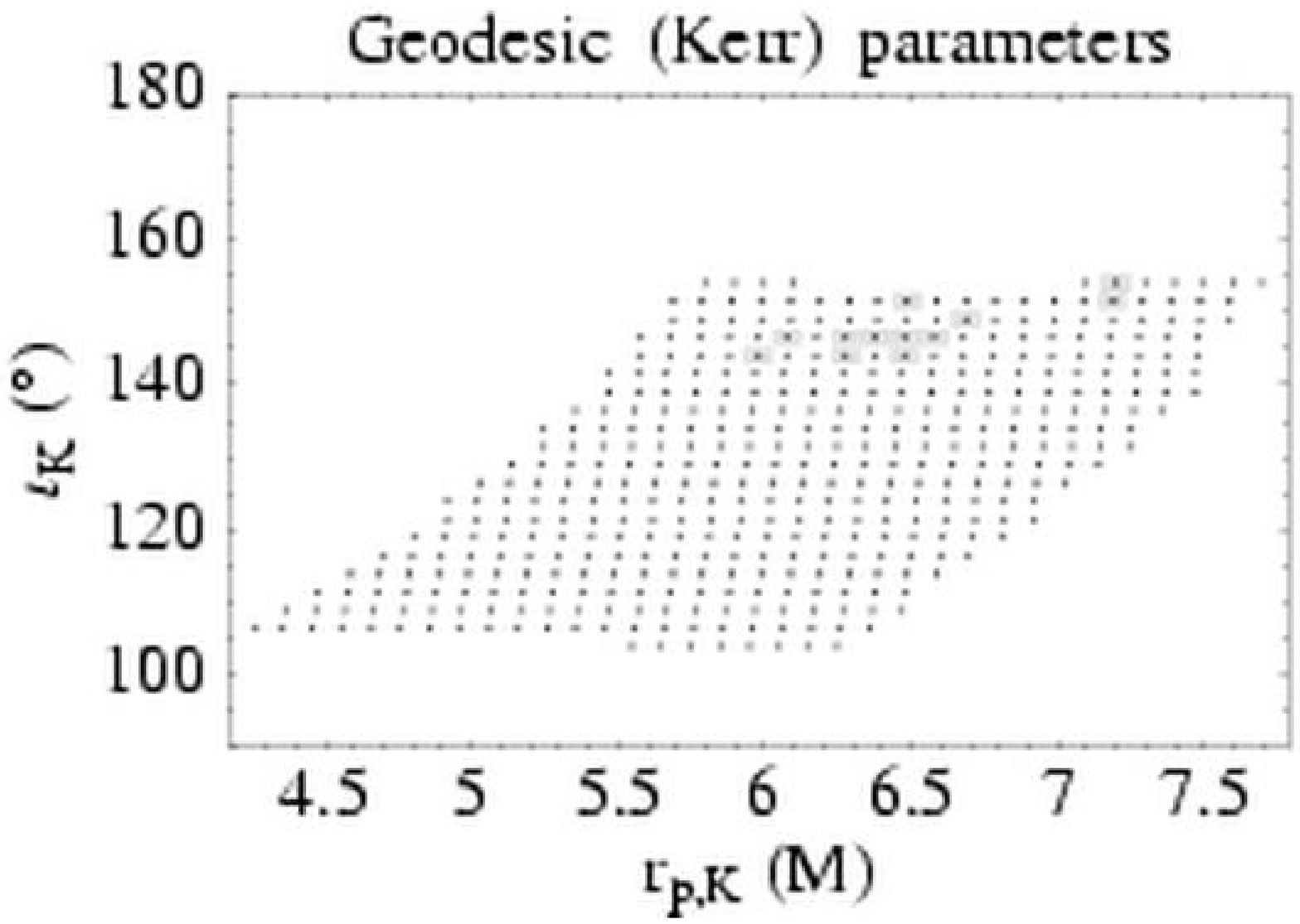}
	& \includegraphics[width=3in]{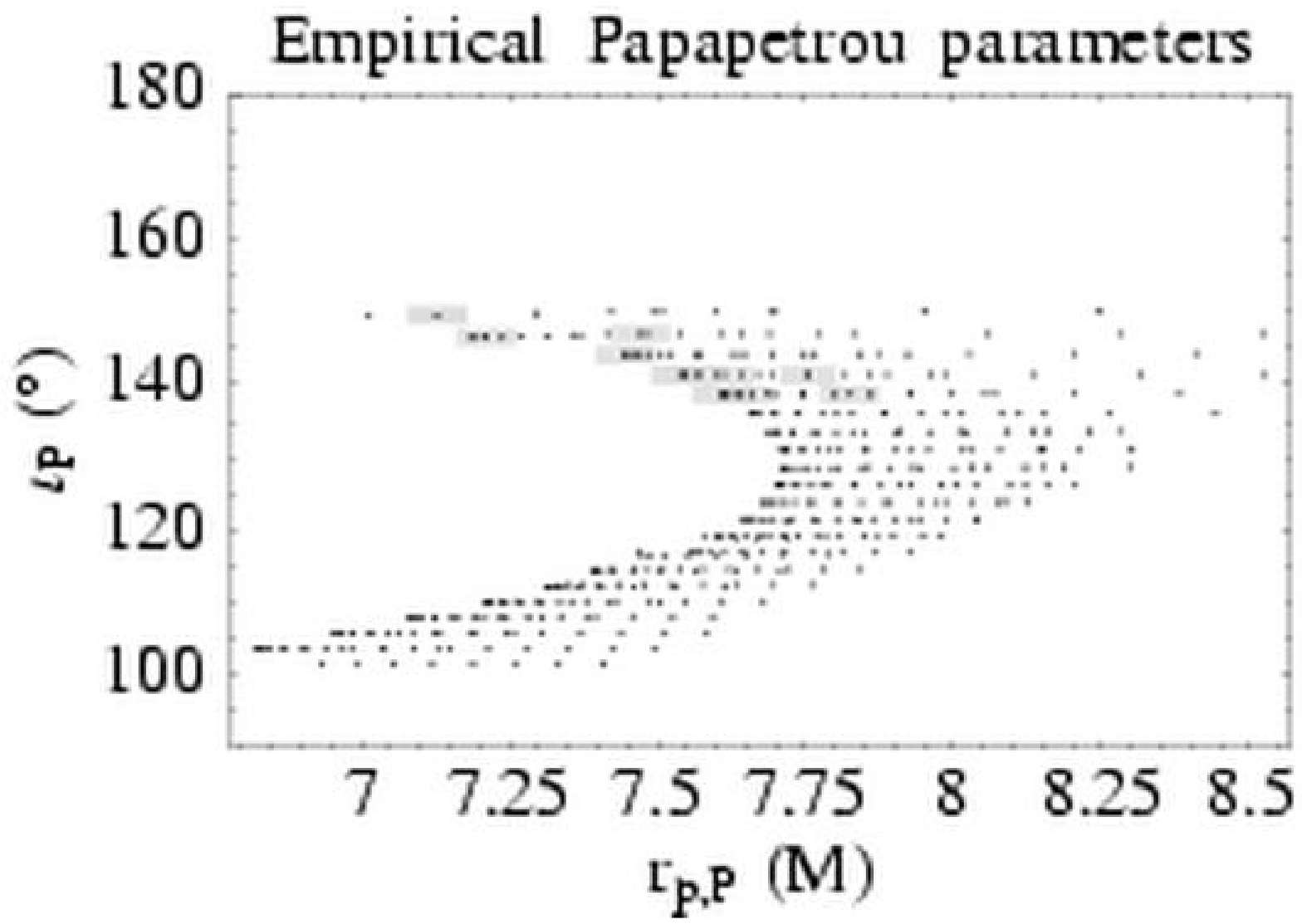}
    & \includegraphics[height=2in]{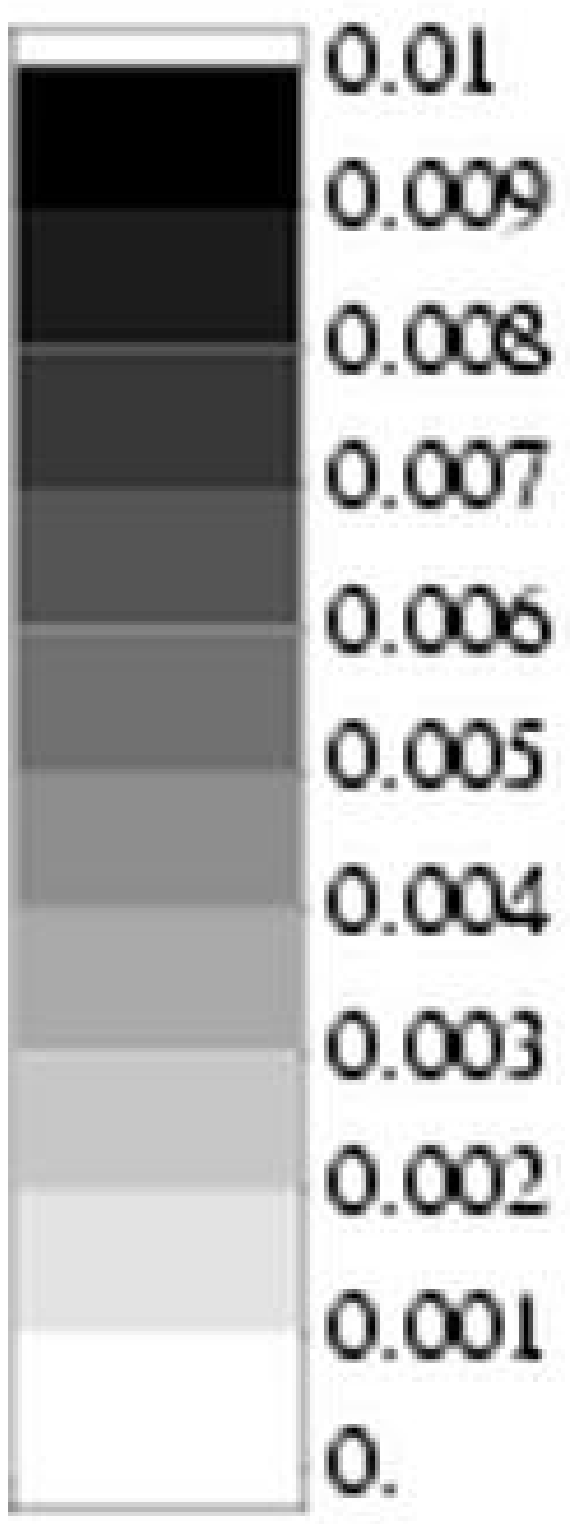}\\
(a) & (b) & \medskip\\
\end{tabular}
\caption{\label{fig:retrograde} $r_p$-$\iota$ map of retrograde orbits
($\iota>90^\circ$)
for $a=1$, $e=0.5$, $S=1$:
chaos strength as a function of pericenter and orbital inclination angle.
(a)~Requested parameters; (b)~empirical parameters.
The rectangles are shaded according to the Lyapunov exponent for the
initial condition represented by each point, with darker shades of gray
representing larger exponents and hence stronger chaos.  
The scaling is the same
as Fig.~\ref{fig:e=0.5_S=1}; an exponent of $\lambda = 0.01\,M^{-1}$ 
would appear
black.  The chaos is weak for these retrograde orbits: 
the largest Lyapunov exponent in
the plot is $\lambda=3.5\times10^{-4}\,M^{-1}$.}
\end{figure*}

\subsection{Spin cutoffs for chaos}
\label{sec:cutoff}

In this section we provide \emph{spin cutoff values} for chaos, i.e., the
minimum spin values for which chaos exists.  For a given initial condition
defined in terms of fixed orbital parameters (as described in
Sec.~\ref{sec:orbital_param}), we vary the spin parameter~$S$ and find the
maximum value for which chaos occurs.  The smaller this cutoff value, the
stronger the chaos: nonchaotic orbits have a cutoff value of $S=1$, i.e., they
are not chaotic even in the extreme $S=1$ limit; conversely, a cutoff value of
$S=10^{-5}$ would indicate chaos for the (physically realistic) value
$S=10^{-5}$, but not for any smaller values.\footnote{Implicit in this scheme
is an assumption of monotonicity, i.e., monotonically decreasing chaos as $S$
decreases.  While not strictly true (as discussed in~\cite{Hartl_2002_1}), this
assumption is still broadly applicable, and exceptions are rare.}  

Fig.~\ref{fig:e=0.5_cutoff} is an example of a spin cutoff map.  The procedure
for producing such a map is similar to the method used to make the Lyapunov
$r_p$-$\iota$ maps: we consider a grid of points in $r_p$-$\iota$ space, and
for each point we find an approximate value for the spin cutoff.  We begin by
finding out if the system is chaotic for $S=1$, using the Lyapunov map as a
start.  If the orbit for~$S=1$ is chaotic, we halve the spin value and
calculate the Lyapunov exponent for $S=0.5$. If the system is still chaotic, we
consider $S=0.25$; otherwise, we consider $S=0.75$.  The procedure continues
until the difference between chaotic and nonchaotic spin values is smaller than
some threshold, which we choose to be $0.05$.  (This value is chosen to achieve
reasonable accuracy while still completing the calculations in a tolerable
amount of time.)

Figs.~\ref{fig:e=0.5_cutoff}--\ref{fig:a=0.5_cutoff} show  the spin cutoff
values for several parameter combinations corresponding to Lyapunov maps from
Sec.~\ref{sec:rp_i} above.  The plots are color-coded with grayscale so
that the most chaotic points---those with the \emph{smallest} spin cutoff
values---appear darkest.  The points surrounded by white are not chaotic even
for $S=1$.  The cutoffs for the darkest points depend on the parameter values:
for Fig.~\ref{fig:e=0.5_cutoff} ($a=1$ and $e=0.5$), the same as
Fig.~\ref{fig:e=0.5_S=1} above), we find $S_\mathrm{cutoff} = 0.18457$; 
Fig.~\ref{fig:e=0.01_cutoff} ($a=1$ and $e=0.01$, the same as
Fig.~\ref{fig:e=0.01_S=1} above) has $S_\mathrm{cutoff} = 0.65625$; and
Fig.~\ref{fig:a=0.5_cutoff} ($a=0.5$ and $e=0.5$, the same as
Fig.~\ref{fig:a=0.5}) has $S_\mathrm{cutoff} = 0.28125$. We
should not take the digits of precision seriously, since these values are only
accurate to within $0.05$, but \emph{in all cases the spin cutoff values are
significantly above the physically realistic range of
$S\sim10^{-4}$--$10^{-7}$}.

\subsection{Retrograde orbits}
\label{sec:retrograde}

We have considered a wide variety of orbits---varying eccentricity and Kerr
parameter for different pericenter, orbital inclination, and spin
parameters---but so far all orbits have satisfied $0<\iota<\pi/2$, i.e., they
have all been prograde orbits, moving in the same direction as the central
black hole's spin.  We investigate now the case of retrograde orbits, and show
that they are poor candidates for chaos.

It is evident from looking at an $r_p$-$\iota$ plot  of a retrograde orbit
(Fig.~\ref{fig:retrograde}) that the pericenters are much larger than their
prograde counterparts.  For the $S=1$ particle illustrated in
Fig.~\ref{fig:retrograde}, the minimum empirical pericenter is larger than
$6\,M$, in contrast to prograde orbits, which get below $1.5\,M$.  Furthermore,
although it is clear from Fig.~\ref{fig:retrograde}(b) that the parameter space
is severely distorted, the chaos is extremely weak.  The largest Lyapunov
exponent, even in this extreme $S=1$ case, is $\lambda_\mathrm{max} =
3.5\times10^{-4}$, two orders of magnitude smaller than the prograde
case.  Unsurprisingly, all chaos disappears when $S\ll1$. The smallest value of
$S_\mathrm{cutoff}$ is $0.65265$ for the parameter values shown in
Fig.~\ref{fig:retrograde}; we find no evidence of chaos below this value
of~$S$.

\subsection{Varying spin inclination}
\label{sec:Sincl}

So far we have always used the same values for the two spin components passed
to the parameterization procedure (scaled by the total spin $S$):  $S^{\hat r}
= S^{\hat z} = 0.2\,S$.  We consider now the effect of varying these
parameters, and also discuss the corresponding initial spin inclination angles
in a fiducial rest frame.

\begin{figure}
\begin{tabular}{cl}
\includegraphics[width=2.75in]{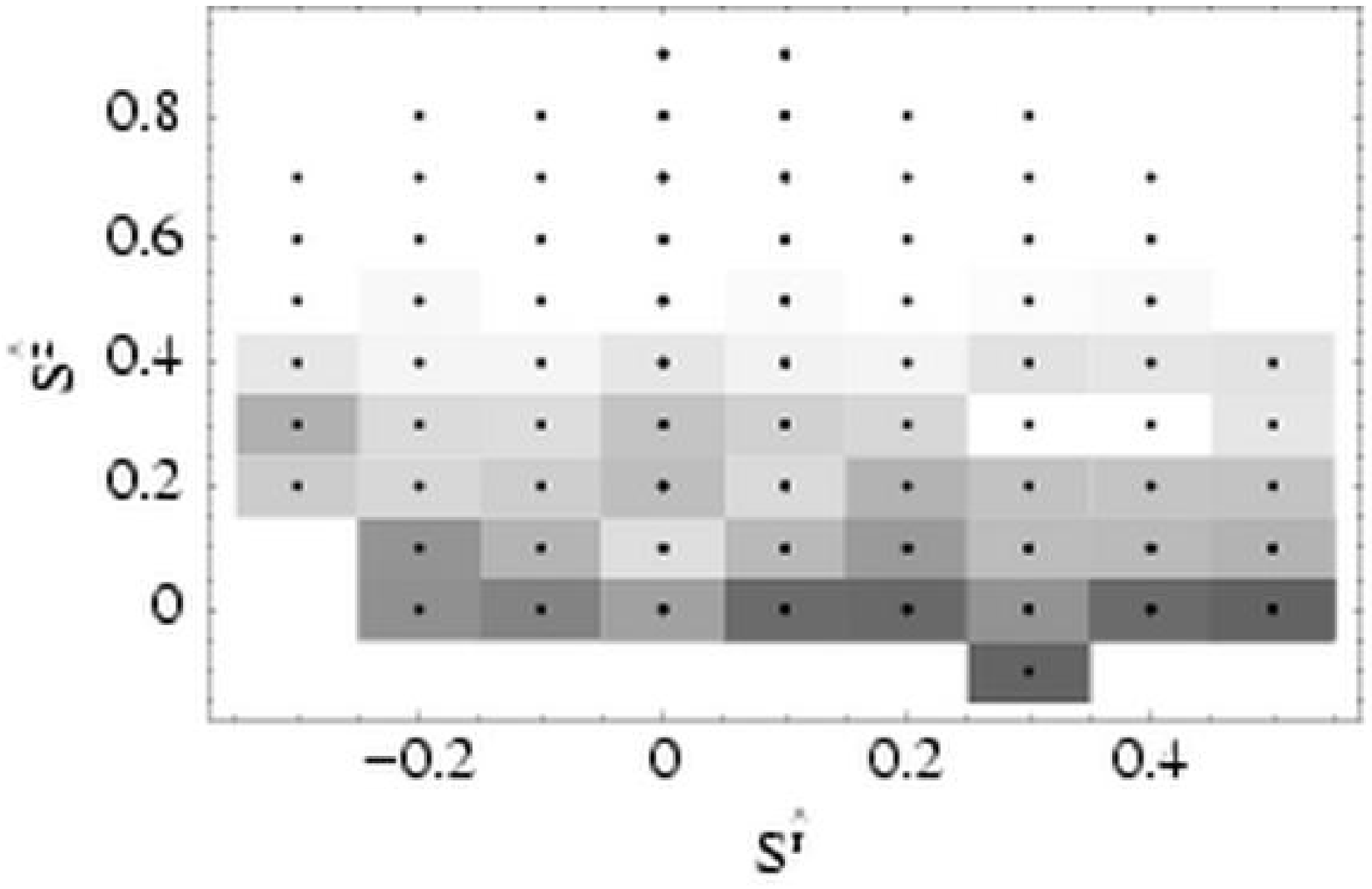}
	& \includegraphics[height=2in]{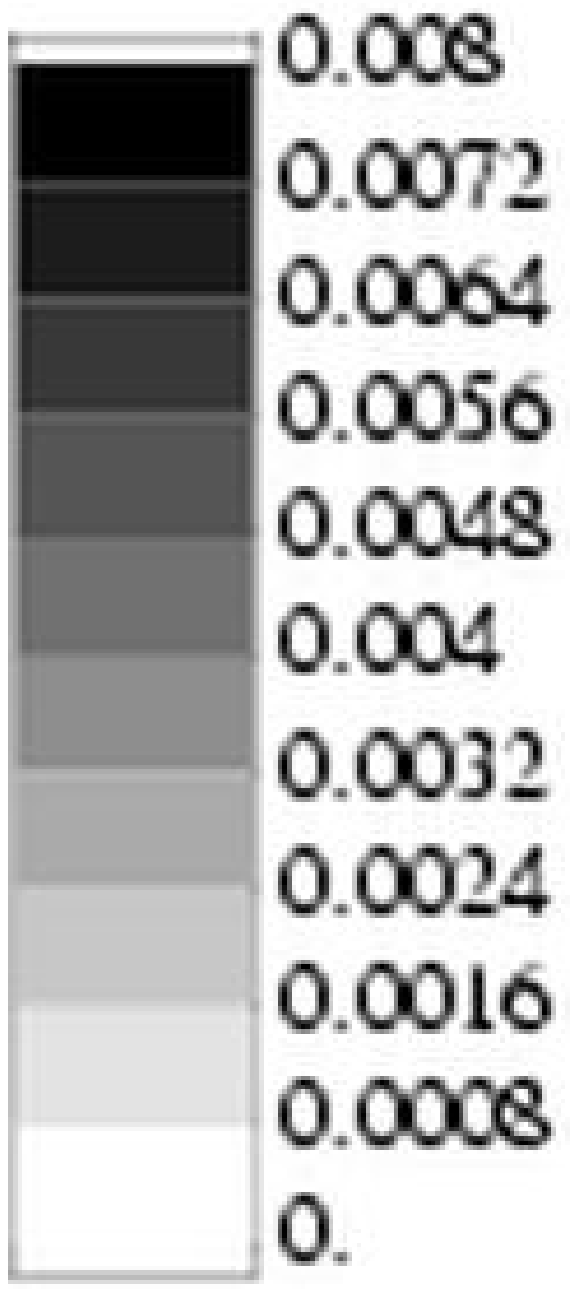}\\
(a) & \medskip\\
\end{tabular}
\caption{\label{fig:Sincl}
Lyapunov exponents for varying values of initial 
spin components $S^{\hat r}$ and $S^{\hat z}$.  The parameter values $S=1$,
$r_{p,\,K}=2.3$, $e_K=0.5$, and $\iota_K=20^\circ$ are held fixed.
The point with $S^{\hat r}=S^{\hat z} = 0.2$ appears in 
Fig.~\ref{fig:e=0.5_S=1} above.
The actual local spin
inclination angles in a fiducial (ZAMO) rest frame appear in
Table~\ref{table:Sincl}. Note that only one valid initial condition exists for
negative initial $S^{\hat z}$.}
\end{figure}

\begin{figure}
\includegraphics[width=3in]{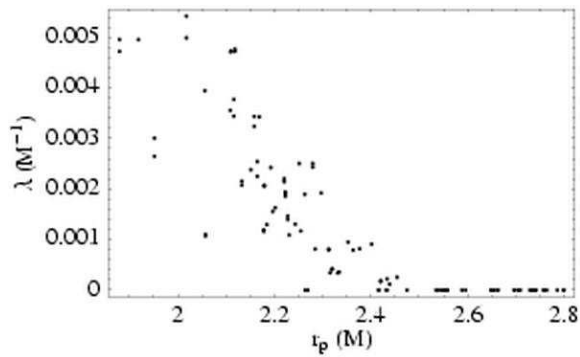}
\caption{\label{fig:Sinclr}
Scatter plot of empirical pericenters $r_{p,\,P}$ 
vs.\ Lyapunov exponent
for the spin inclinations in Fig.~\ref{fig:Sincl}.  The Lyapunov exponent is
primarily a function of pericenter, regardless of spin inclination.}
\end{figure}

We begin with an initial condition that is chaotic for $S=1$ but is otherwise
arbitrary.  We then vary $S^{\hat r}$ and $S^{\hat z}$ and calculate the
Lyapunov exponent for each configuration.  The result for $a=1$, $e_K=0.5$,
$r_{p,\,K}=2.3$ and $\iota_K=20^\circ$ appears in Fig.~\ref{fig:Sincl}.   When
$S^{\hat r} = S^{\hat z} = 0.2$, the parameter values chosen correspond to a
point from Fig.~\ref{fig:e=0.5_S=1}.  There are not valid initial conditions
for all choices of parameter; in particular, negative values of $S^{\hat z}$
are often unstable or are unable to satisfy the spin constraints. 
Nevertheless, there is a large variety of parameters that do give rise to valid
orbits, and many of them are chaotic.  The strongest chaos exists for orbits
that have small values of $S^{\hat z}$, but this appears to be mainly because
such orbits are able to achieve small empirical pericenters.  As
Fig.~\ref{fig:Sinclr} clearly shows, the Lyapunov exponent generally decreases
as pericenter increases, with no chaotic orbits above $r_{p,\,P} =
2.5\,M$.

While we are forced by the constraints to parameterize the equations of motion
in terms of spin components, it is more convenient to visualize the spin in a
fiducial rest frame that is hypersurface-orthogonal to the particle's
trajectory.  In Kerr spacetime, the hypersurface-orthogonal observers are the
zero angular momentum observers (ZAMO), which is the same frame we use when
calculating the Lyapunov exponents.  By projecting the components of the spin
vector $S^{\mu}$ into the ZAMO frame in the same way as we do for the projected
norm (Sec.~\ref{sec:papa_lyap}), we can find the \emph{local} value of the 
spin inclination angle~$\theta_\mathrm{local}$ (i.e., the angle between the
spin axis and the axis of the central black hole).  The results appear in
Table~\ref{table:Sincl}. It is clear that our variation of spin components
samples a large variety of initial spin inclination angles.

\begin{table*}
\caption{\label{table:Sincl}
Local spin inclination angles $\theta_\mathrm{local}$
in a fiducial (ZAMO) rest frame as a function of $S^{\hat r}$ and $S^{\hat z}$.
The parameter values $S=1$,
$r_{p,\,K}=2.3$, $e_K=0.5$, and $\iota_K=20^\circ$ are held fixed.
An illustration of their Lyapunov exponents appears in Fig.~\ref{fig:Sincl}.}
\begin{ruledtabular}
\begin{tabular}{c|ccccccccccc}
\multicolumn{10}{c}{$S^{\hat r}$} \\
\hline
$S^{\hat z}$  &  -0.3  &  -0.2  &  -0.1  &  0  &  0.1  &  0.2  &  0.3  &  0.4  &  0.5 \\
\hline
0.9 &    &     &     &  $22.7^\circ$  &  $28.7^\circ$  &     &     &     &    \\
0.8 &    &  $48.6^\circ$  &  $40.4^\circ$  &  $36.1^\circ$  &  $40.2^\circ$  &  $48.4^\circ$  &  $55.9^\circ$  &     &    \\
0.7 & $60.7^\circ$  &  $54.4^\circ$  &  $47.8^\circ$  &  $44.7^\circ$  &  $47.9^\circ$  &  $54.5^\circ$  &  $60.8^\circ$  &  $65.7^\circ$  &    \\
0.6 & $64.9^\circ$  &  $59.6^\circ$  &  $54.2^\circ$  &  $51.7^\circ$  &  $54.5^\circ$  &  $59.9^\circ$  &  $65.2^\circ$  &  $69.3^\circ$  &    \\
0.5 & $69.1^\circ$  &  $64.6^\circ$  &  $60.2^\circ$  &  $58.2^\circ$  &  $60.5^\circ$  &  $65.0^\circ$  &  $69.4^\circ$  &  $72.8^\circ$  &    \\
0.4 & $73.2^\circ$  &  $69.6^\circ$  &  $66.0^\circ$  &  $64.4^\circ$  &  $66.3^\circ$  &  $70.0^\circ$  &  $73.5^\circ$  &  $76.3^\circ$  &  $78.4^\circ$ \\
0.3 & $77.3^\circ$  &  $74.6^\circ$  &  $71.8^\circ$  &  $70.5^\circ$  &  $72.0^\circ$  &  $74.9^\circ$  &  $77.6^\circ$  &  $79.7^\circ$  &  $81.3^\circ$ \\
0.2 & $81.5^\circ$  &  $79.6^\circ$  &  $77.7^\circ$  &  $76.8^\circ$  &  $77.9^\circ$  &  $79.8^\circ$  &  $81.7^\circ$  &  $83.1^\circ$  &  $84.2^\circ$ \\
0.1 &    &  $84.7^\circ$  &  $83.7^\circ$  &  $83.3^\circ$  &  $83.9^\circ$  &  $84.9^\circ$  &  $85.8^\circ$  &  $86.5^\circ$  &  $87.1^\circ$ \\
0.0 &    &  $90.0^\circ$  &  $90.0^\circ$  &  $90.0^\circ$  &  $90.0^\circ$  &  $90.0^\circ$  &  $90.0^\circ$  &  $90.0^\circ$  &  $90.0^\circ$ \\
-0.1 &    &     &     &     &     &     &  $94.2^\circ$  &     &    \\
\end{tabular}
\end{ruledtabular}
\end{table*}

\subsection{Deep integrations}
\label{sec:deep}

Since we adopted a maximum time of $10^5\,M$ for the Lyapunov integrations, it
is reasonable to ask whether chaos might manifest itself on a longer
timescale.  This is certainly possible, but it appears that most initial
conditions are either chaotic on a timescale of order $10^2$--$10^4\,M$ or are
not chaotic at all, as discussed in Sec.~\ref{sec:chaos_detector}.  An example
appears in Fig.~\ref{fig:lyap_compare_e=0.6}, where one initial condition is
unambiguously chaotic, while a second located close by is not chaotic, even on
a much longer timescale.
  
\begin{figure}
\includegraphics[width=3in]{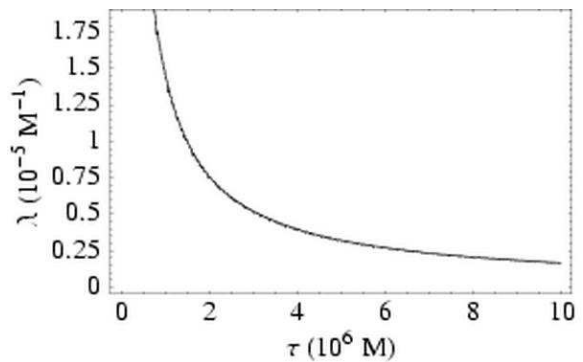}
\caption{\label{fig:deep_S=0.1} Approximate Lyapunov exponent vs.\ time for a
nonchaotic deep ($\tau_\mathrm{final}=10^7\,M$) integration.  The
parameter values are $a=1$, $S=0.1$, $e=0.5$, $r_p = 1.32\,M$, and $\iota =
28.5^\circ$, corresponding to one of the inner orbits from
Fig.~\ref{fig:e=0.5_S=0.1}.  The Lyapunov exponent appears to be zero; its
time-evolution has the characteristic hyperbolic shape expected as
$\log{[r_e(\tau)]}/\tau$ approaches zero for large times.  A least-squares fit of
$\log{[r_e(\tau)]}$ vs.\ $\tau$ gives a value of
$\lambda\approx2.8\times10^{-7}\,M^{-1}$.
}
\end{figure}

\begin{figure}
\includegraphics[width=3in]{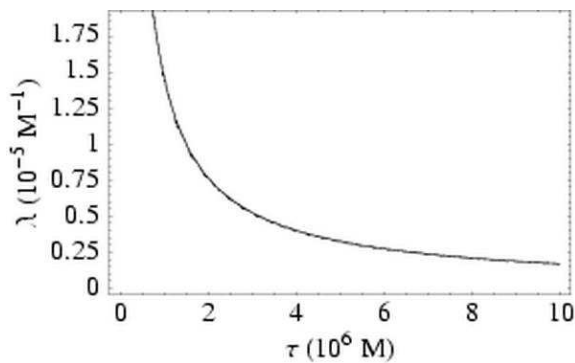}
\caption{\label{fig:deep_S=1e-4} Approximate Lyapunov exponent vs.\ time for a
nonchaotic deep ($\tau_\mathrm{final}=10^7\,M$) integration.  The
parameter values are $a=1$, $S=10^{-4}$, $e=0.5$, $r_p = 1.32\,M$, and $\iota =
28.5^\circ$, corresponding to one of the inner orbits from
Fig.~\ref{fig:e=0.5_S=1e-4}.  As in Fig.~\ref{fig:deep_S=0.1}, 
the Lyapunov exponent appears to be zero.  A least-squares fit of
$\log{[r_e(\tau)]}$ vs.\ $\tau$ gives a value of
$\lambda\approx3.0\times10^{-7}\,M^{-1}$.}
\end{figure}

\begin{figure}
\includegraphics[width=3in]{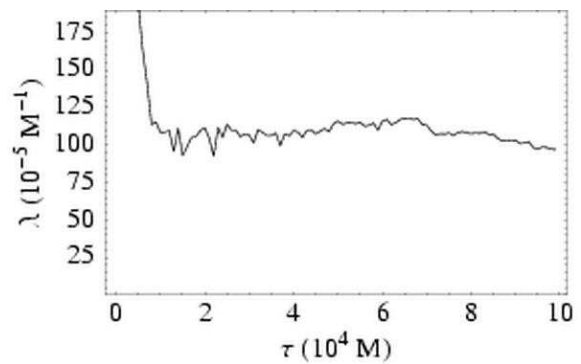}
\caption{\label{fig:chaotic_deep_S=0.1} 
Approximate Lyapunov exponent vs.\ time for a
chaotic deep ($\tau_\mathrm{final}=10^5\,M$) integration.  The simulation data
is identical to that shown in Fig.~\ref{fig:lyap_long_e=0.6};
the
parameter values are $a=1$, $S=0.1$, $e_K=0.6$, $r_{p,\,K} = 1.21\,M$, and 
$\iota_K =
31^\circ$, corresponding to the chaotic orbit from
Fig.~\ref{fig:e=0.6_S=0.1}.  The Lyapunov exponent for this chaotic orbit levels
off after less than $10^5\,M$, in contrast to Figs.~\ref{fig:deep_S=0.1} 
and~\ref{fig:deep_S=1e-4}, where $\lambda$ continues to decrease even after
$10^7\,M$.}
\end{figure}

To convince ourselves that slow chaos is not lurking in the apparently
nonchaotic regions, we performed a few longer-time integrations. In particular,
we calculated the Lyapunov exponents using $\tau_\mathrm{final} = 10^7\,M$ for
all the innermost ($r_p = 1.32$) orbits from Fig.~\ref{fig:e=0.5_S=0.1}
($S=0.1$) and Fig.~\ref{fig:e=0.5_S=1e-4} ($S=10^{-4}$), which are strongly
chaotic when $S=1$ but are apparently without chaos below $S=0.1$.  The largest
exponent occurs for $\iota=28.5^\circ$, which is therefore the worst-case
scenario.  Plots of the Lyapunov exponents vs.\ time appear in
Fig.~\ref{fig:deep_S=0.1} ($S=0.1$) and Fig.~\ref{fig:deep_S=1e-4}
($S=10^{-4}$); their magnitudes are on the order of $\lambda_\mathrm{max} =
3\times10^{-7}\,M^{-1}$, corresponding to an $e$-folding timescale of
approximately $3.3\times10^6\,M$. For comparison, we show $\lambda$ vs.\ $\tau$
for a chaotic initial condition in Fig.~\ref{fig:chaotic_deep_S=0.1}; it is
clear that the Lyapunov exponent asymptotes to a nonzero value in much less
than $10^{5}\,M$, even for weak chaos.  (The initial condition in 
Fig.~\ref{fig:chaotic_deep_S=0.1} is the $S=0.1$ orbit illustrated in
Figs.~\ref{fig:lyap_e=0.6} and~\ref{fig:lyap_long_e=0.6}, whose Lyapunov
exponent is actually quite small compared to analogous $S=1$ orbits.)  The long
integrations thus provide strong evidence that the  disappearance of nearly all
chaotic orbits below $S=0.1$ is a real effect.

\begin{figure*}
\begin{tabular}{ccccc}
\includegraphics[height=2in]{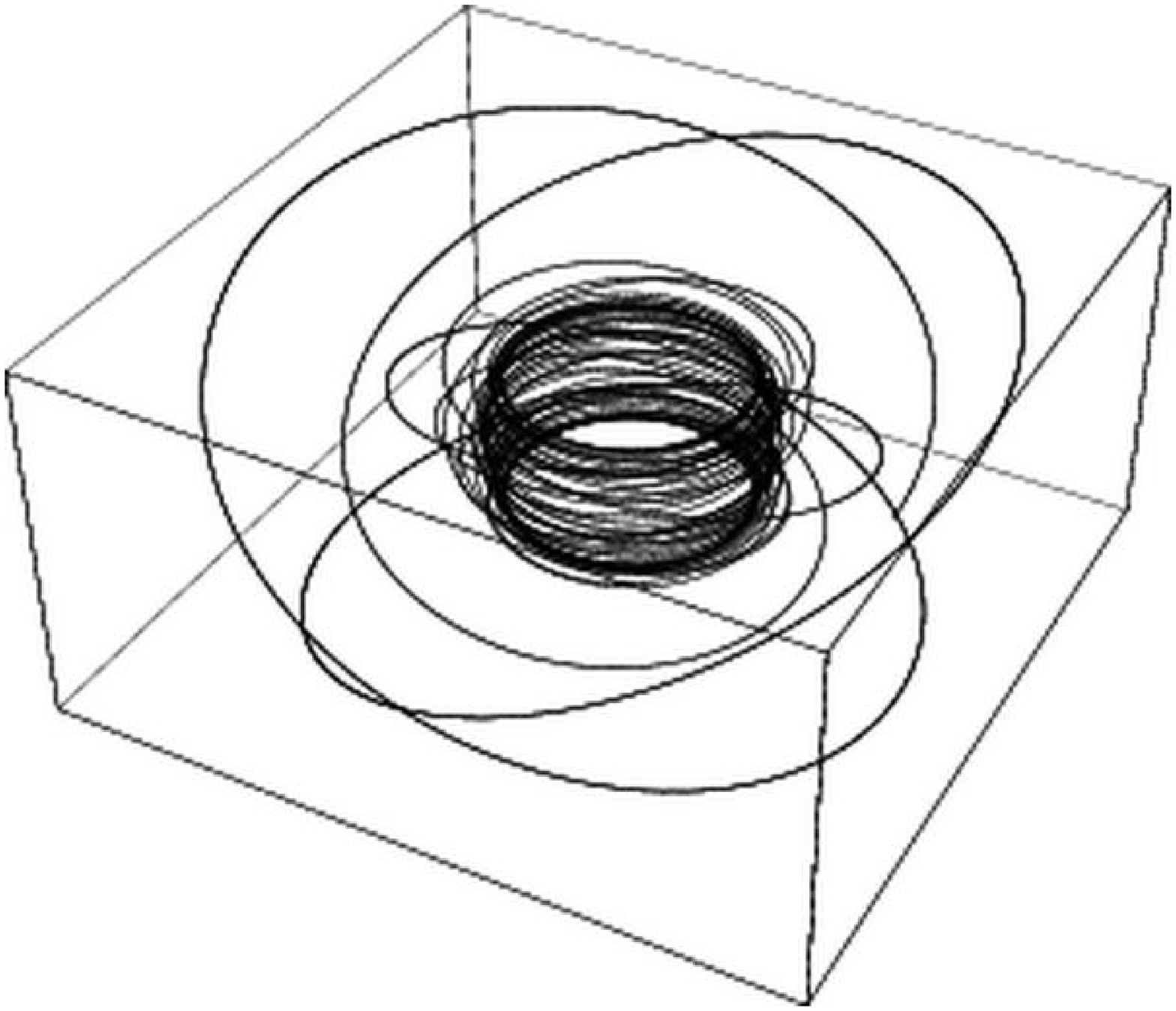}
    & \hspace{0.25in}
	& \includegraphics[height=2in]{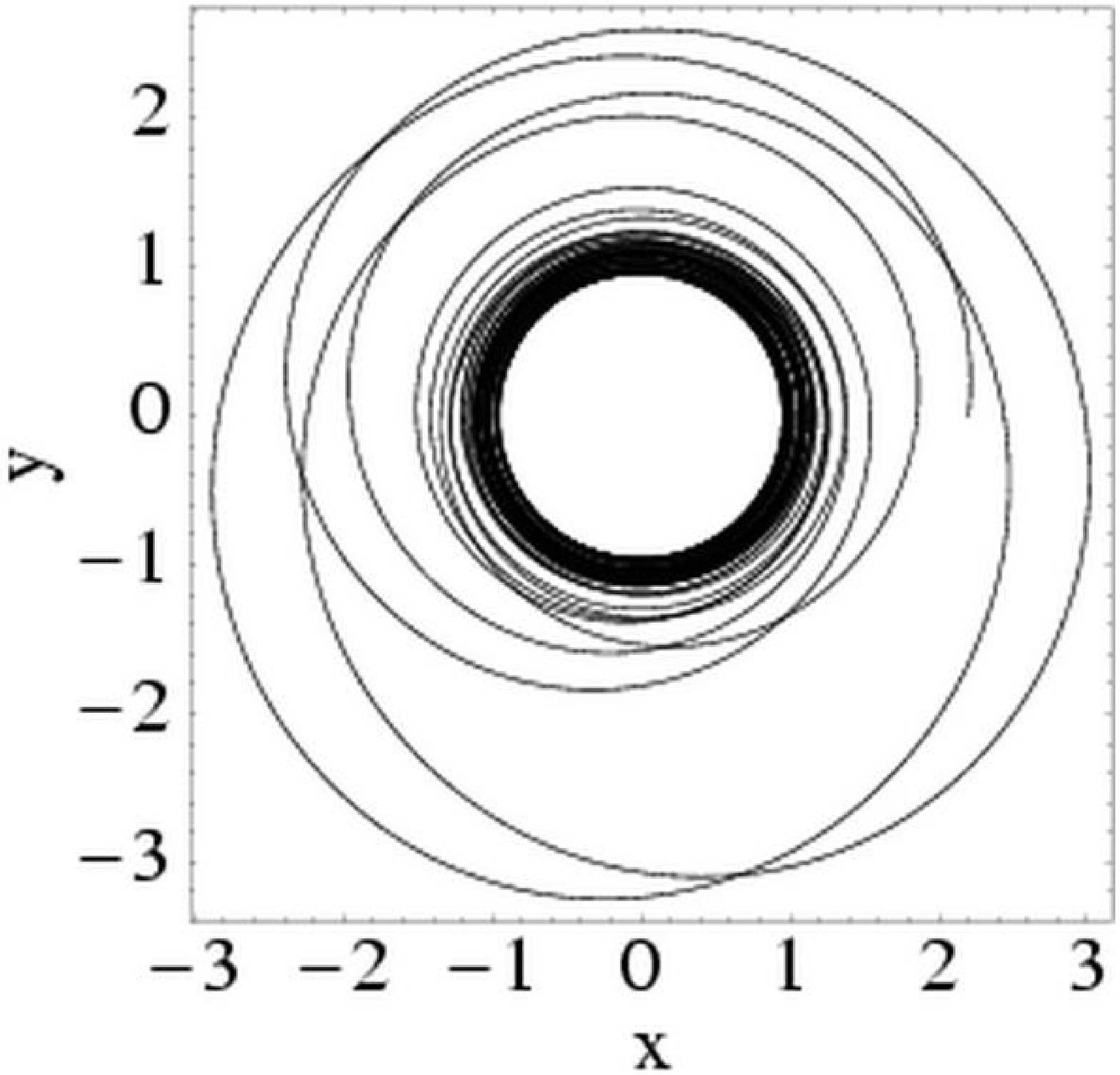}
    & \hspace{0.25in}
    & \includegraphics[height=1.75in]{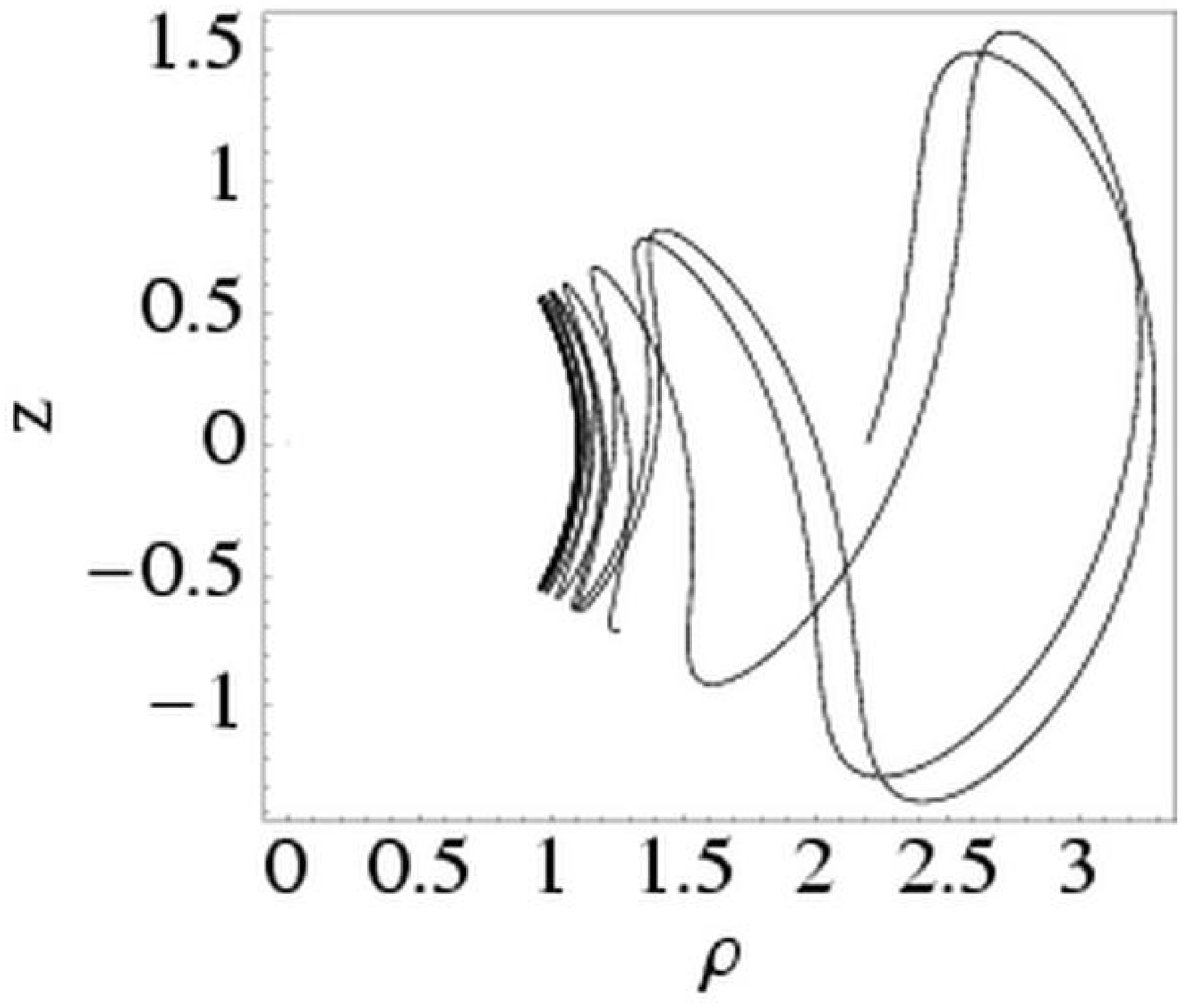}\\
(a) & & (b) & & (c)\medskip\\
\end{tabular}
\caption{\label{fig:mimic_orbit} The orbit of a chaos mimic:
a non-spinning ($S=0$)
test particle in maximal ($a=1$) Kerr spacetime, 
plotted in Boyer-Lindquist coordinates.  A Lyapunov plot of these initial
conditions appears in Fig.~\ref{fig:mimic} from Sec.~\ref{sec:chaos_detector}.
(a) The orbit embedded in three-dimensional space, treating the Boyer-Lindquist
coordinates as ordinary spherical polar coordinates;
(b) $y=r\sin\theta\sin\phi$ vs.\ $x=r\sin\theta\cos\phi$; 
(c) $z$ vs.\ $\rho=\sqrt{x^2+y^2}$.  The inclination angle is $\iota=31^\circ$,
while the pericenter is $r_p=1.1\,M$ (just $0.1\,M$ above the horizon at
$r_H=1\,M$).}
\end{figure*}

\begin{figure*}
\begin{tabular}{ccccc}
\includegraphics[height=2in]{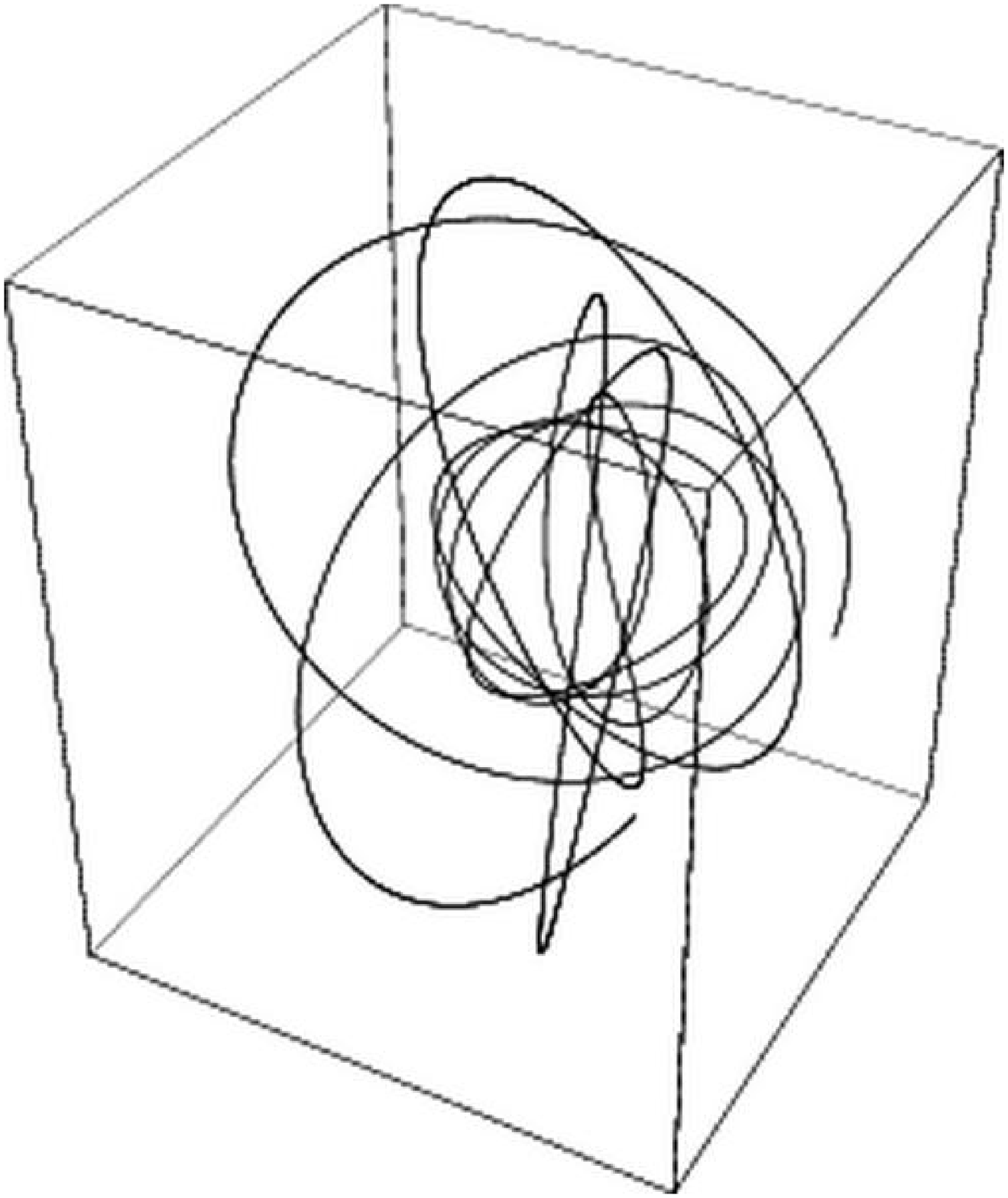}
    & \hspace{0.25in}
	& \includegraphics[height=2in]{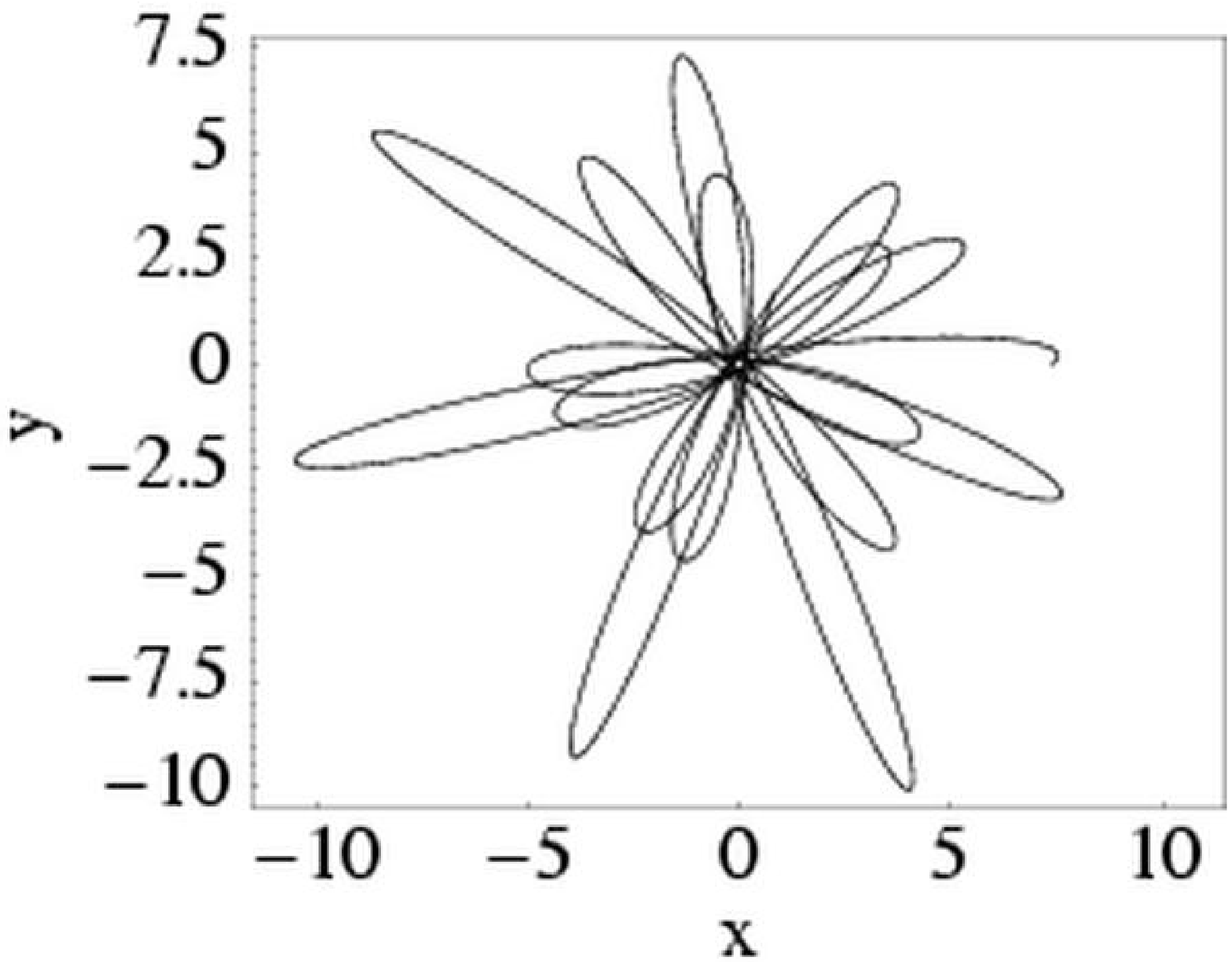}
    & \hspace{0.25in}
    & \includegraphics[height=2.5in]{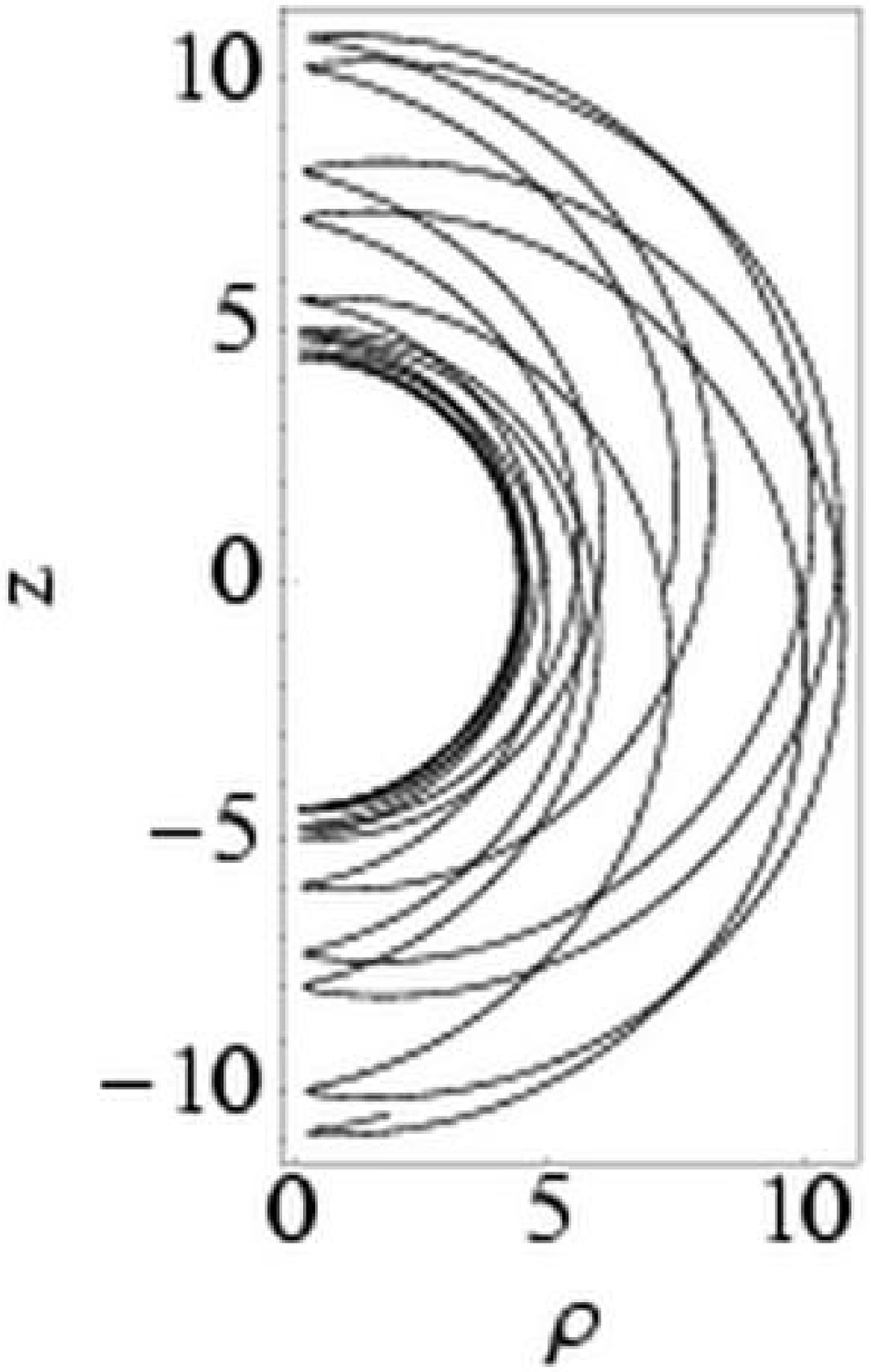}\\
(a) & & (b) & & (c)\medskip\\
\end{tabular}
\caption{\label{fig:mimic_orbit_2} Another chaos mimic:
a non-spinning ($S=0$)
test particle in maximal ($a=1$) Kerr spacetime, 
plotted in Boyer-Lindquist coordinates.  A Lyapunov plot of these initial
conditions appears in Fig.~\ref{fig:mimic_2}.
(a) The orbit embedded in three-dimensional space, treating the Boyer-Lindquist
coordinates as ordinary spherical polar coordinates;
(b) $y=r\sin\theta\sin\phi$ vs.\ $x=r\sin\theta\cos\phi$; 
(c) $z$ vs.\ $\rho=\sqrt{x^2+y^2}$.  The inclination angle is $\iota=88.5^\circ$,
while the pericenter is $r_p=4.4\,M$.}
\end{figure*}

\subsection{$S = 0$ and chaos mimics}
\label{sec:chaos_mimics}

Since the case of $S = 0$ corresponds exactly to geodesic orbits in Kerr
spacetime, such systems cannot be chaotic---the return of the Carter
constant~$Q$ and the loss of the spin degrees of freedom make the system fully
integrable.  Nevertheless, even some geodesic orbits can have a large
separation of nearby initial conditions, which can appear to be chaotic.  These
chaos mimics typically spend many orbital periods whirling around deep in the
strongly relativistic zone near the horizon, only occasionally zooming out to
higher radii.  These so-called \emph{zoom-whirl} orbits may provide significant
challenges to detection despite their formal integrability.

\begin{figure}
\includegraphics[width=3in]{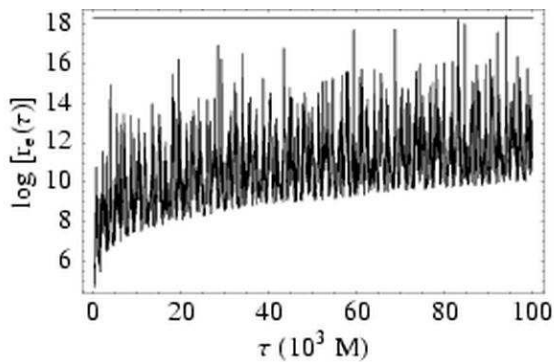}
\caption{\label{fig:mimic_2} A chaos mimic: the natural logarithm of the 
principal ellipsoid axis ($\log{[r_e(\tau)]}$) vs.\ $\tau$ for an
$S=0$ orbit. 
The size of the initial deviation vector is $\epsilon_0 = 10^{-8}$.
The value of $\log{[r_e(\tau)]}$
briefly rises up to the saturation level at 
$\log(0.9/\epsilon_0)$ [so that $\epsilon(\tau) = 0.9$], 
but the orbit is not chaotic since its spin
satisfies~$S=0$.  A plot of the corresponding
orbit appears in Fig.~\ref{fig:mimic_orbit_2}.}
\end{figure}

An example of how much divergence an $S=0$ orbit can experience appears in
Fig.~\ref{fig:mimic}.  A picture of the corresponding orbit (visualized in
Boyer-Lindquist coordinates embedded in ordinary space) appears in
Fig.~\ref{fig:mimic_orbit}, which makes clear the large number of low-radius
$\phi$-periods characteristic of zoom-whirl orbits.  A second example of a
chaos mimic appears in Figs.~\ref{fig:mimic_2} and~\ref{fig:mimic_orbit_2}. 
This orbit, in contrast to the previous one, does not have a particularly small
pericenter, but its high inclination angle and zoom-whirliness allow it to
mimic chaotic orbits.  

The chaos mimics can exhibit large growth of the initial deviation vector,
approximately a factor of $10^6$--$10^8$, after $10^5\,M$.  The principal means
for detecting them is by requiring several high-separation points in a row (on
a time-$T\approx 100\,M$ basis) as mentioned in Sec.~\ref{sec:chaos_detector};
the mimics have oscillations with high amplitudes due to their zoom-whirliness,
but they do not represent true saturations of the separation vector.

\section{Conclusions}
\label{sec:conclusions}

The Papapetrou equations, which model a spinning test particle, exhibit chaotic
solutions in Kerr spacetime for a wide range of parameters.  In terms of the
mass~$M$ of the central (Kerr) black hole, the largest Lyapunov exponents are
of order $\lambda_\mathrm{max} = 0.01\,M^{-1}$, which represents an exponential
divergence of trajectories on a timescale of $\tau_\lambda = 1/\lambda =
100\,M$.  Furthermore, there are many chaotic orbits with exponents in the
range $10^{-3}$--$10^{-4}\,M^{-1}$. Despite the large number of chaotic orbits,
we find that values of~$\lambda$ corresponding to unambiguous chaos occur
exclusively when the spin parameter~$S$ is not small compared to unity. In
particular, we find virtually no chaos for spin values below $S=0.1$, and no
evidence of any chaos for spins below $S=10^{-4}$.  

The strongest determinant of chaotic behavior, apart from the spin parameter,
is the pericenter of the orbit in question.  The most highly chaotic orbits are
those that reach pericenters near the horizon of the black hole.  This is due
to the high spacetime curvature in these regions, which maximizes the size of
the coupling of the spin to the Riemann curvature tensor
[Eq.~(\ref{eq:Dixon}].  When the Kerr parameter $a$ is small, so that the Kerr
metric differs only slightly from the Schwarzschild metric, the pericenters are
much higher than in the extreme Kerr ($a=1$) case.  Chaos in the Papapetrou
system is therefore weak when $a$ is small.  The prevalence of chaos is also
dependent on orbital eccentricity.  Near-circular ($e=0.01$) orbits have many
fewer regions of chaotic orbits than those with higher eccentricities ($e=0.5$
or $e=0.6$).  This seems due primarily to the lower pericenters accessible to
high-eccentricity orbits.

The dependence of the Lyapunov exponents on~$S$ is our most important result: in
all cases considered, physically realistic values of~$S$ (satisfying $S\ll1$)
are not chaotic.  We have shown conclusively that the Papapetrou equations
admit many solutions that are formally chaotic, but without exception such
chaotic solutions occur only for relatively large values of $S$.  Below the
upper limit for physically realistic spins ($S\sim10^{-4}$), we find no
evidence of chaotic solutions.  As a practical matter, this means that chaos
will not manifest itself in the gravitational radiation from extreme mass-ratio
binary inspirals.  

\begin{acknowledgments}

I would like to thank Scott Hughes and Teviet Creighton for providing notes on
parameterizing Kerr geodesics in terms of orbital parameters.  I also thank
Sterl Phinney for his careful reading of the manuscript and perceptive
comments.  This work was supported in part by NASA grant NAG5-10707.

\end{acknowledgments}

\appendix*
\section{Spin vector formulation}

We summarize here the formulation of the Papapetrou equations in terms of the
spin 1-form $S_\mu$ (often referred to loosely as the ``spin vector''), as
mentioned in Sec.~\ref{sec:eom}. In this paper we consider a spinning particle
of rest mass~$\mu$ orbiting a central Kerr black hole of mass~$M$, and it is
convenient to  measure all times and lengths in terms of~$M$ and all momenta in
terms of~$\mu$.  In these normalized units, the equations of motion in terms of
the spin 1-form are
\begin{eqnarray}
\label{eq:Papapetrou}
\frac{dx^\mu}{d\tau}&=&v^\mu\nonumber\\ \nabla_{\vec v}\,p_\mu&=&-R^{*\
\alpha\beta}_{\mu\nu}v^\nu p_\alpha S_\beta\\ \nabla_{\vec
v}\,S_\mu&=&-p_\mu \left(R^{*\alpha\ \gamma\delta}_{\ \ \beta}
	S_\alpha v^\beta p_\gamma S_\delta\right)\nonumber
\end{eqnarray}
where \begin{equation} R^{*\alpha\ \mu\nu}_{\
\ \beta}=\textstyle{{1\over2}}R^{\alpha}_{\ \beta\rho\sigma}
\epsilon^{\rho\sigma\mu\nu}.  
\end{equation}

The supplementary condition Eq.~(\ref{eq:pSortho}) allows for an 
explicit solution for the 4-velocity $v^\mu$ in terms of~$p^\mu$:
\begin{equation}
\label{eq:v}
v^\mu=N(p^\mu+w^\mu),
\end{equation}
where
\begin{equation}
\label{eq:w}
w^\mu=-{^*}R^{*\mu\alpha\beta\gamma} S_\alpha p_\beta S_\gamma
\end{equation}
and
\begin{equation}
{^*}R^{*\alpha\beta\mu\nu}=\textstyle{{1\over2}}R^{*\alpha\beta\rho\sigma}
\epsilon_{\rho\sigma}^{\ \ \ \mu\nu}.
\end{equation}
The normalization constant~$N$ is fixed by the constraint $v_\mu v^\mu=-1$.

The spin 1-form satisfies two orthogonality constraints:
\begin{equation}
\label{eq:constraint_1}
p^\mu S_\mu = 0
\end{equation}
and
\begin{equation}
\label{eq:constraint_2}
v^\mu S_\mu = 0.
\end{equation}
These two constraints are equivalent as long as~$v^\mu$ is given by 
Eq.~(\ref{eq:v}): $0 = v^\mu S_{\mu}\propto p^\mu S_\mu + w^\mu S_\mu = p^\mu
S_\mu$, since by definition of $w^\mu$ [Eq.~(\ref{eq:w})] the second term
involves the contraction of a symmetric tensor with an antisymmetric tensor and
therefore vanishes.   We enforce Eq.~(\ref{eq:constraint_1}) in our
parameterization scheme, and we use  Eq.~(\ref{eq:v}) in the equations of
motion, so Eq.~(\ref{eq:constraint_2}) is then automatically satisfied.

\bibliography{mdh_2002_2}% uses BibTeX to make bibliography; see mdh_2002_2.bib

\begin{figure*}
\begin{tabular}{ccl}
\includegraphics[width=3in]{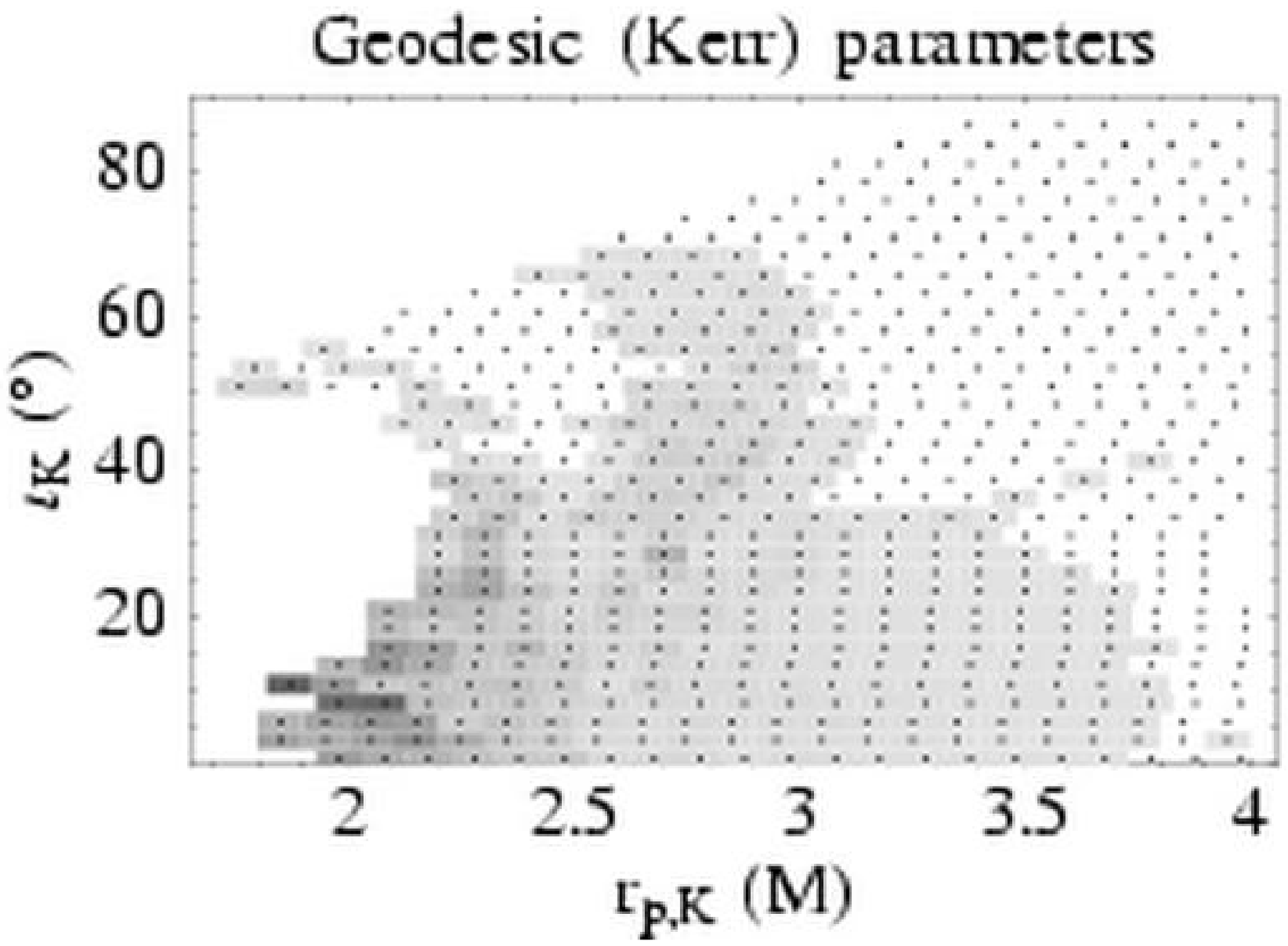}
	& \includegraphics[width=3in]{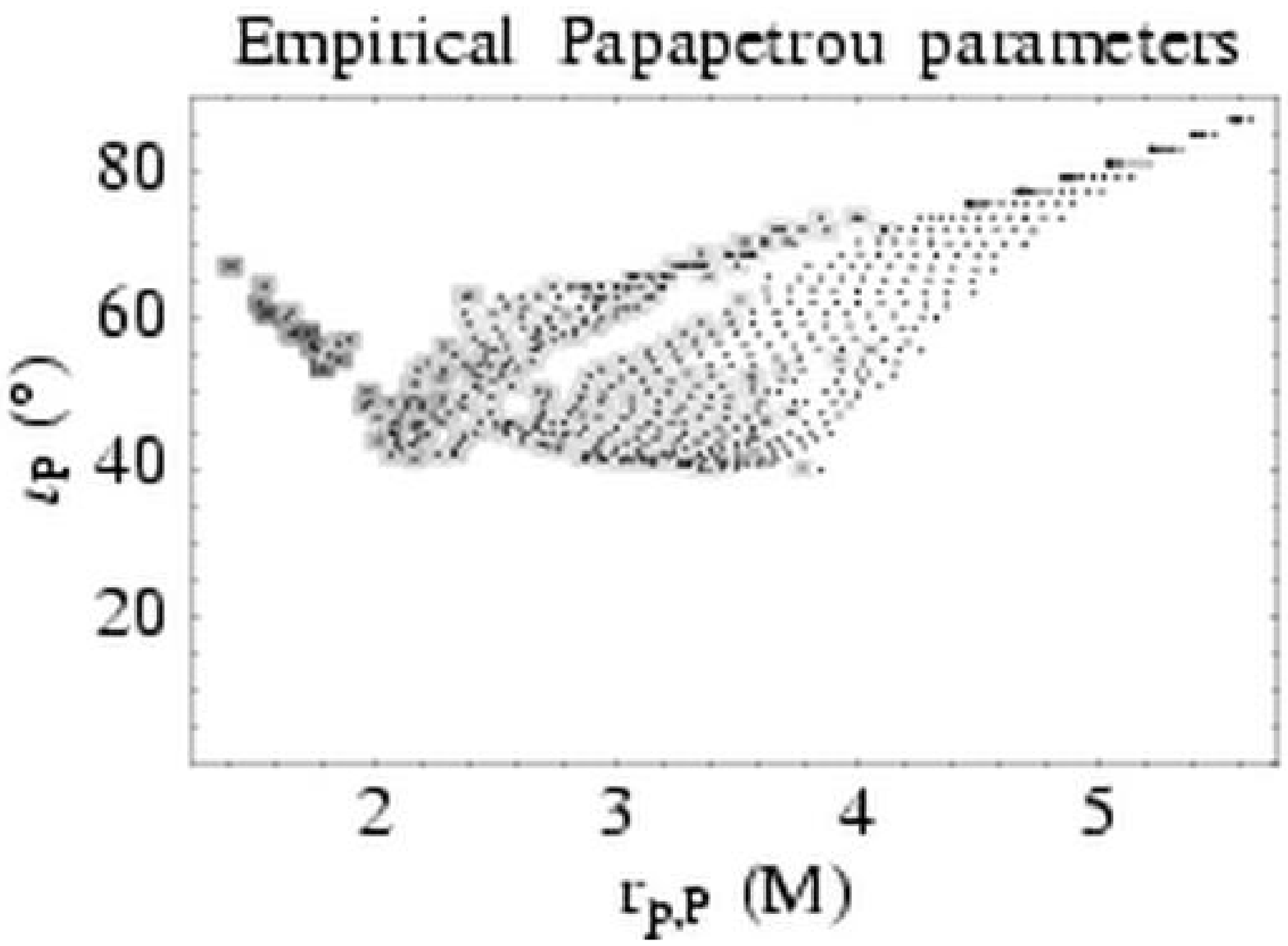}
    & \includegraphics[height=2in]{r_iota_1_0.5_1_0.2_-0.2_0_0.dat_sb.eps}\\
(a) & (b) & \medskip\\
\end{tabular}
\caption{\label{fig:e=0.5_S=0.9} $r_p$-$\iota$ map for $S=0.9$, $a=1$, 
and $e=0.5$.  (a)~Requested parameters; (b)~empirical parameters.  The shading is
scaled to the same maximum Lyapunov exponent as Fig.~\ref{fig:e=0.5_S=1}.
Chaotic orbits are widespread.}
\end{figure*}

\begin{figure*}
\begin{tabular}{ccl}
\includegraphics[width=3in]{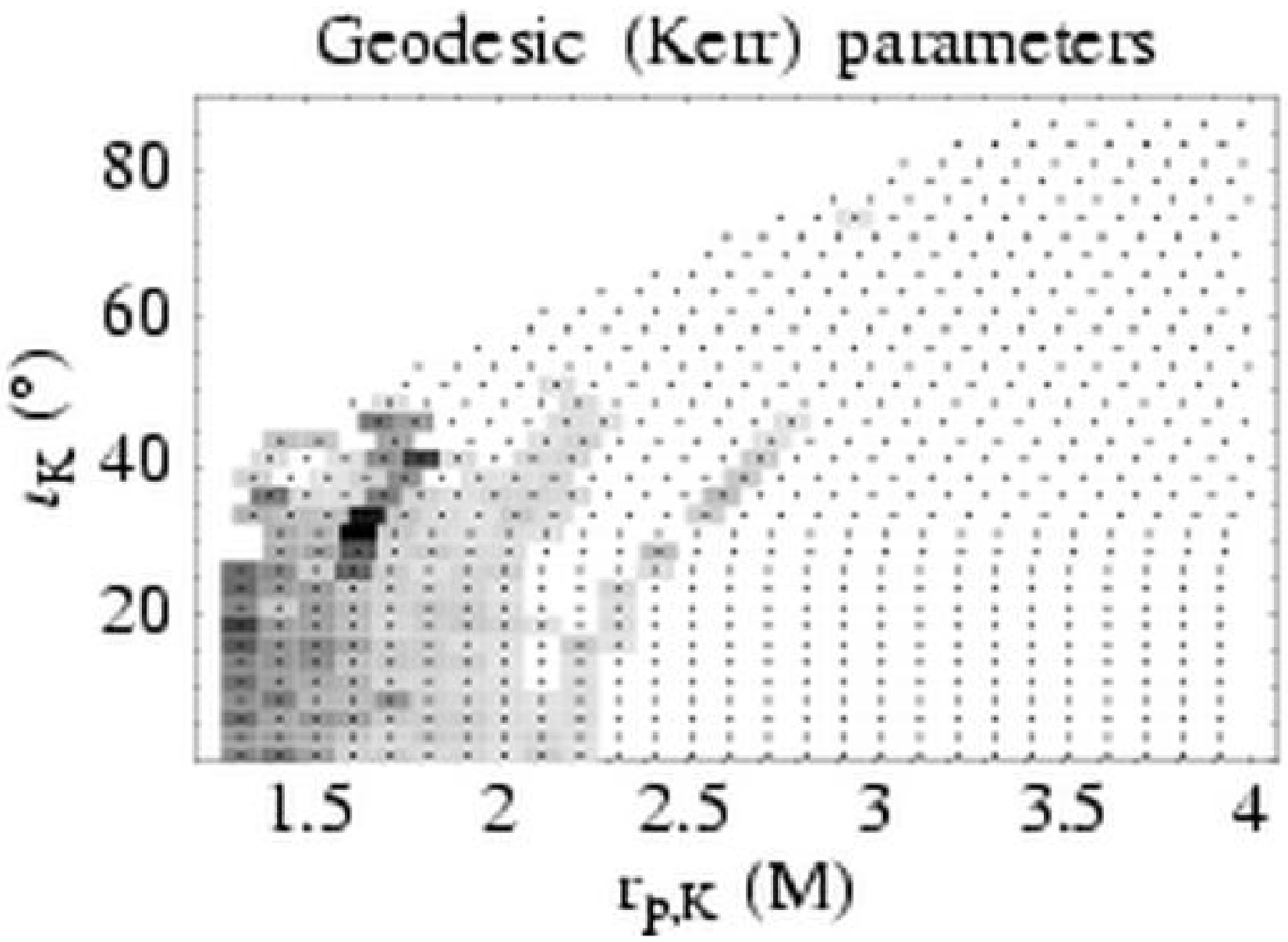}
	& \includegraphics[width=3in]{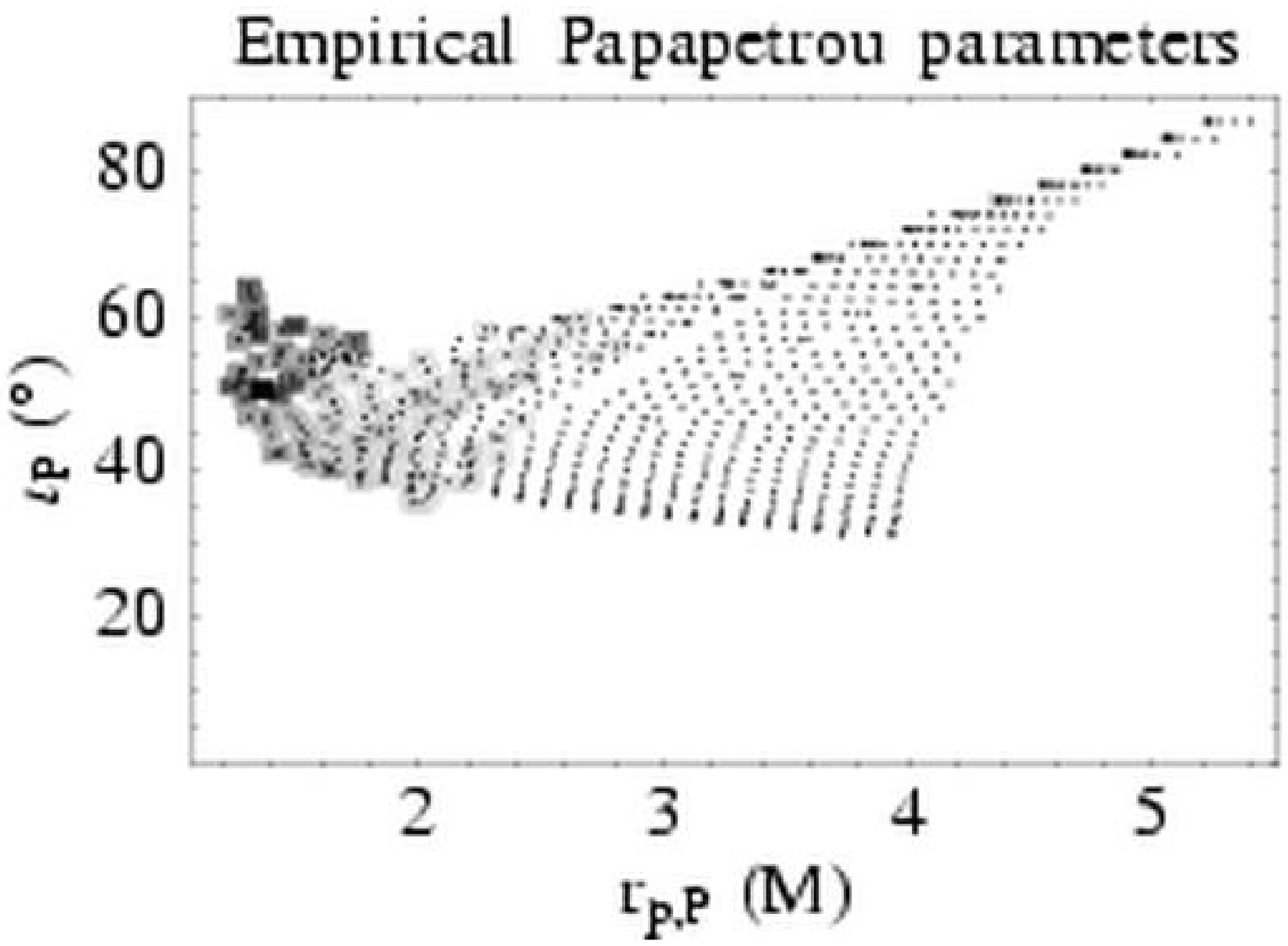}
    & \includegraphics[height=2in]{r_iota_1_0.5_1_0.2_-0.2_0_0.dat_sb.eps}\\
(a) & (b) & \medskip\\
\end{tabular}
\caption{\label{fig:e=0.5_S=0.5} $r_p$-$\iota$ map for $S=0.5$, $a=1$, 
and $e=0.5$.  (a)~Requested parameters; (b)~empirical parameters.  The shading is
scaled to the same maximum Lyapunov exponent as Fig.~\ref{fig:e=0.5_S=1}.
Because of the extremely low pericenters accessible at this value of $S$ (which
are excluded when $S=1$), the chaos for $S=0.5$ is the strongest we find.  The
largest Lyapunov exponent is just over $\lambda=0.01\,M^{-1}$.}
\end{figure*}

\begin{figure*}
\begin{tabular}{ccl}
\includegraphics[width=3in]{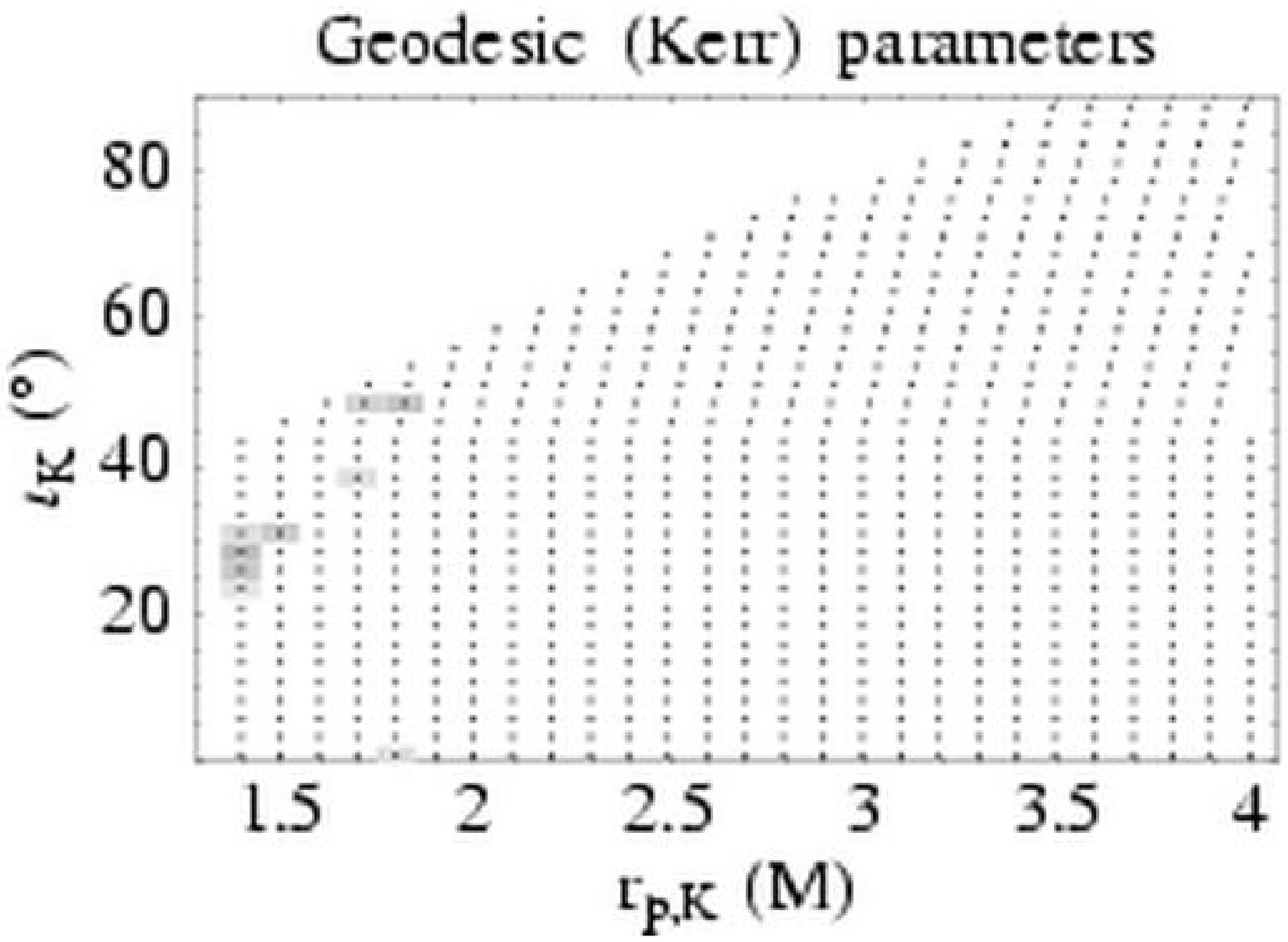}
	& \includegraphics[width=3in]{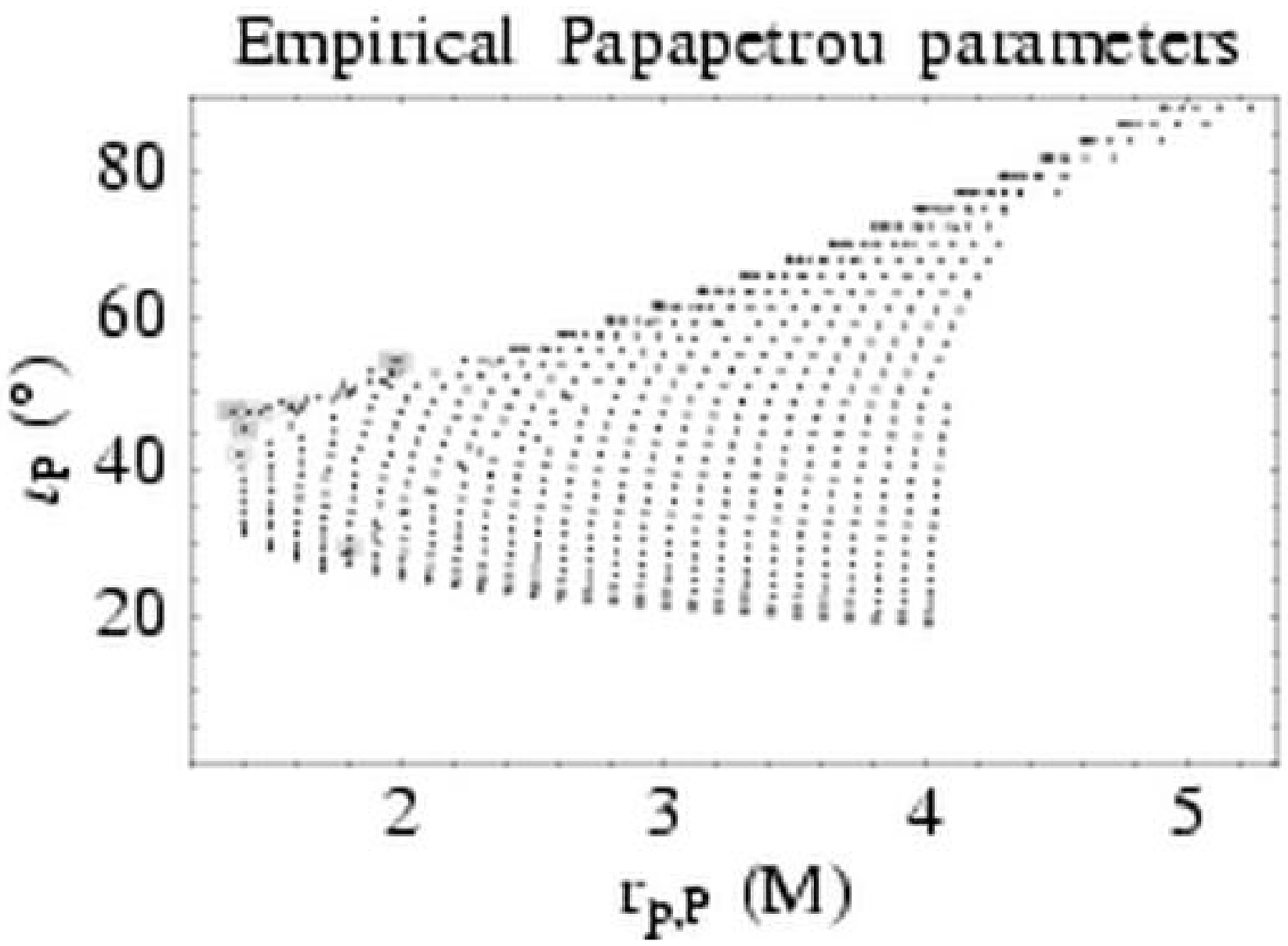}
    & \includegraphics[height=2in]{r_iota_1_0.5_1_0.2_-0.2_0_0.dat_sb.eps}\\
(a) & (b) & \medskip\\
\end{tabular}
\caption{\label{fig:e=0.5_S=0.2} $r_p$-$\iota$ map for $S=0.2$, $a=1$, 
and $e=0.5$.  (a)~Requested parameters; (b)~empirical parameters.  The shading is
scaled to the same maximum Lyapunov exponent as Fig.~\ref{fig:e=0.5_S=1}.
Only a few orbits are chaotic.}
\end{figure*}

\begin{figure*}
\begin{tabular}{ccl}
\includegraphics[width=3in]{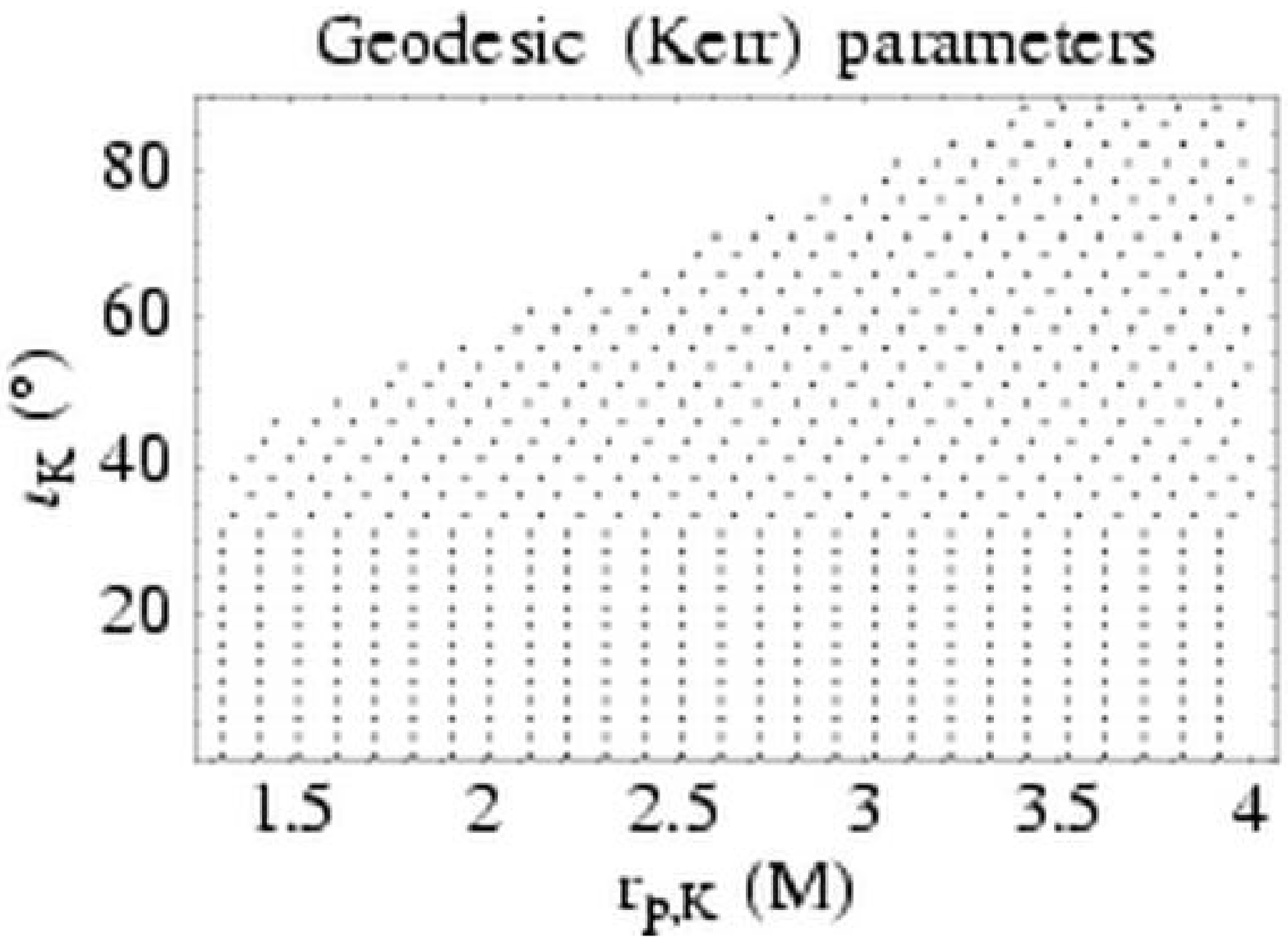}
	& \includegraphics[width=3in]{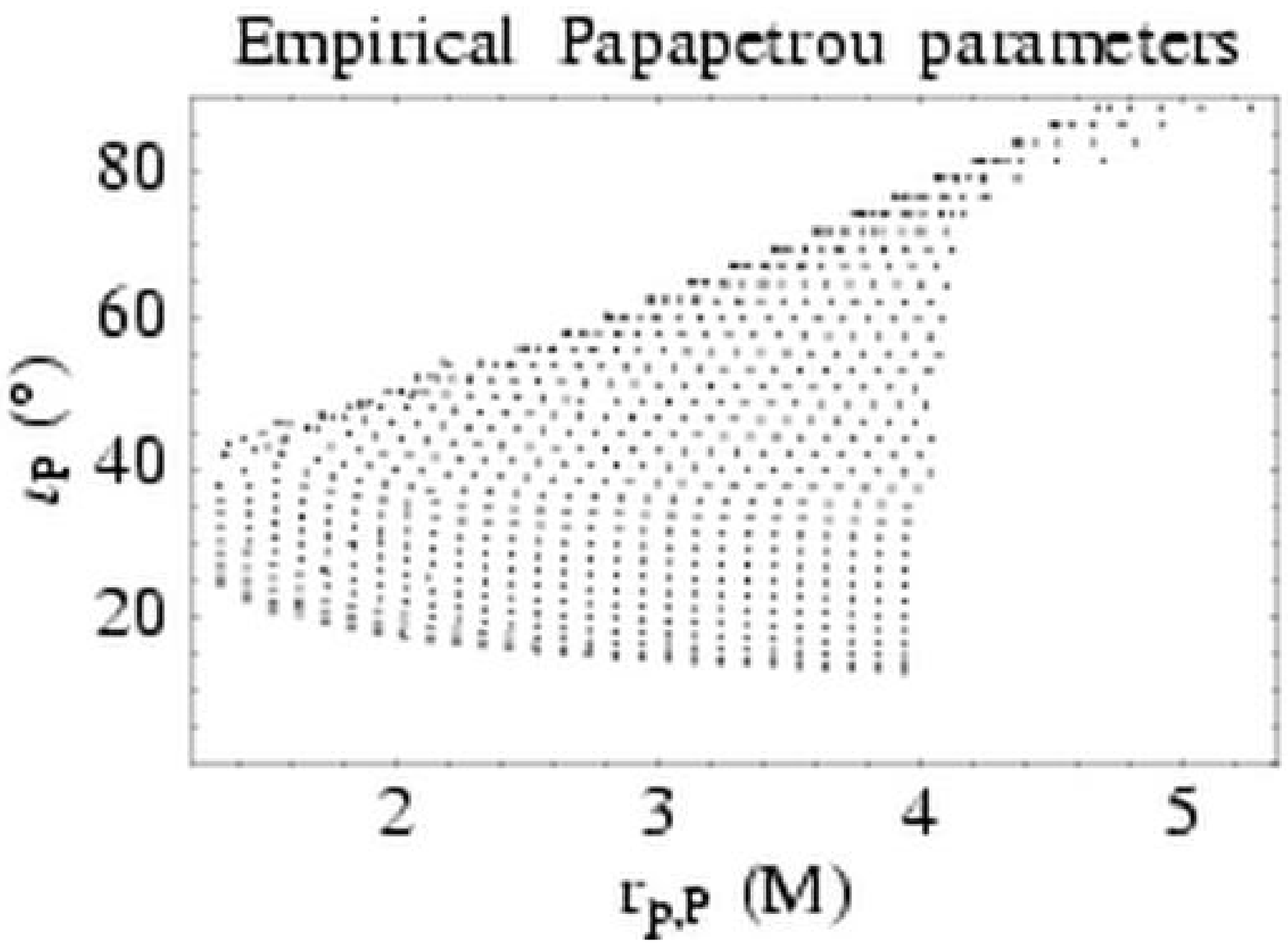}
    & \includegraphics[height=2in]{r_iota_1_0.5_1_0.2_-0.2_0_0.dat_sb.eps}\\
(a) & (b) & \medskip\\
\end{tabular}
\caption{\label{fig:e=0.5_S=0.1} $r_p$-$\iota$ map for $S=0.1$, $a=1$, 
and $e=0.5$.  (a)~Requested parameters; (b)~empirical parameters.  The shading is
scaled to the same maximum Lyapunov exponent as Fig.~\ref{fig:e=0.5_S=1}.
All chaos is gone, although the parameter space is still somewhat compressed.}
\end{figure*}

\begin{figure*}
\begin{tabular}{ccl}
\includegraphics[width=3in]{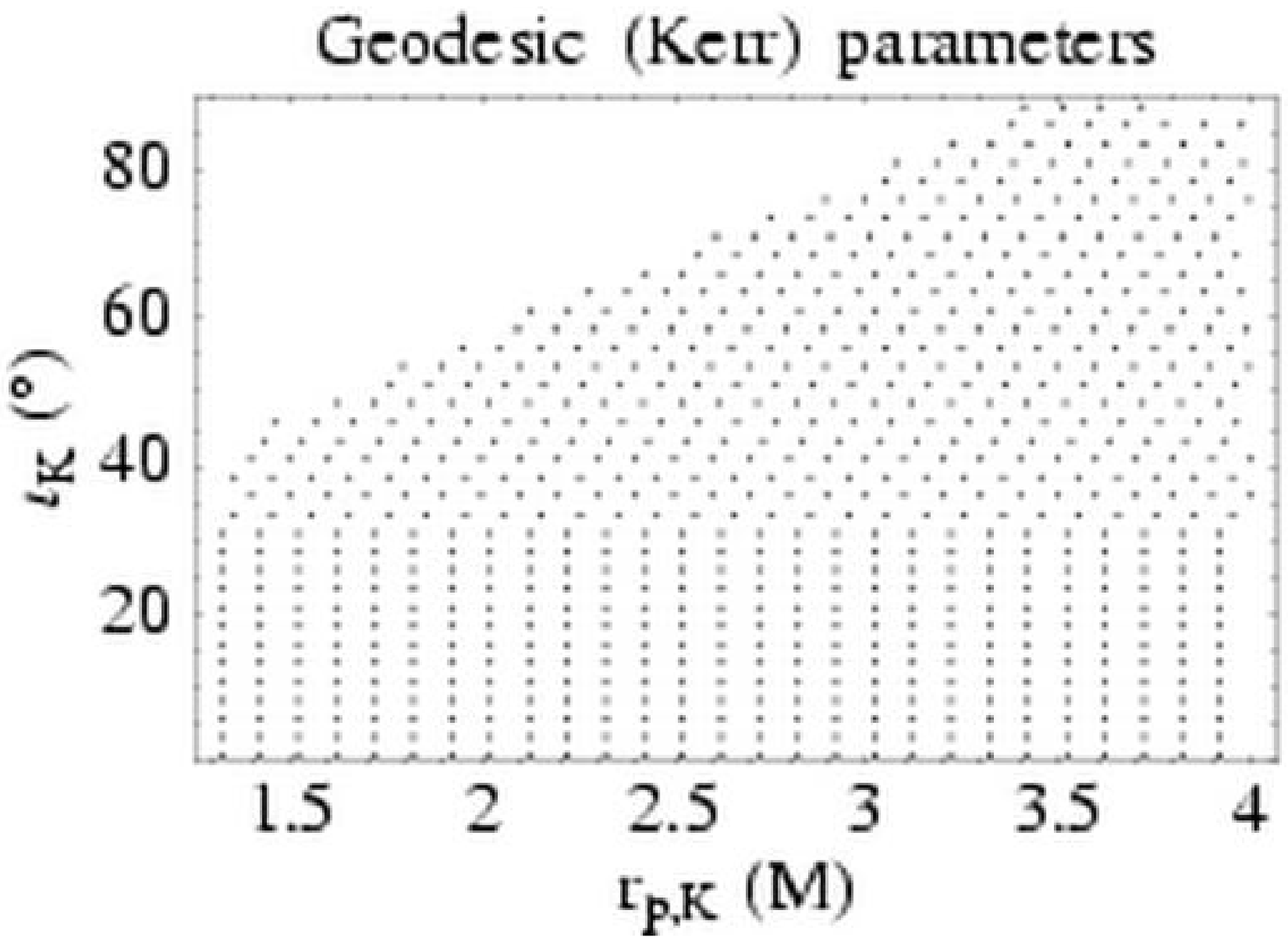}
	& \includegraphics[width=3in]{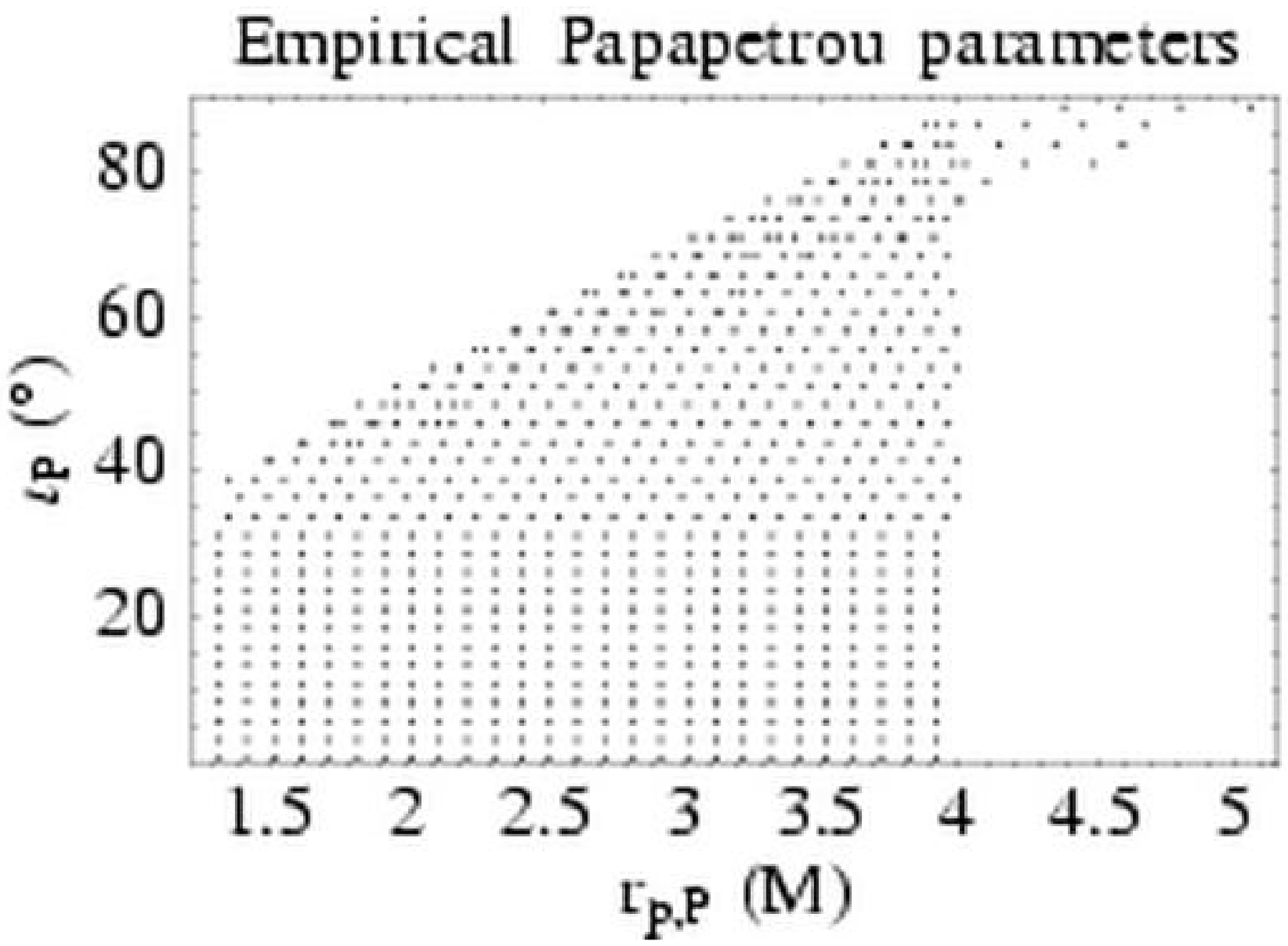}
    & \includegraphics[height=2in]{r_iota_1_0.5_1_0.2_-0.2_0_0.dat_sb.eps}\\
(a) & (b) & \medskip\\
\end{tabular}
\caption{\label{fig:e=0.5_S=1e-4} $r_p$-$\iota$ map for $S=10^{-4}$, $a=1$, 
and $e=0.5$.  (a)~Requested parameters; (b)~empirical parameters.  The shading is
scaled to the same maximum Lyapunov exponent as Fig.~\ref{fig:e=0.5_S=1}.
There are no chaotic orbits.  The empirical parameter values in~(b) are
indistinguishable from the requested values except for initial conditions that
specify values of~$\iota$ corresponding to
unstable orbits (as discussed in Sec.~\ref{sec:empirical_orbit}).}
\end{figure*}

\begin{figure*}
\begin{tabular}{ccl}
\includegraphics[width=3in]{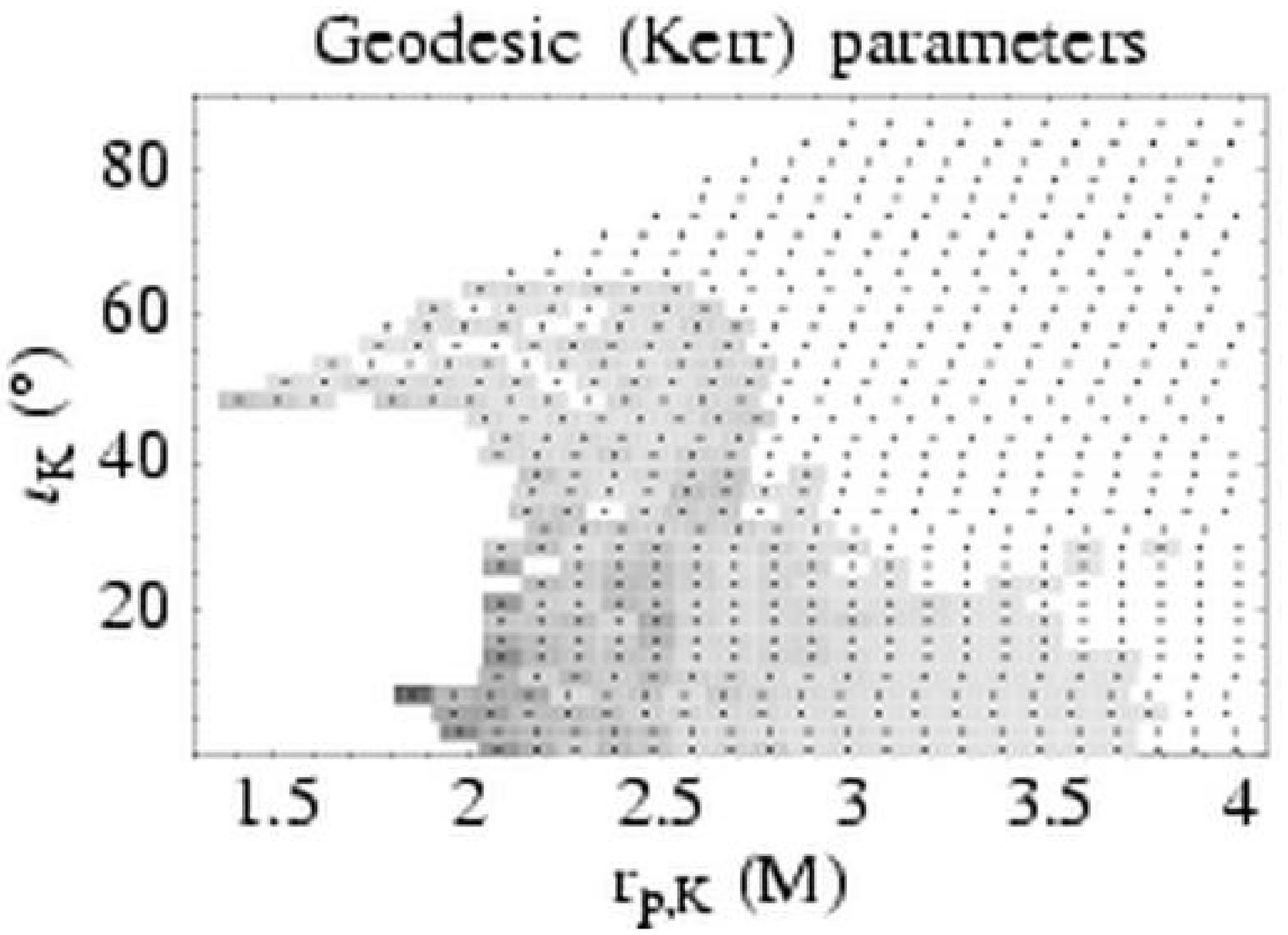}
	& \includegraphics[width=3in]{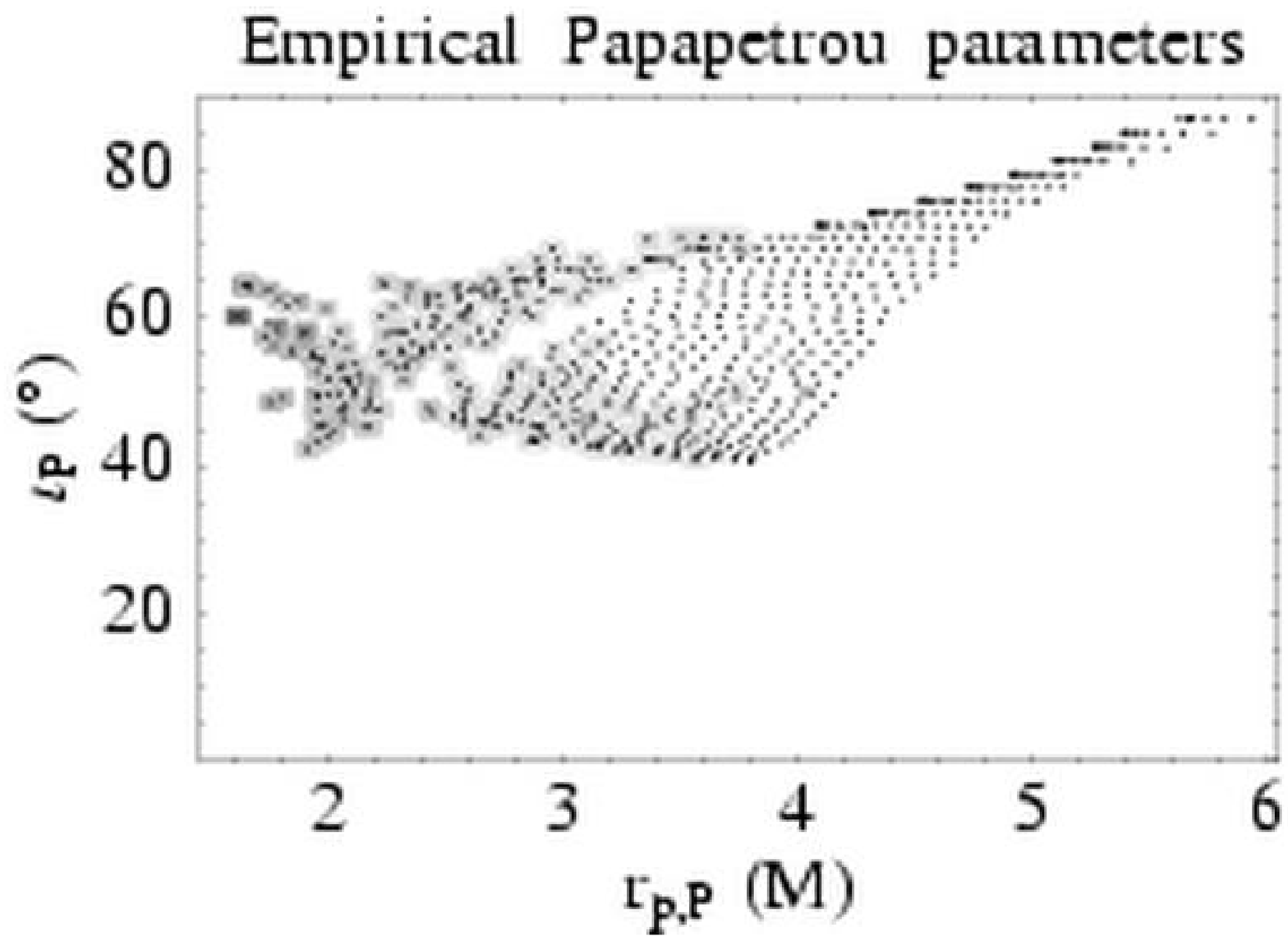}
    & \includegraphics[height=2in]{r_iota_1_0.5_1_0.2_-0.2_0_0.dat_sb.eps}\\
(a) & (b) & \medskip\\
\end{tabular}
\caption{\label{fig:e=0.6_S=1} $r_p$-$\iota$ map for $S=1$, $a=1$, 
and $e=0.6$.  (a)~Requested parameters; (b)~empirical parameters.  The shading is
scaled to the same maximum Lyapunov exponent as Fig.~\ref{fig:e=0.5_S=1}.
Chaotic orbits are widespread.  The largest Lyapunov exponent is
$\lambda=5.5\times10^{-3}\,M^{-1}$.}
\end{figure*}

\begin{figure*}
\begin{tabular}{ccl}
\includegraphics[width=3in]{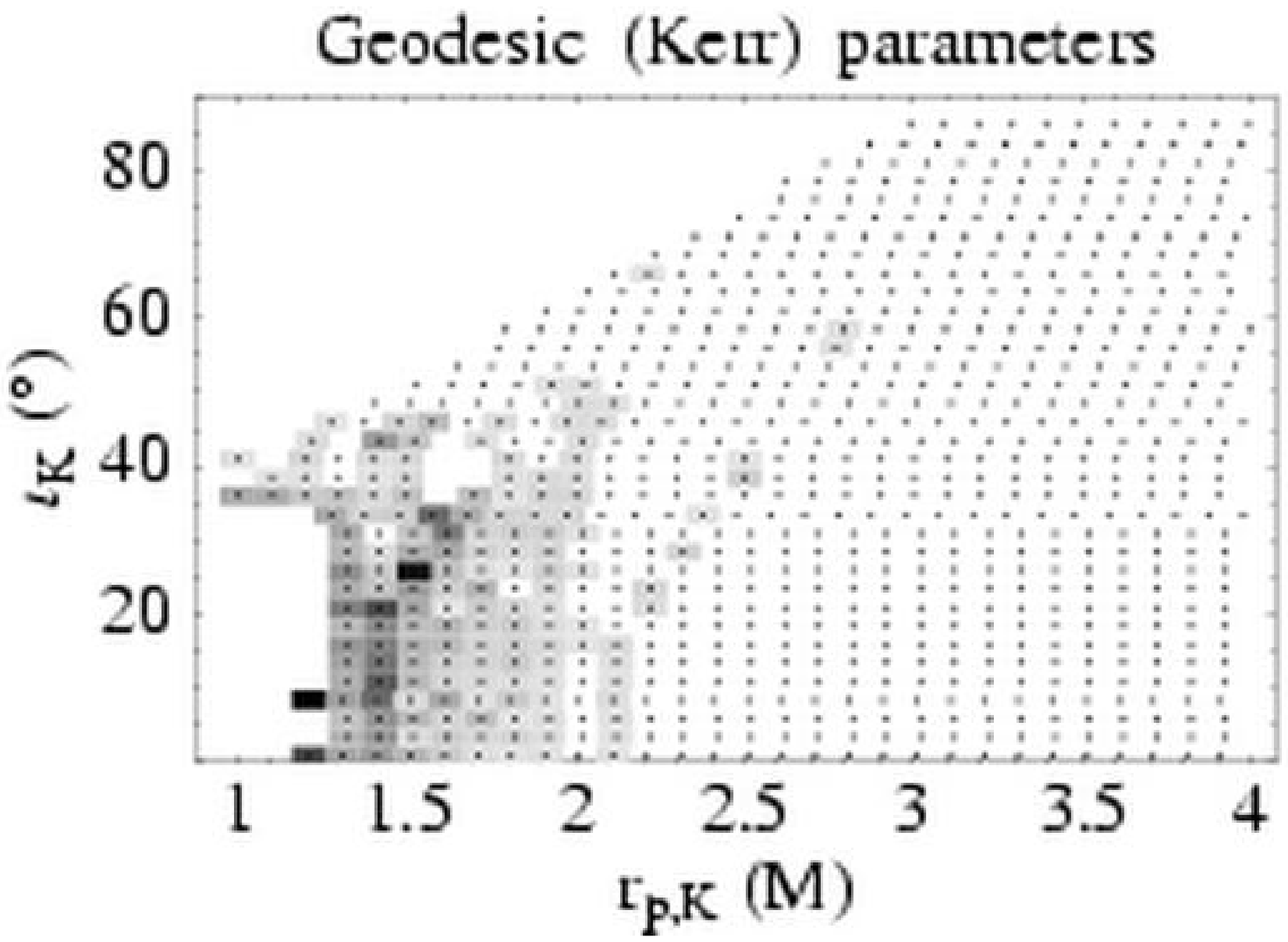}
	& \includegraphics[width=3in]{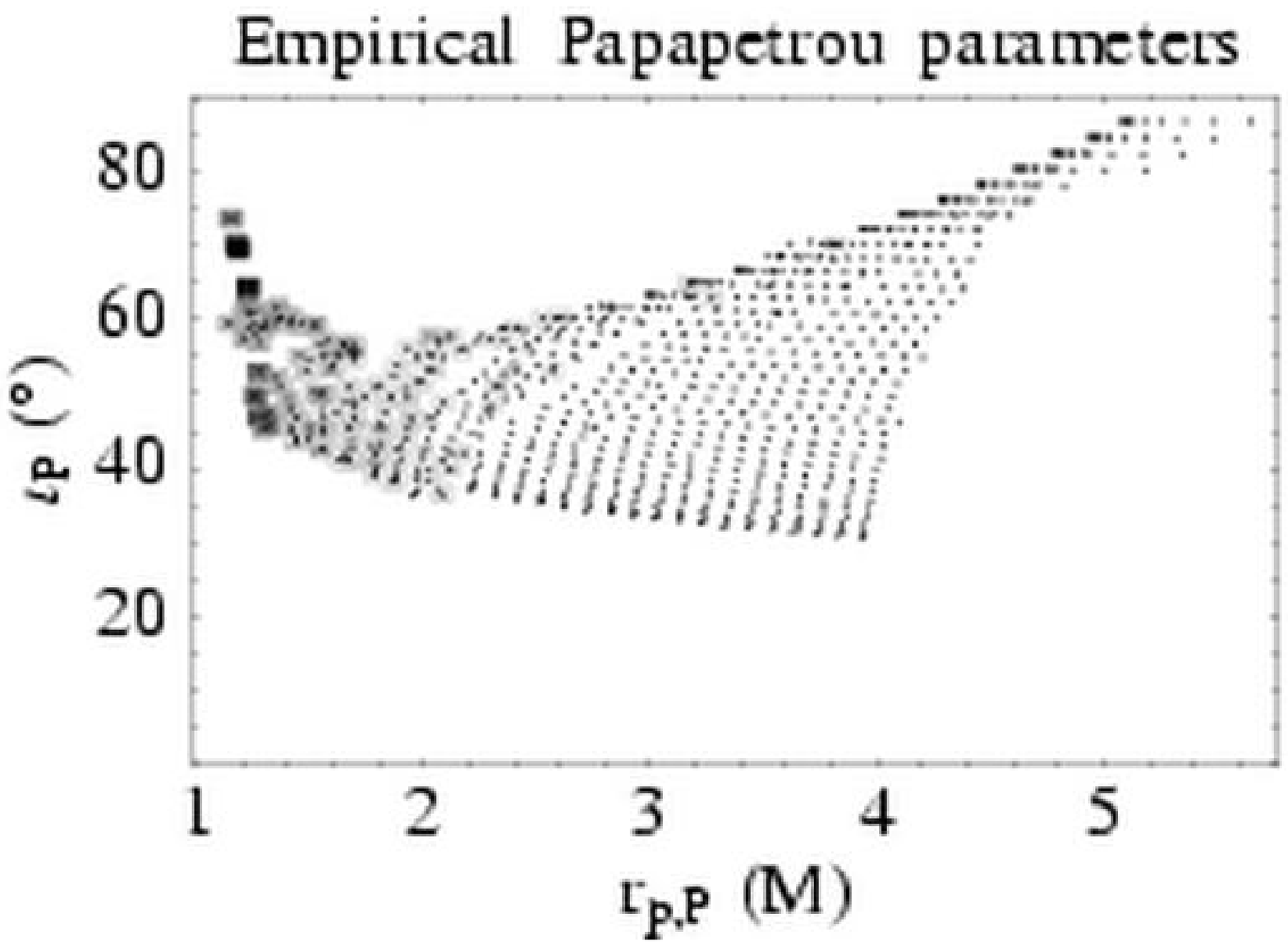}
    & \includegraphics[height=2in]{r_iota_1_0.5_1_0.2_-0.2_0_0.dat_sb.eps}\\
(a) & (b) & \medskip\\
\end{tabular}
\caption{\label{fig:e=0.6_S=0.5} $r_p$-$\iota$ map for $S=0.5$, $a=1$, 
and $e=0.6$.  (a)~Requested parameters; (b)~empirical parameters.  The shading is
scaled to the same maximum Lyapunov exponent as Fig.~\ref{fig:e=0.5_S=1}.
As in the case of Fig.~\ref{fig:e=0.5_S=0.5}, the low accessible pericenters
give rise to strong chaos.  The largest Lyapunov exponent is
$\lambda=9.2\times10^{-3}\,M^{-1}$.}
\end{figure*}

\begin{figure*}
\begin{tabular}{ccl}
\includegraphics[width=3in]{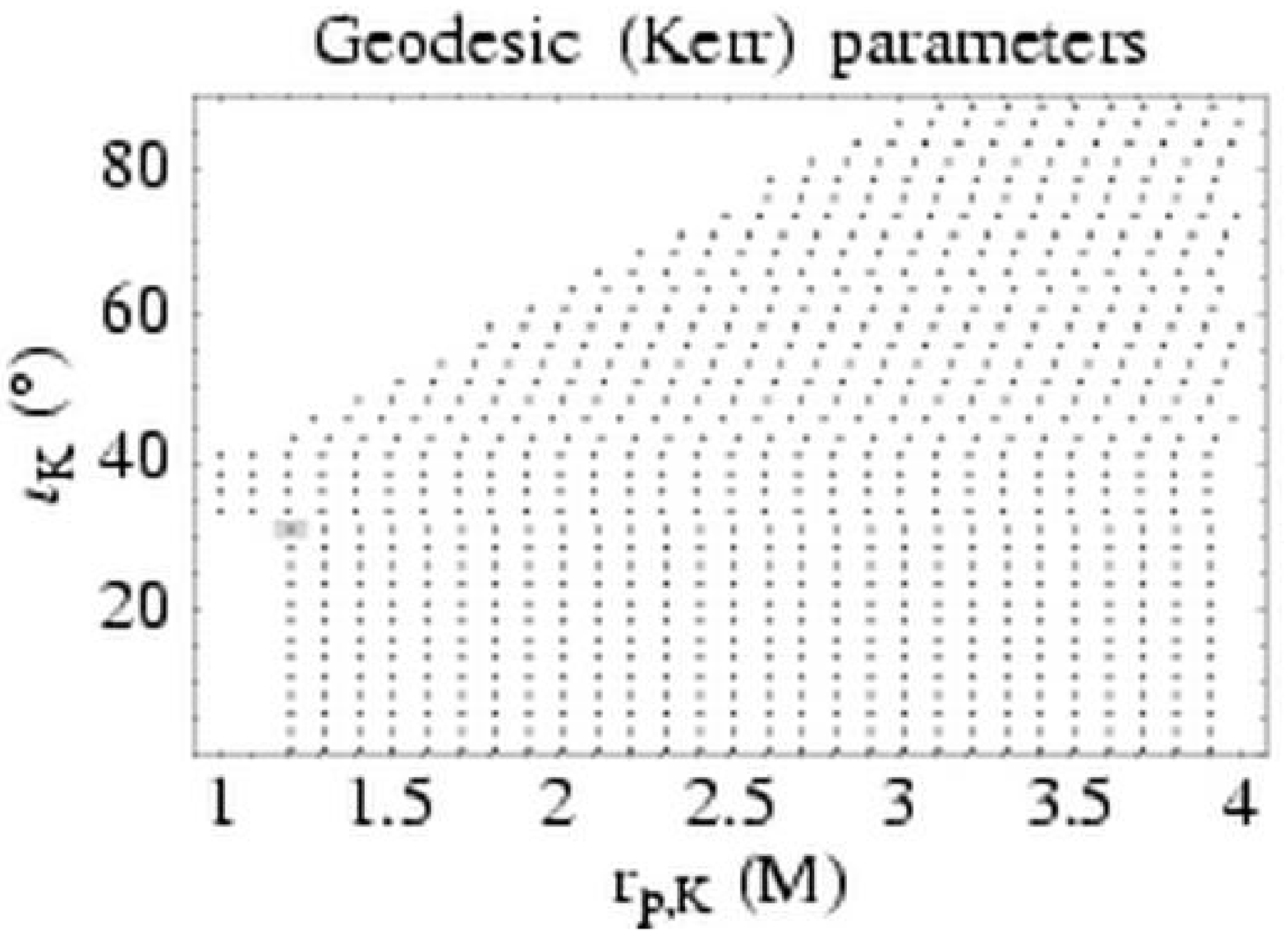}
	& \includegraphics[width=3in]{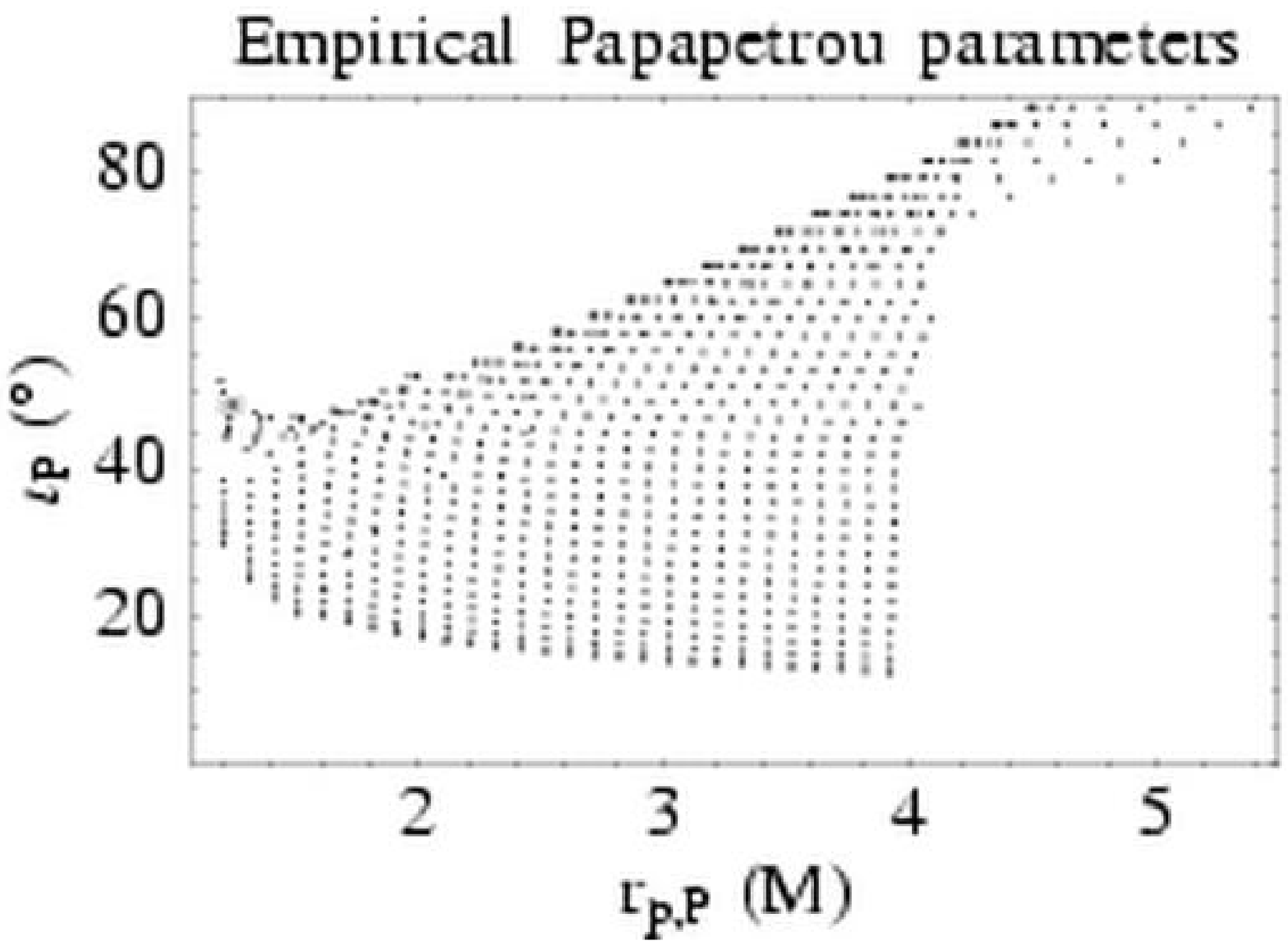}
    & \includegraphics[height=2in]{r_iota_1_0.5_1_0.2_-0.2_0_0.dat_sb.eps}\\
(a) & (b) & \medskip\\
\end{tabular}
\caption{\label{fig:e=0.6_S=0.1} $r_p$-$\iota$ map for $S=0.1$, $a=1$, 
and $e=0.6$.  (a)~Requested parameters; (b)~empirical parameters.  The shading is
scaled to the same maximum Lyapunov exponent as Fig.~\ref{fig:e=0.5_S=1}.
There is only one chaotic initial condition (which is in fact the only $S=0.1$
chaos we find), but the chaos is real, as discussed in Sec.~\ref{sec:e} and
illustrated in Fig.~\ref{fig:lyap_long_e=0.6}.
The Lyapunov exponent is
$\lambda=1.0\times10^{-3}\,M^{-1}$.}
\end{figure*}

\begin{figure*}
\begin{tabular}{ccl}
\includegraphics[width=3in]{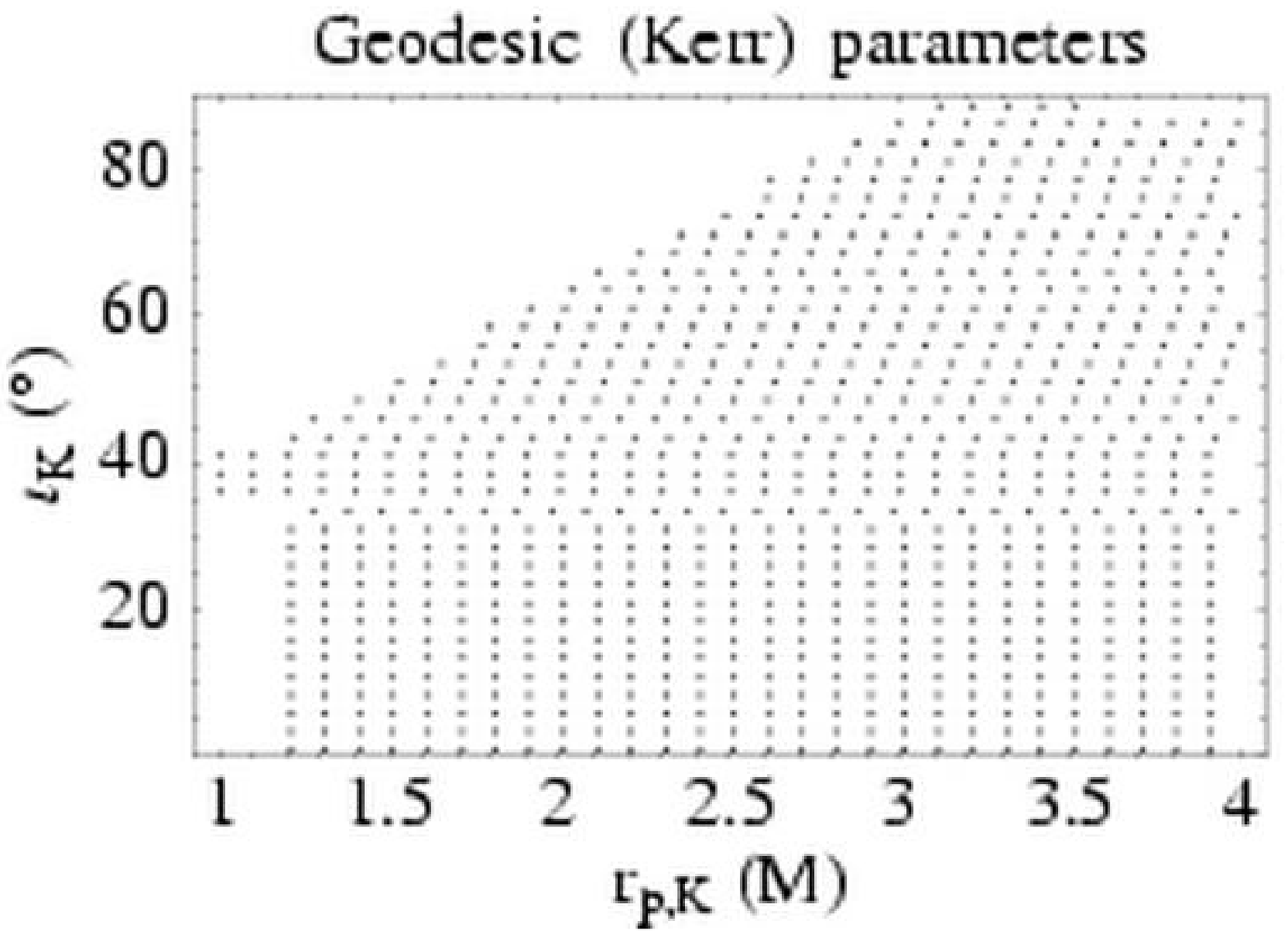}
	& \includegraphics[width=3in]{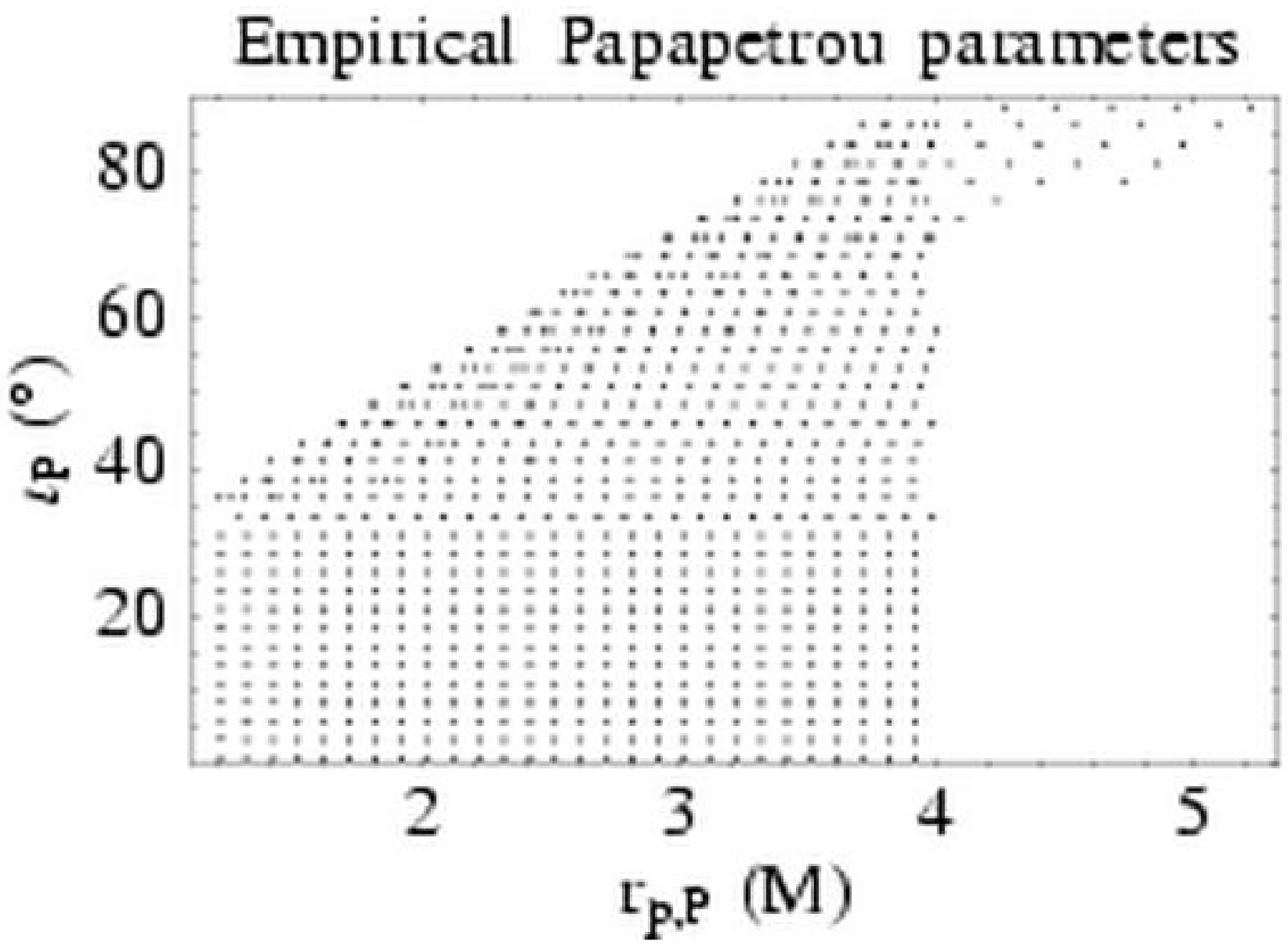}
    & \includegraphics[height=2in]{r_iota_1_0.5_1_0.2_-0.2_0_0.dat_sb.eps}\\
(a) & (b) & \medskip\\
\end{tabular}
\caption{\label{fig:e=0.6_S=1e-4} $r_p$-$\iota$ map for $S=10^{-4}$, $a=1$, 
and $e=0.6$.  (a)~Requested parameters; (b)~empirical parameters.  
The shading is
scaled to the same maximum Lyapunov exponent as Fig.~\ref{fig:e=0.5_S=1}.
All chaos has disappeared.  As in Fig.~\ref{fig:e=0.5_S=1e-4}, the empirical 
parameter values in~(b) are
indistinguishable from the requested values except for initial conditions that
specify unstable orbits (Sec.~\ref{sec:empirical_orbit}).}
\end{figure*}

\begin{figure*}
\begin{tabular}{ccl}
\includegraphics[width=3in]{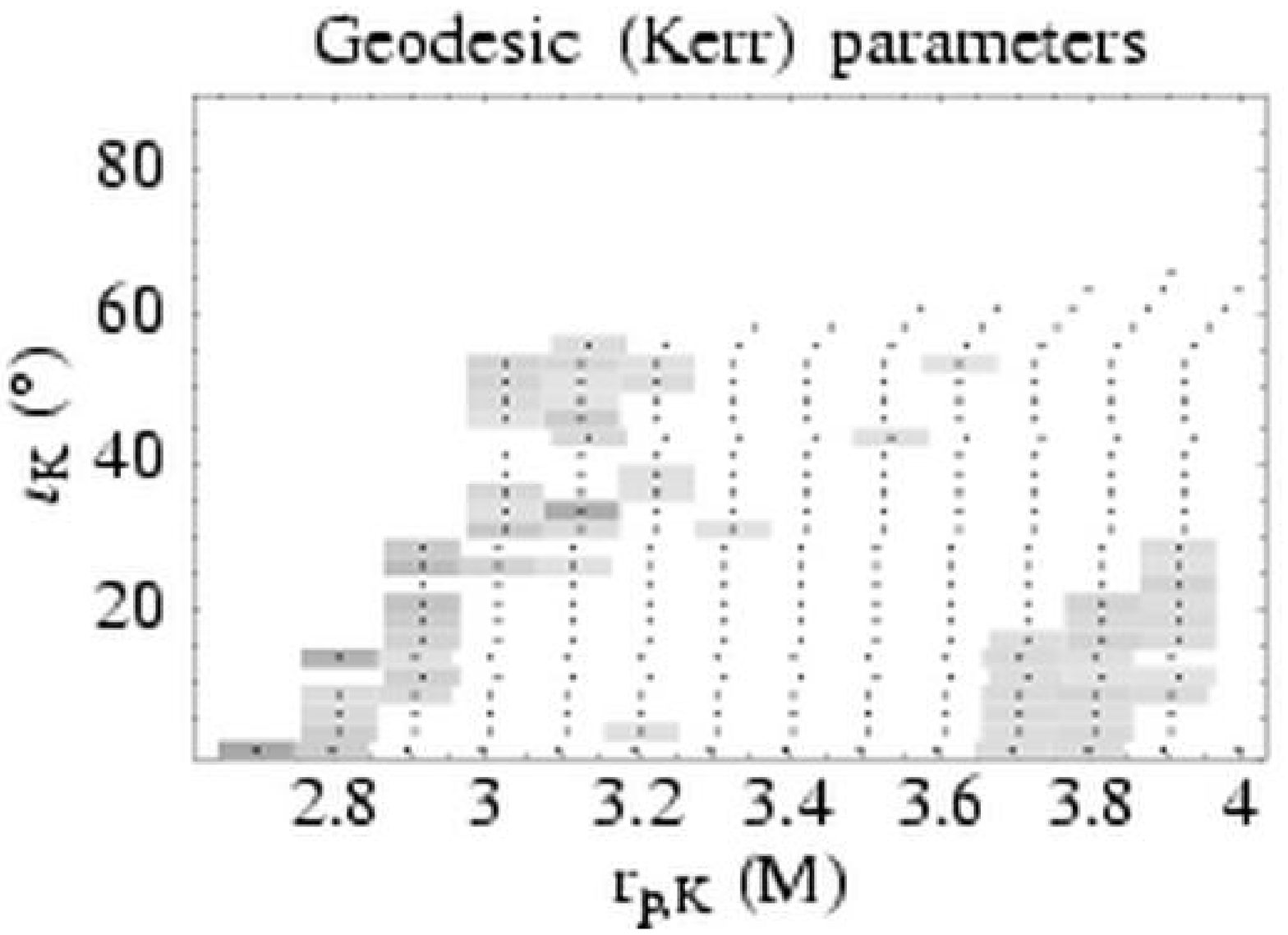}
	& \includegraphics[width=3in]{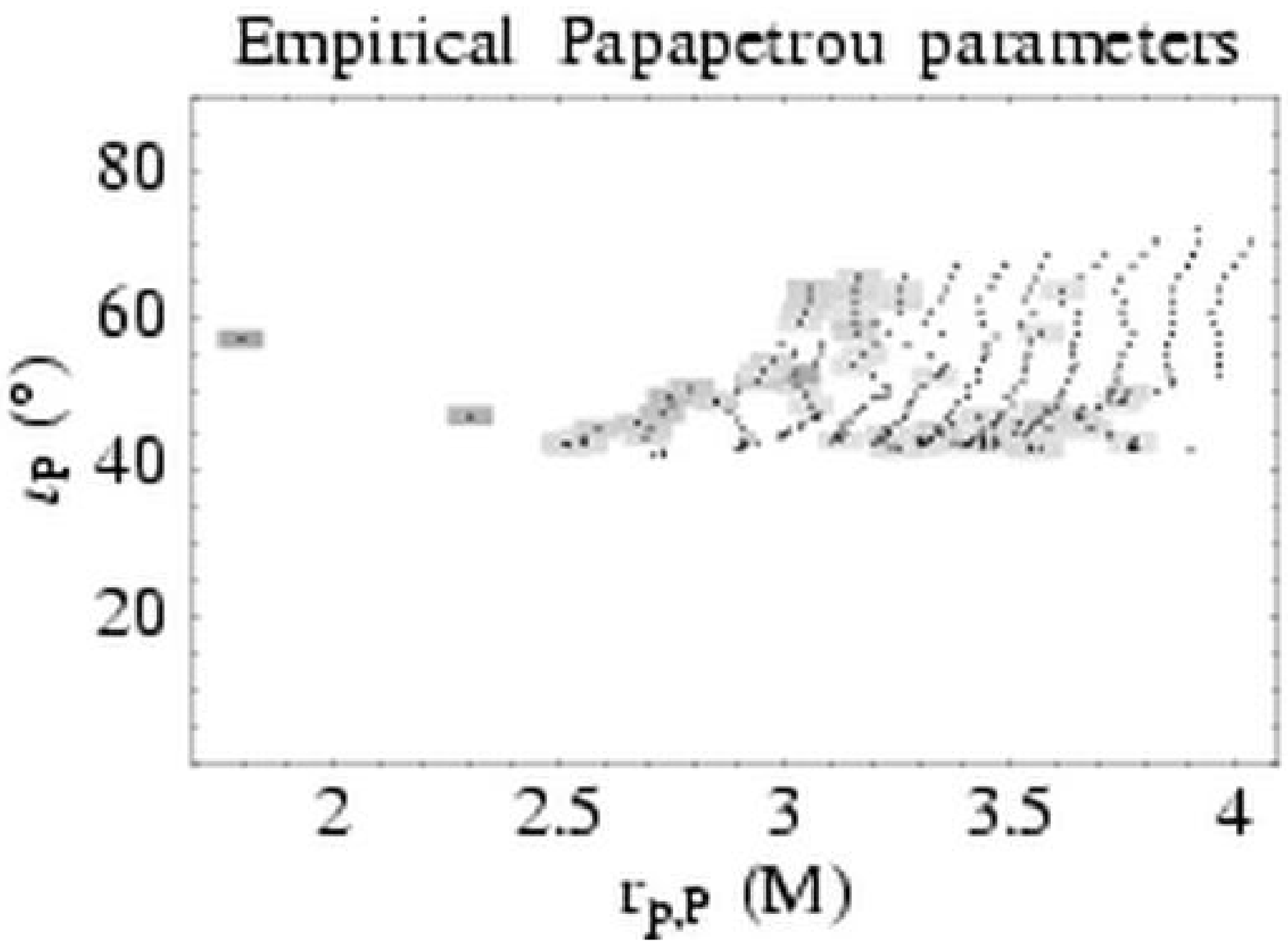}\
    & \includegraphics[height=2in]{r_iota_1_0.5_1_0.2_-0.2_0_0.dat_sb.eps}\\
(a) & (b) & \medskip\\
\end{tabular}
\caption{\label{fig:e=0.01_S=1} $r_p$-$\iota$ map for $S=1$, $a=1$, 
and $e=0.01$.  (a)~Requested parameters; (b)~empirical parameters.  The shading is
scaled to the same maximum Lyapunov exponent as Fig.~\ref{fig:e=0.5_S=1}.
There is some relatively weak chaos.  The largest Lyapunov exponent is
$\lambda=2.7\times10^{-3}\,M^{-1}$. }
\end{figure*}

\begin{figure*}
\begin{tabular}{ccl}
\includegraphics[width=3in]{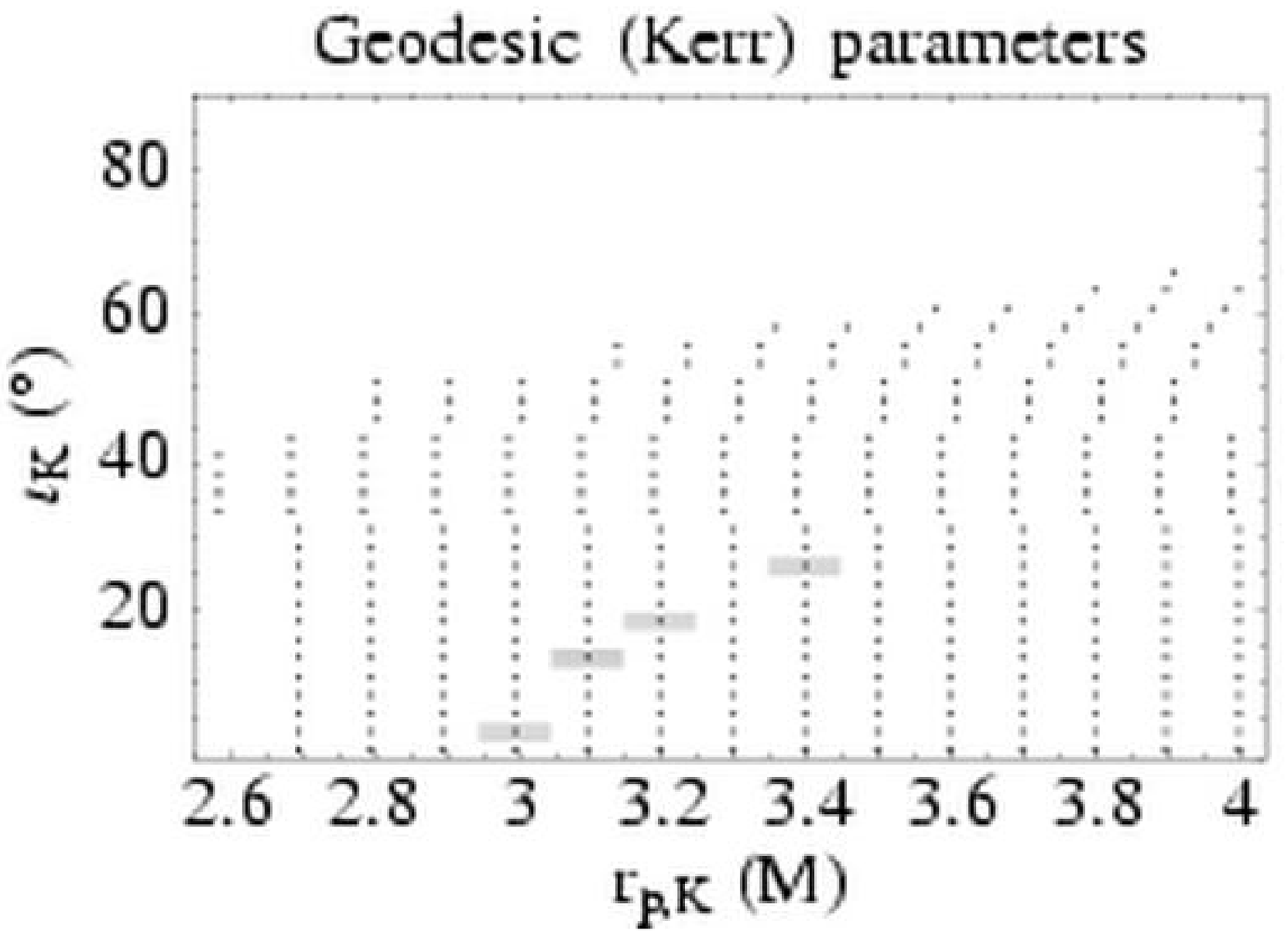}
	& \includegraphics[width=3in]{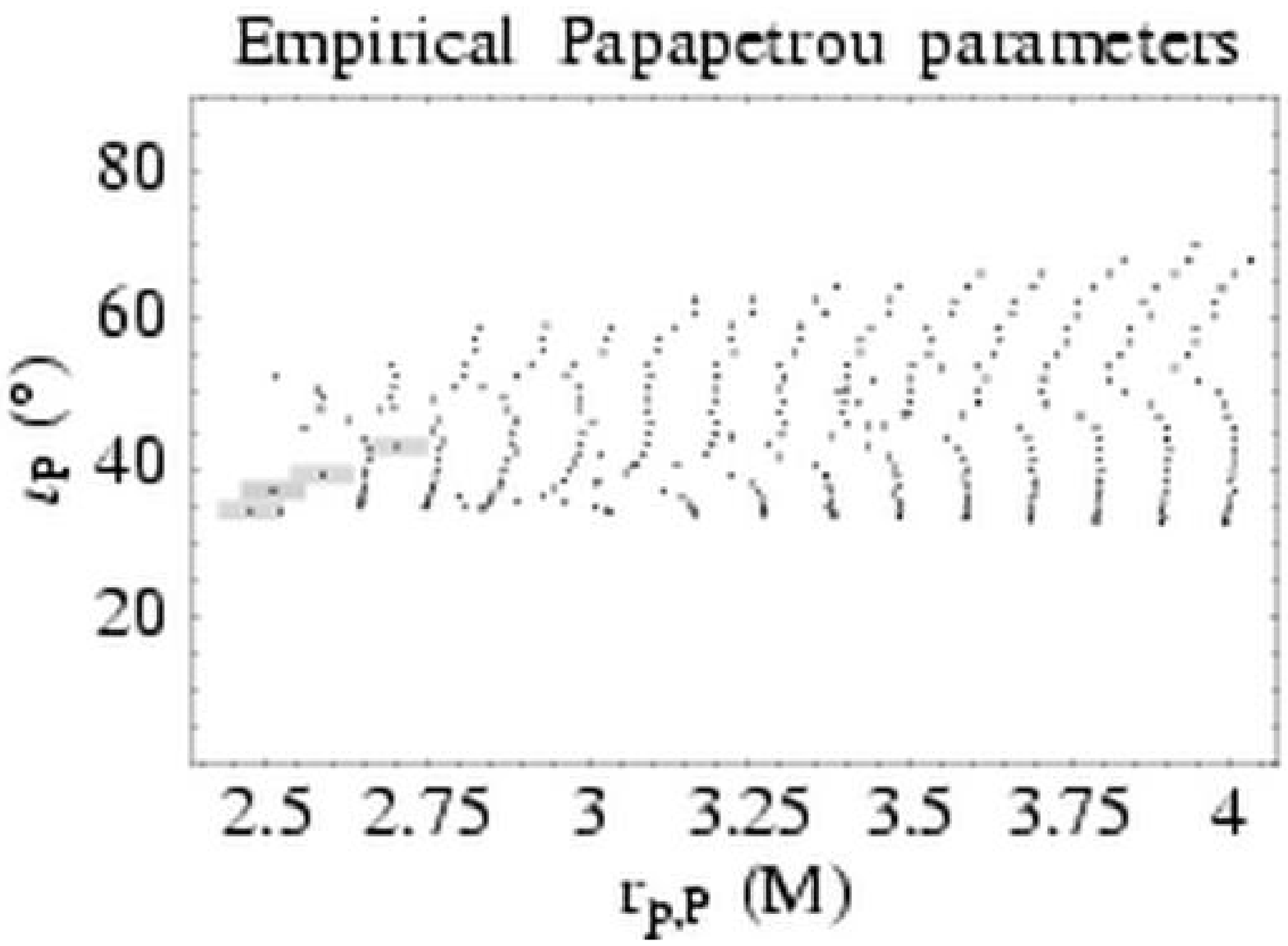}
    & \includegraphics[height=2in]{r_iota_1_0.5_1_0.2_-0.2_0_0.dat_sb.eps}\\
(a) & (b) & \medskip\\
\end{tabular}
\caption{\label{fig:e=0.01_S=0.5} $r_p$-$\iota$ map for $S=0.5$, $a=1$, 
and $e=0.01$.  (a)~Requested parameters; (b)~empirical parameters.  The shading is
scaled to the same maximum Lyapunov exponent as Fig.~\ref{fig:e=0.5_S=1}.
There are a few regions of weak chaos.}
\end{figure*}

\clearpage

\begin{figure*}
\begin{tabular}{ccl}
\includegraphics[width=3in]{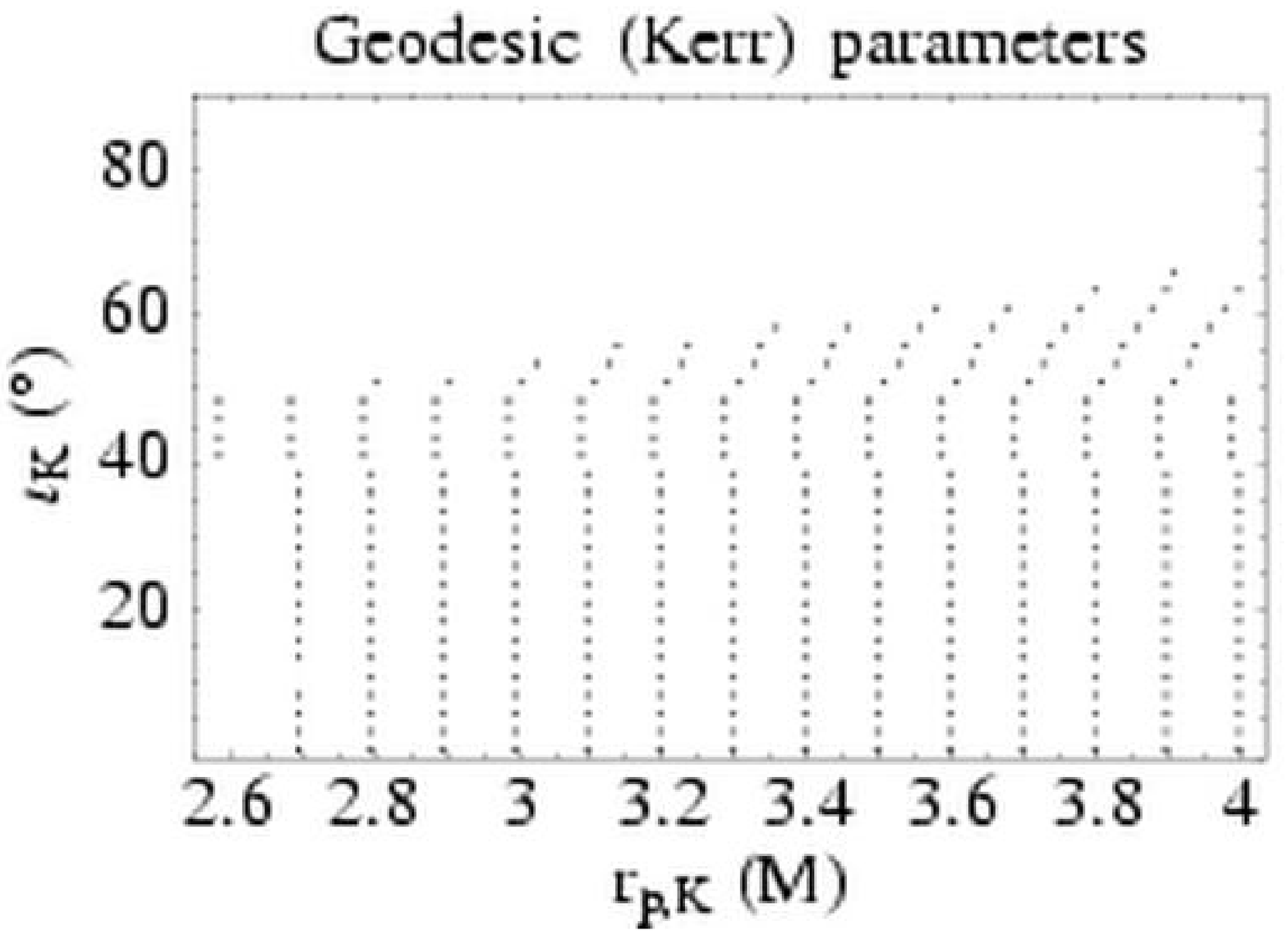}
	& \includegraphics[width=3in]{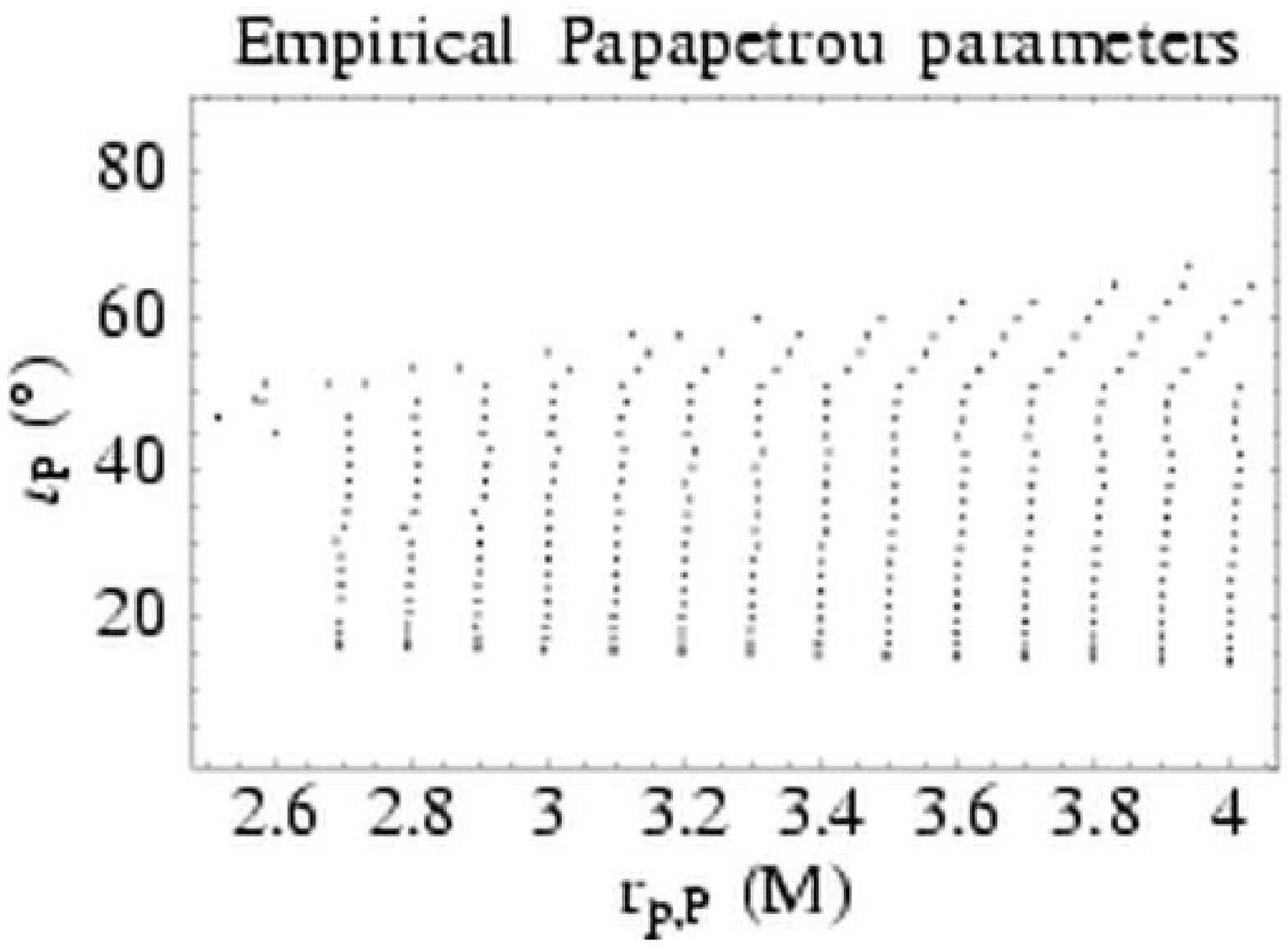}
    & \includegraphics[height=2in]{r_iota_1_0.5_1_0.2_-0.2_0_0.dat_sb.eps}\\
(a) & (b) & \medskip\\
\end{tabular}
\caption{\label{fig:e=0.01_S=0.1} $r_p$-$\iota$ map for $S=0.1$, $a=1$, 
and $e=0.01$.  (a)~Requested parameters; (b)~empirical parameters.  The shading is
scaled to the same maximum Lyapunov exponent as Fig.~\ref{fig:e=0.5_S=1}.
There are apparently no chaotic initial conditions.}
\end{figure*}

\begin{figure*}
\begin{tabular}{ccl}
\includegraphics[width=3in]{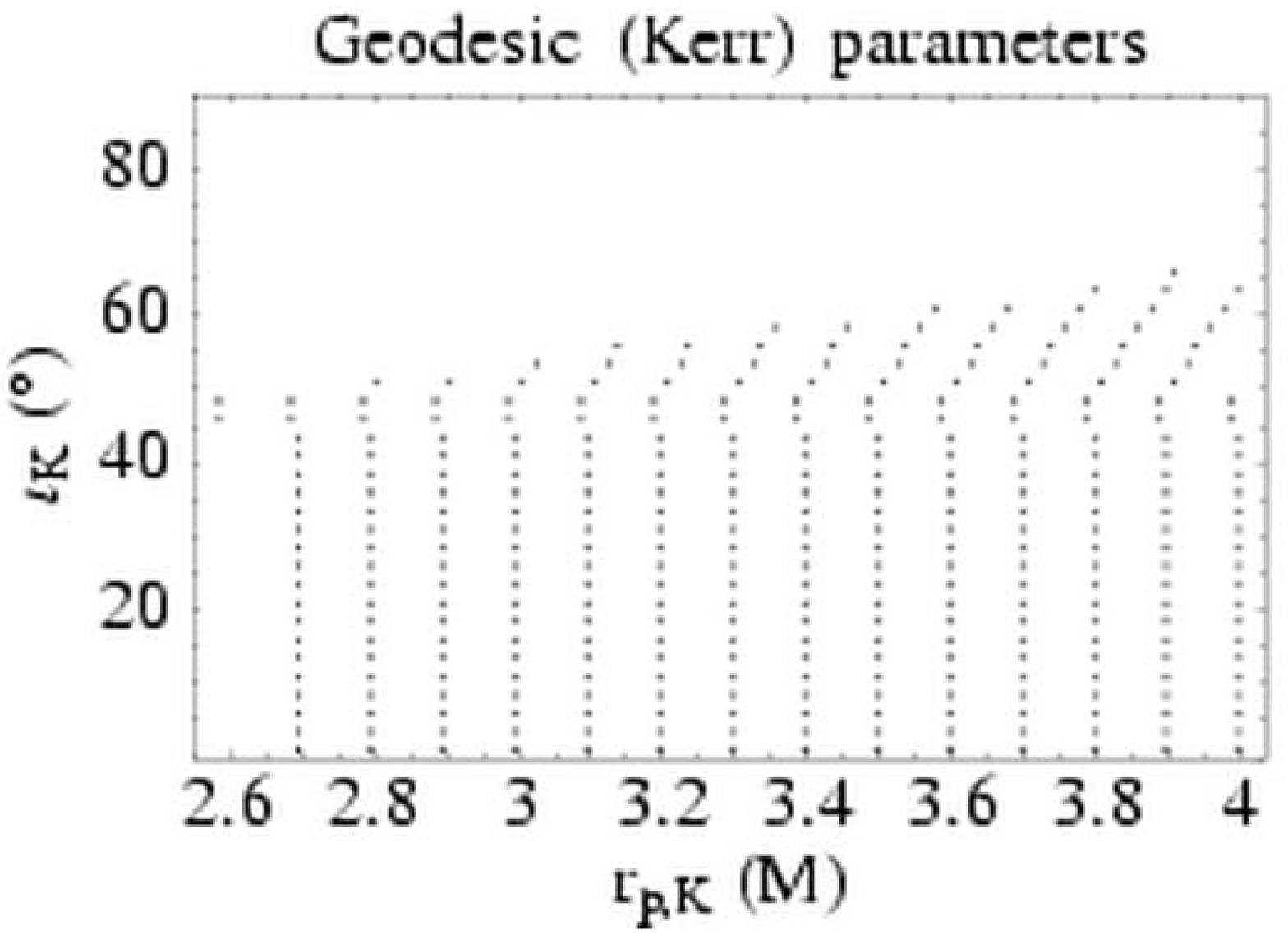}
	& \includegraphics[width=3in]{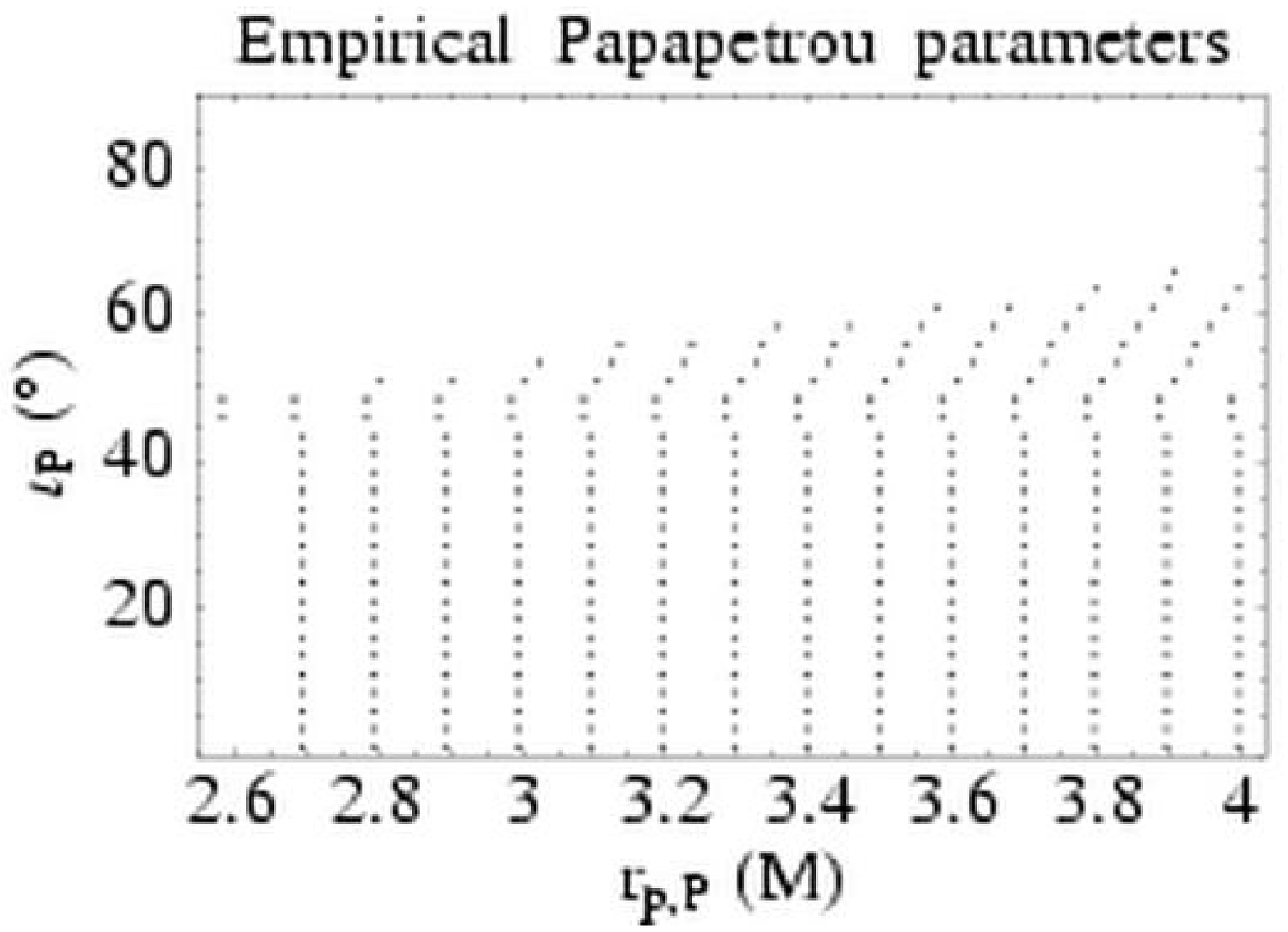}
    & \includegraphics[height=2in]{r_iota_1_0.5_1_0.2_-0.2_0_0.dat_sb.eps}\\
(a) & (b) & \medskip\\
\end{tabular}
\caption{\label{fig:e=0.01_S=1e-4} $r_p$-$\iota$ map for $S=10^{-4}$, $a=1$, 
and $e=0.01$.  (a)~Requested parameters; (b)~empirical parameters.  The shading is
scaled to the same maximum Lyapunov exponent as Fig.~\ref{fig:e=0.5_S=1}.
The chaos had disappeared.}
\end{figure*}

\begin{figure*}
\begin{tabular}{ccl}
\includegraphics[width=3in]{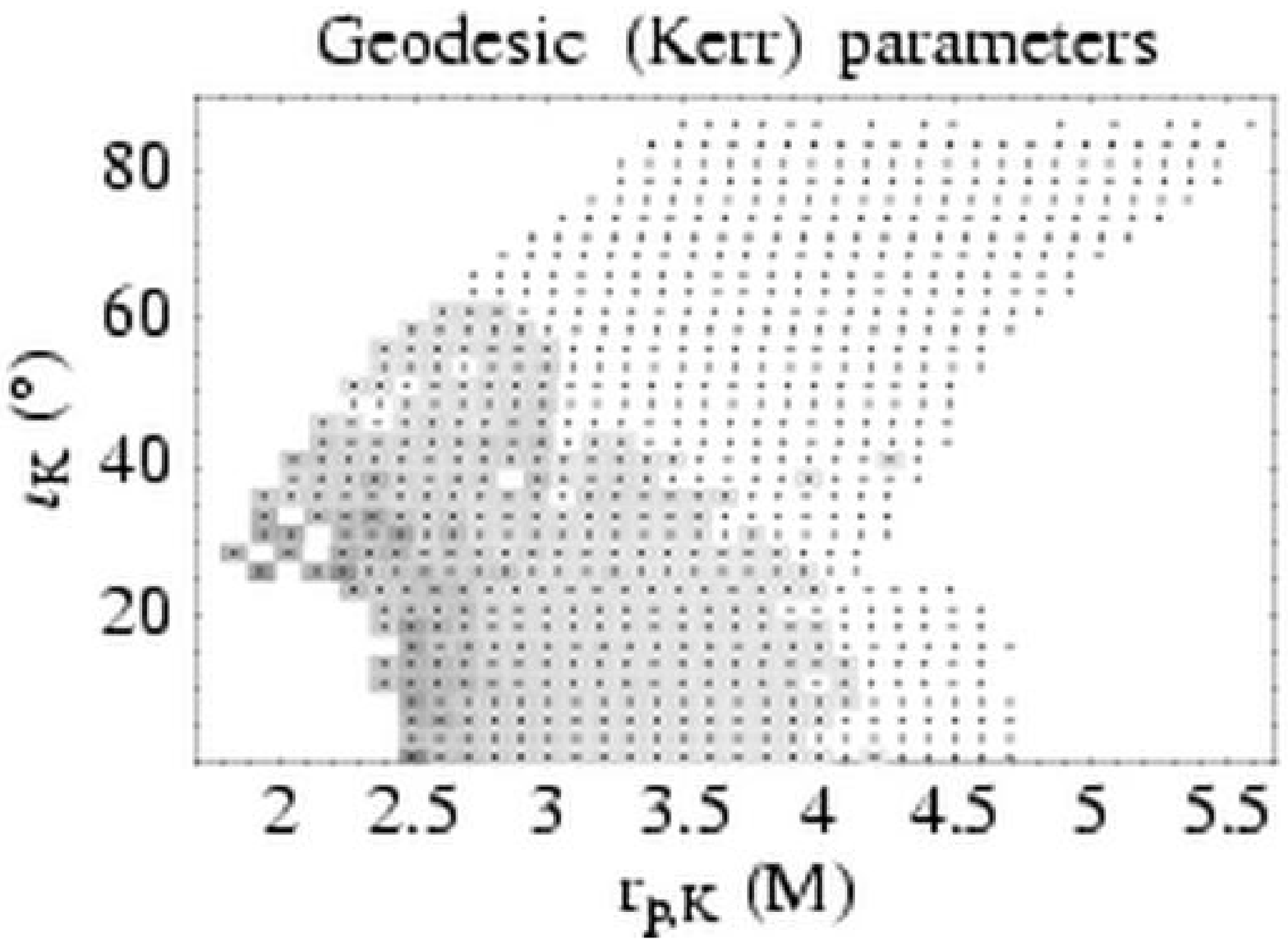}
	& \includegraphics[width=3in]{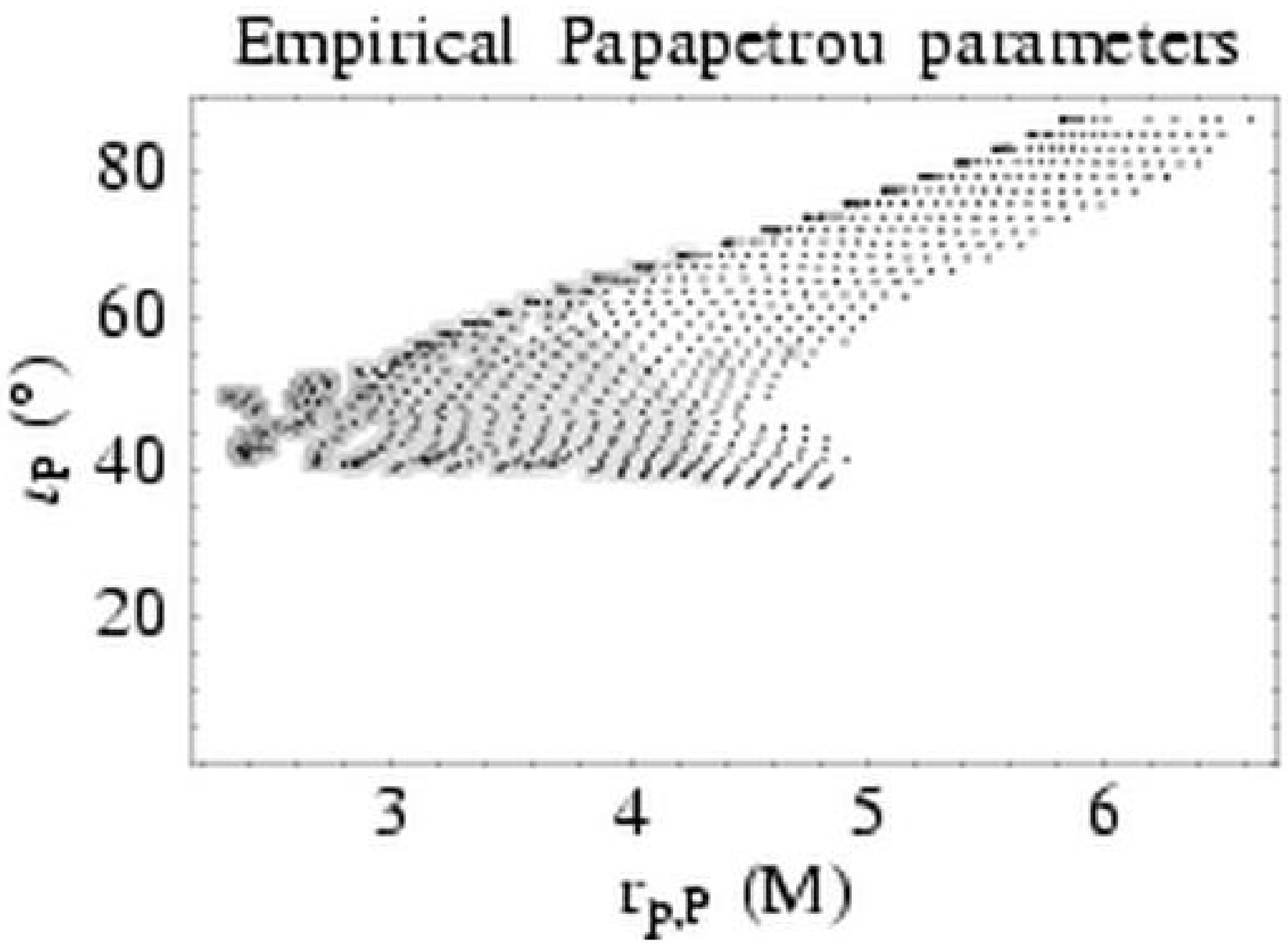}
    & \includegraphics[height=2in]{r_iota_1_0.5_1_0.2_-0.2_0_0.dat_sb.eps}\\
(a) & (b) & \medskip\\
\end{tabular}
\caption{\label{fig:a=0.9} $r_p$-$\iota$ map for $S=1$, $a=0.9$, 
and $e=0.5$.  (a)~Requested parameters; (b)~empirical parameters.  The shading is
scaled to the same maximum Lyapunov exponent as Fig.~\ref{fig:e=0.5_S=1}.
Chaotic orbits are widespread.}
\end{figure*}

\begin{figure*}
\begin{tabular}{ccl}
\includegraphics[width=3in]{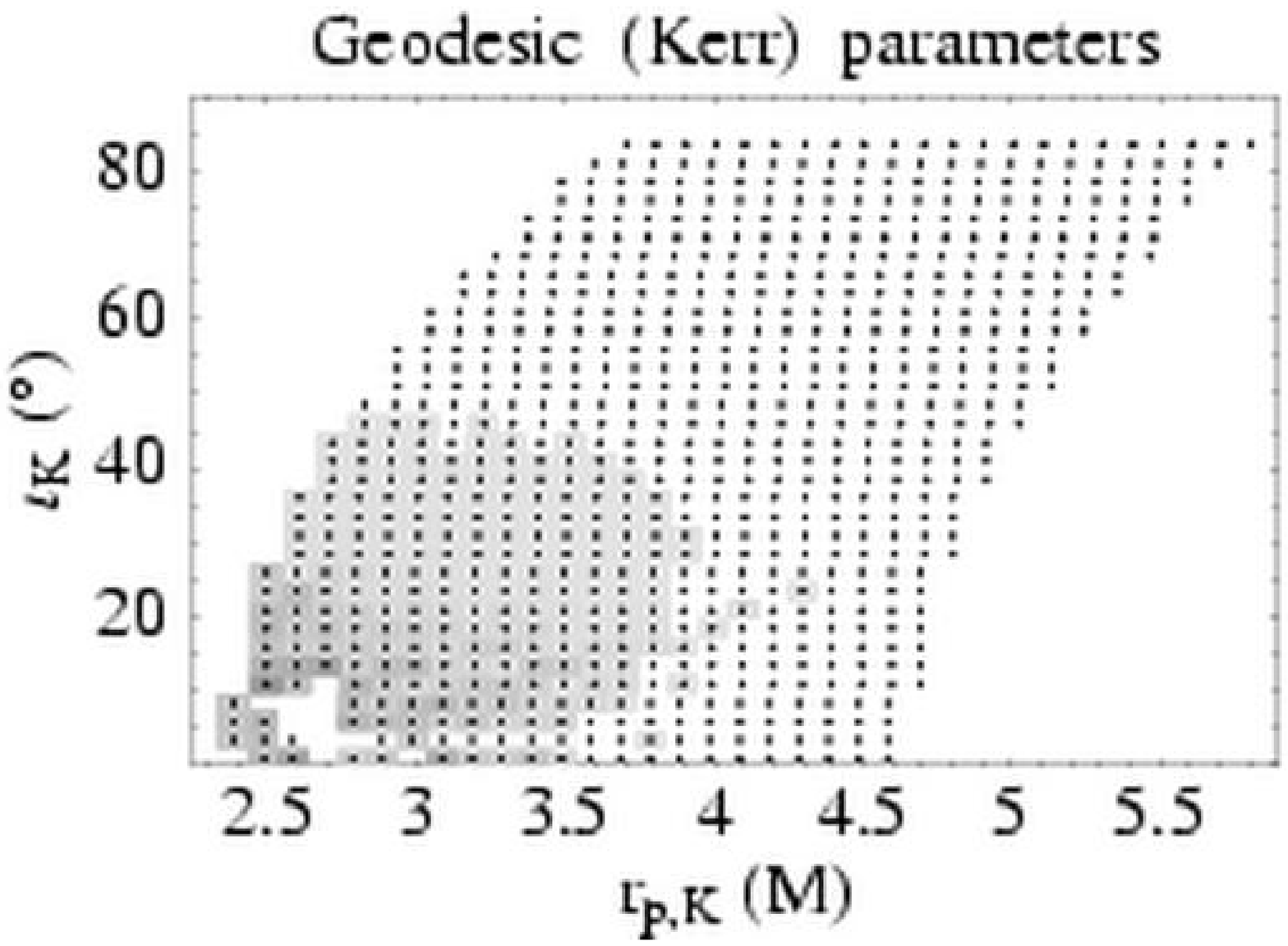}
	& \includegraphics[width=3in]{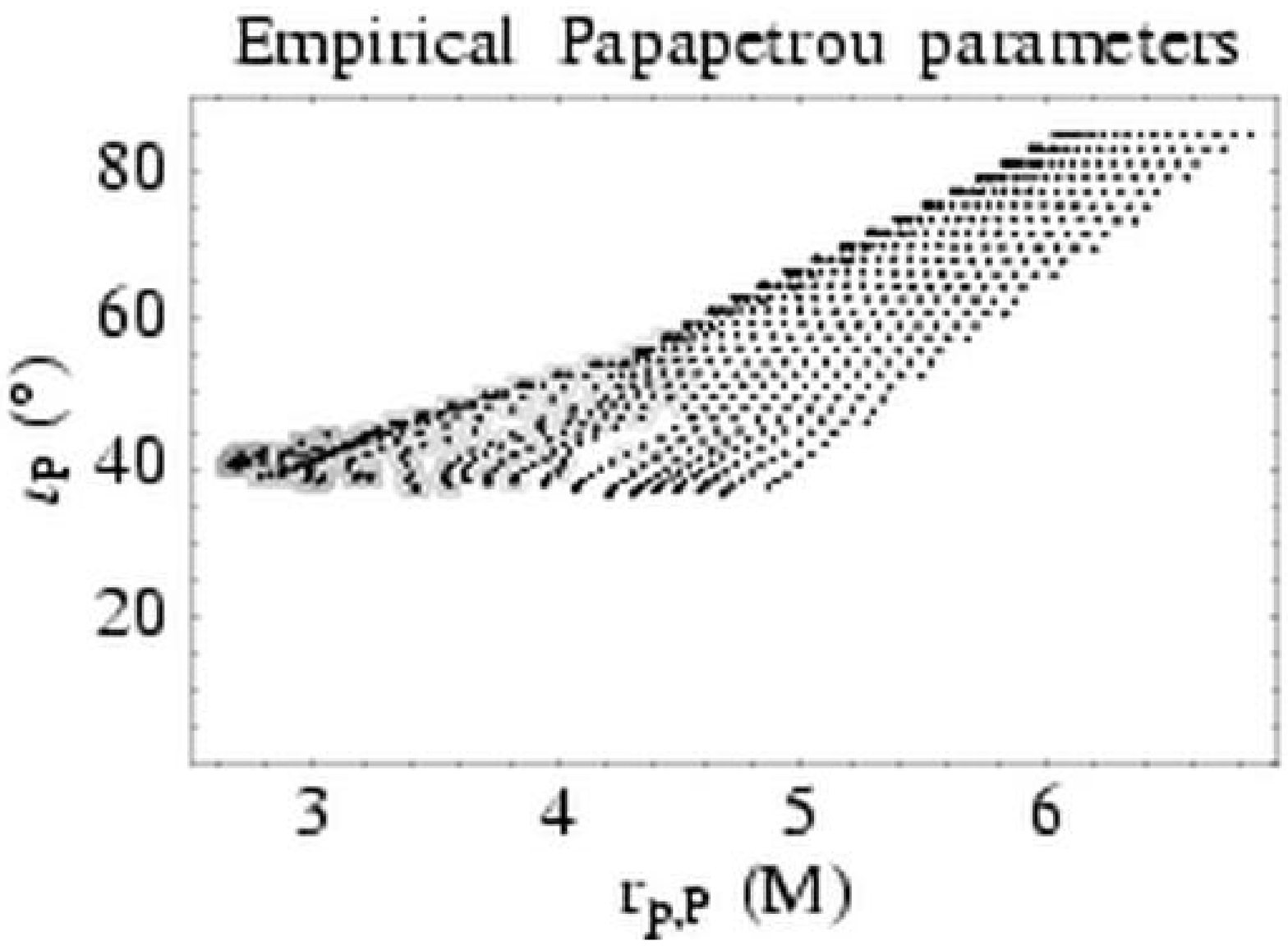}
    & \includegraphics[height=2in]{r_iota_1_0.5_1_0.2_-0.2_0_0.dat_sb.eps}\\
(a) & (b) & \medskip\\
\end{tabular}
\caption{\label{fig:a=0.7} $r_p$-$\iota$ map for $S=1$, $a=0.7$, 
and $e=0.5$.  (a)~Requested parameters; (b)~empirical parameters.  The shading is
scaled to the same maximum Lyapunov exponent as Fig.~\ref{fig:e=0.5_S=1}.
There is still substantial chaos.}
\end{figure*}

\begin{figure*}
\begin{tabular}{ccl}
\includegraphics[width=3in]{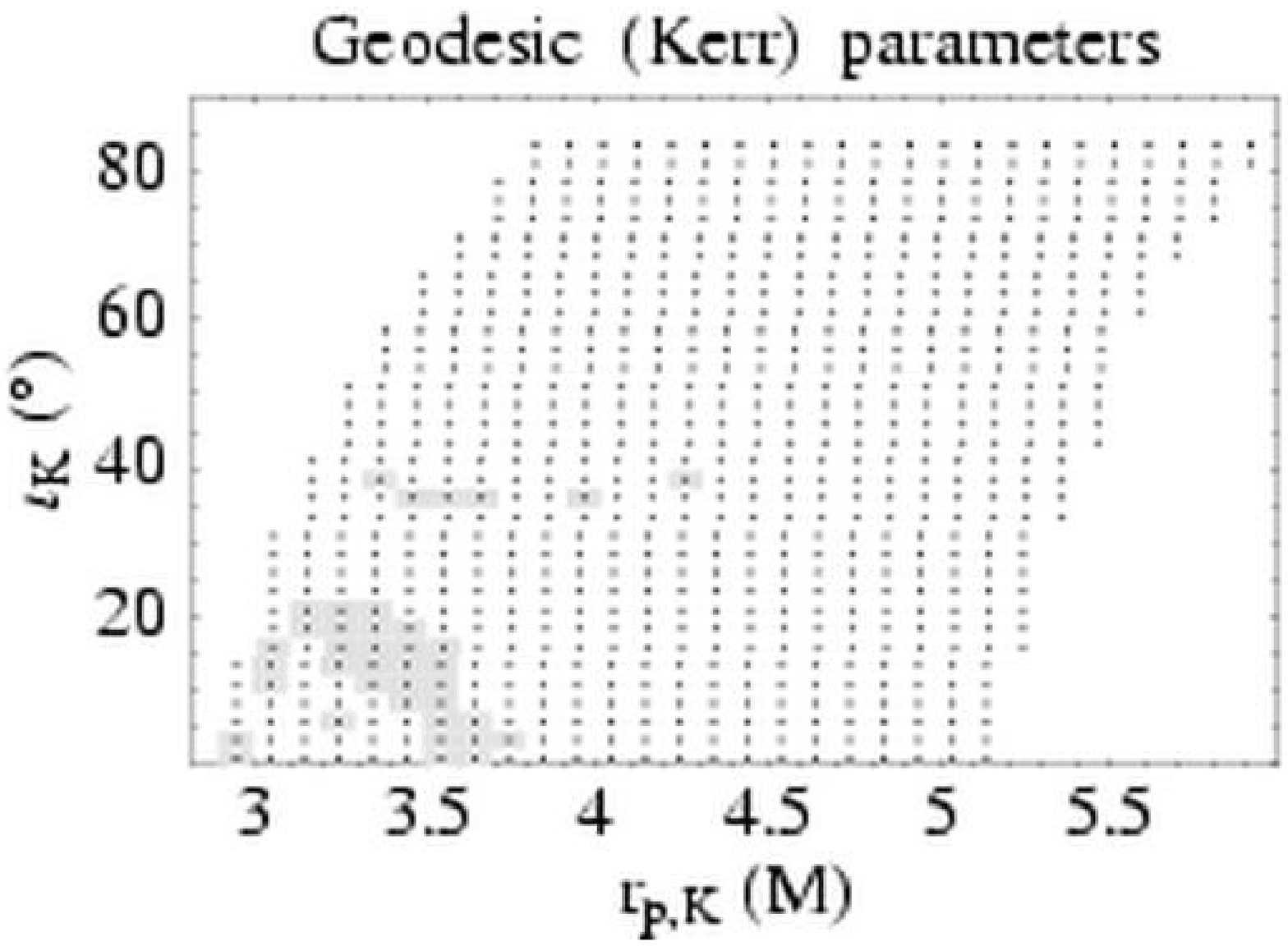}
	& \includegraphics[width=3in]{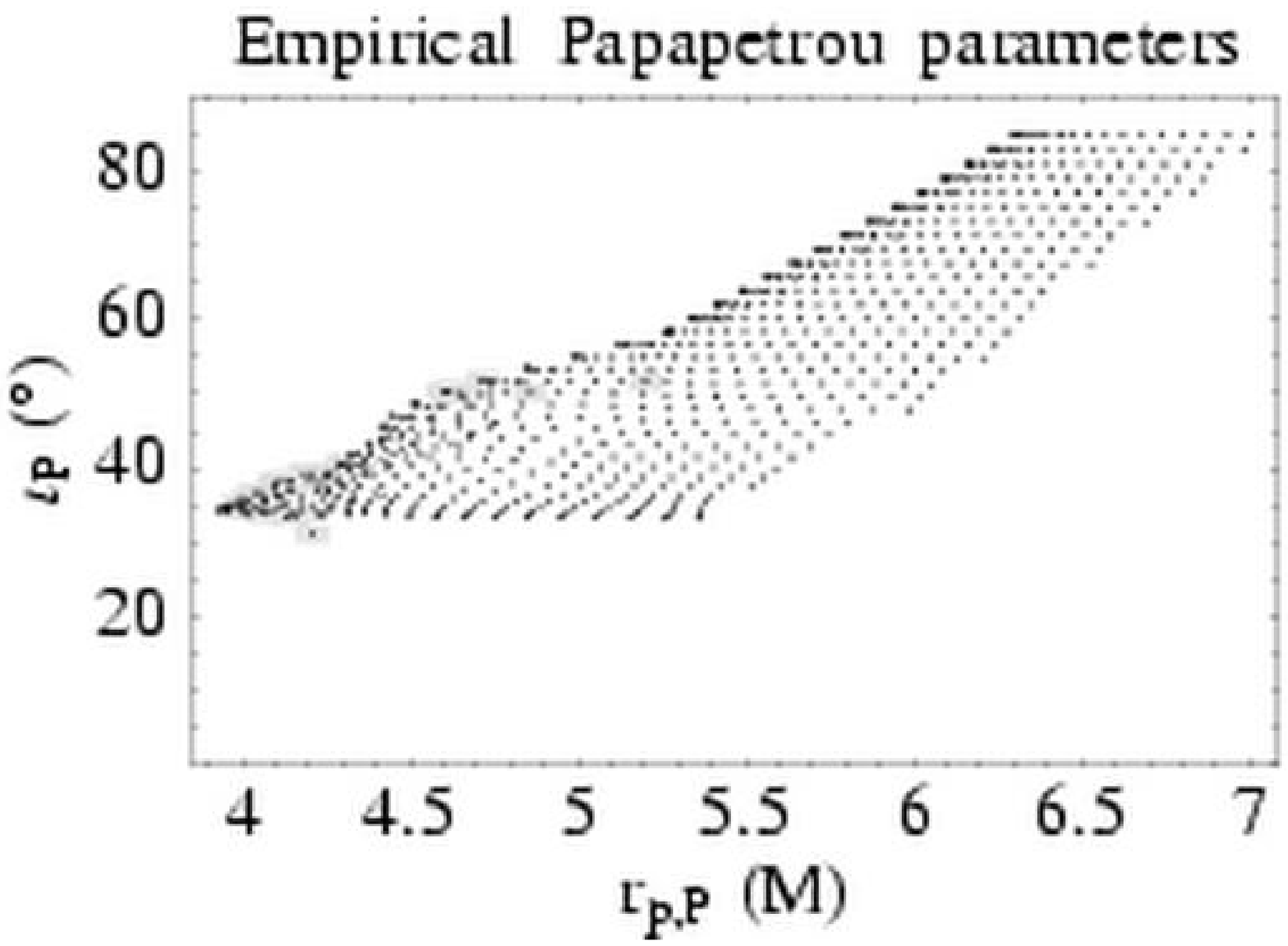}
    & \includegraphics[height=2in]{r_iota_1_0.5_1_0.2_-0.2_0_0.dat_sb.eps}\\
(a) & (b) & \medskip\\
\end{tabular}
\caption{\label{fig:a=0.5} $r_p$-$\iota$ map for $S=1$, $a=0.5$, 
and $e=0.5$.  (a)~Requested parameters; (b)~empirical parameters.  The shading is
scaled to the same maximum Lyapunov exponent as Fig.~\ref{fig:e=0.5_S=1}.
The chaos is largely confined to low pericenter orbits. }
\end{figure*}

\begin{figure*}
\begin{tabular}{ccl}
\includegraphics[width=3in]{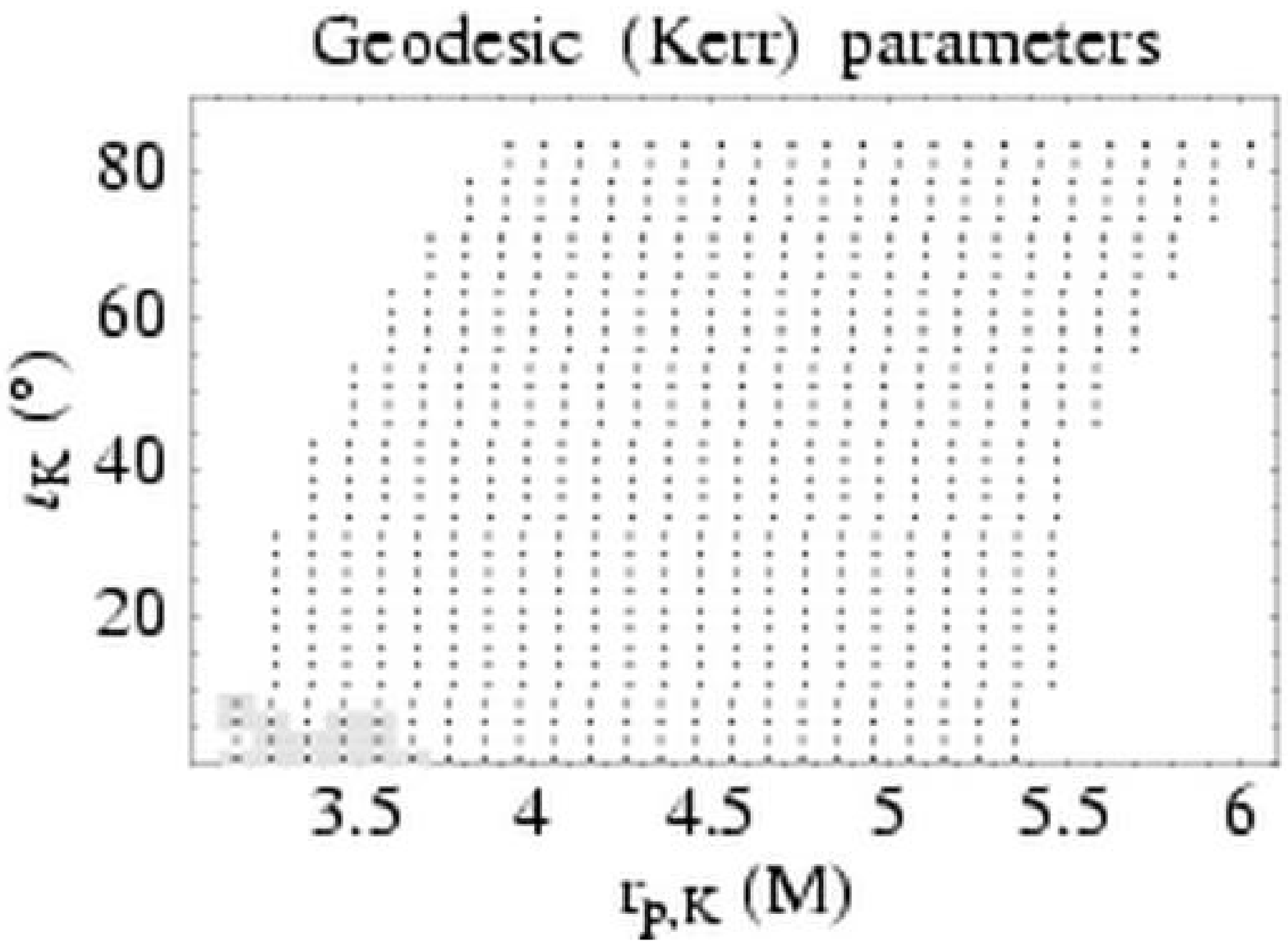}
	& \includegraphics[width=3in]{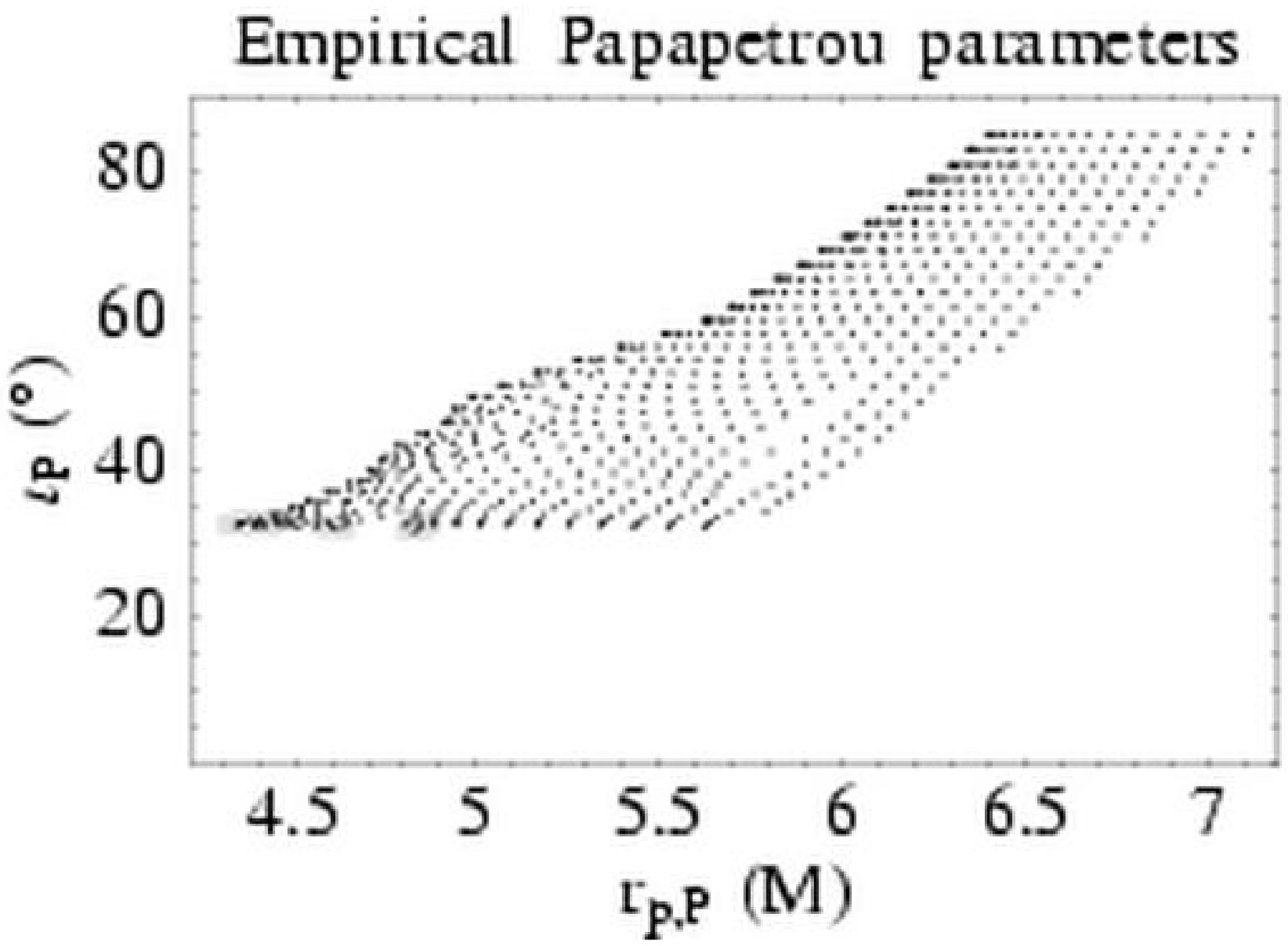}
    & \includegraphics[height=2in]{r_iota_1_0.5_1_0.2_-0.2_0_0.dat_sb.eps}\\
(a) & (b) & \medskip\\
\end{tabular}
\caption{\label{fig:a=0.4} $r_p$-$\iota$ map for $S=1$, $a=0.4$, 
and $e=0.5$.  (a)~Requested parameters; (b)~empirical parameters.  The shading is
scaled to the same maximum Lyapunov exponent as Fig.~\ref{fig:e=0.5_S=1}.
Only a handful of initial conditions are chaotic. }
\end{figure*}

\begin{figure*}
\begin{tabular}{ccl}
\includegraphics[width=3in]{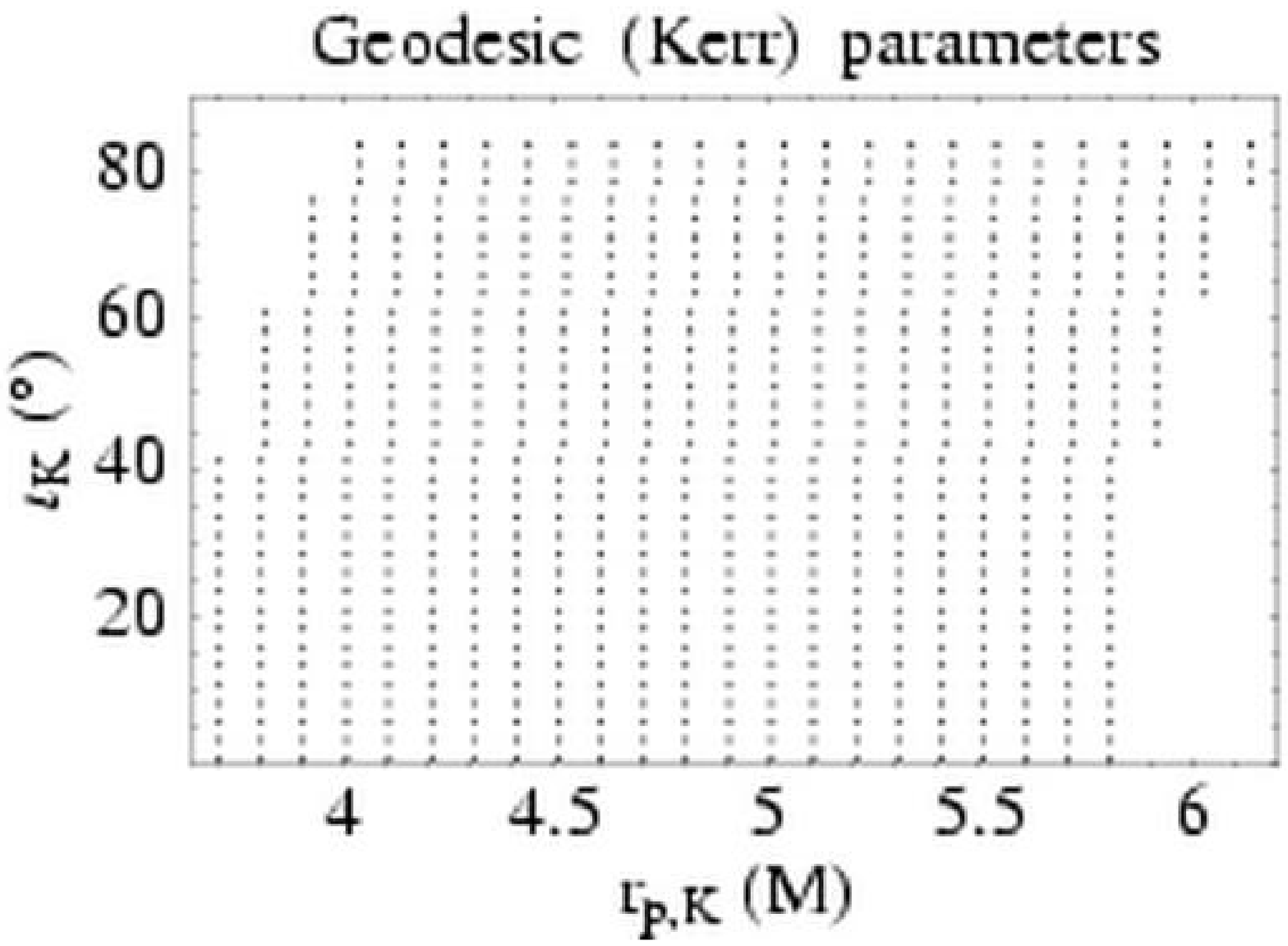}
	& \includegraphics[width=3in]{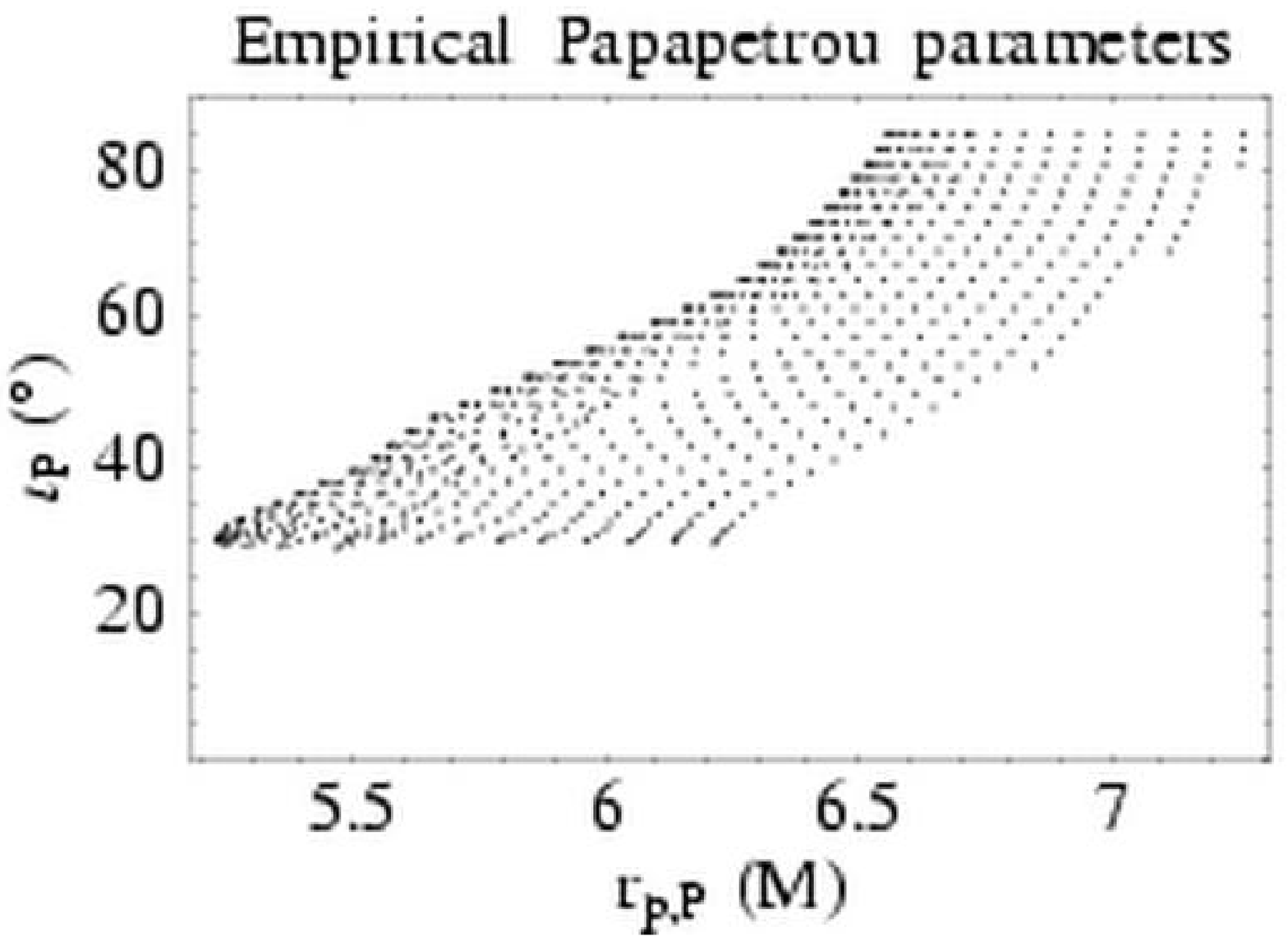}
    & \includegraphics[height=2in]{r_iota_1_0.5_1_0.2_-0.2_0_0.dat_sb.eps}\\
(a) & (b) & \medskip\\
\end{tabular}
\caption{\label{fig:a=0.2} $r_p$-$\iota$ map for $S=1$, $a=0.2$, 
and $e=0.5$.  (a)~Requested parameters; (b)~empirical parameters.  The shading is
scaled to the same maximum Lyapunov exponent as Fig.~\ref{fig:e=0.5_S=1}.
No initial conditions are chaotic. }
\end{figure*}

\begin{figure*}
\begin{tabular}{ccl}
\includegraphics[width=3in]{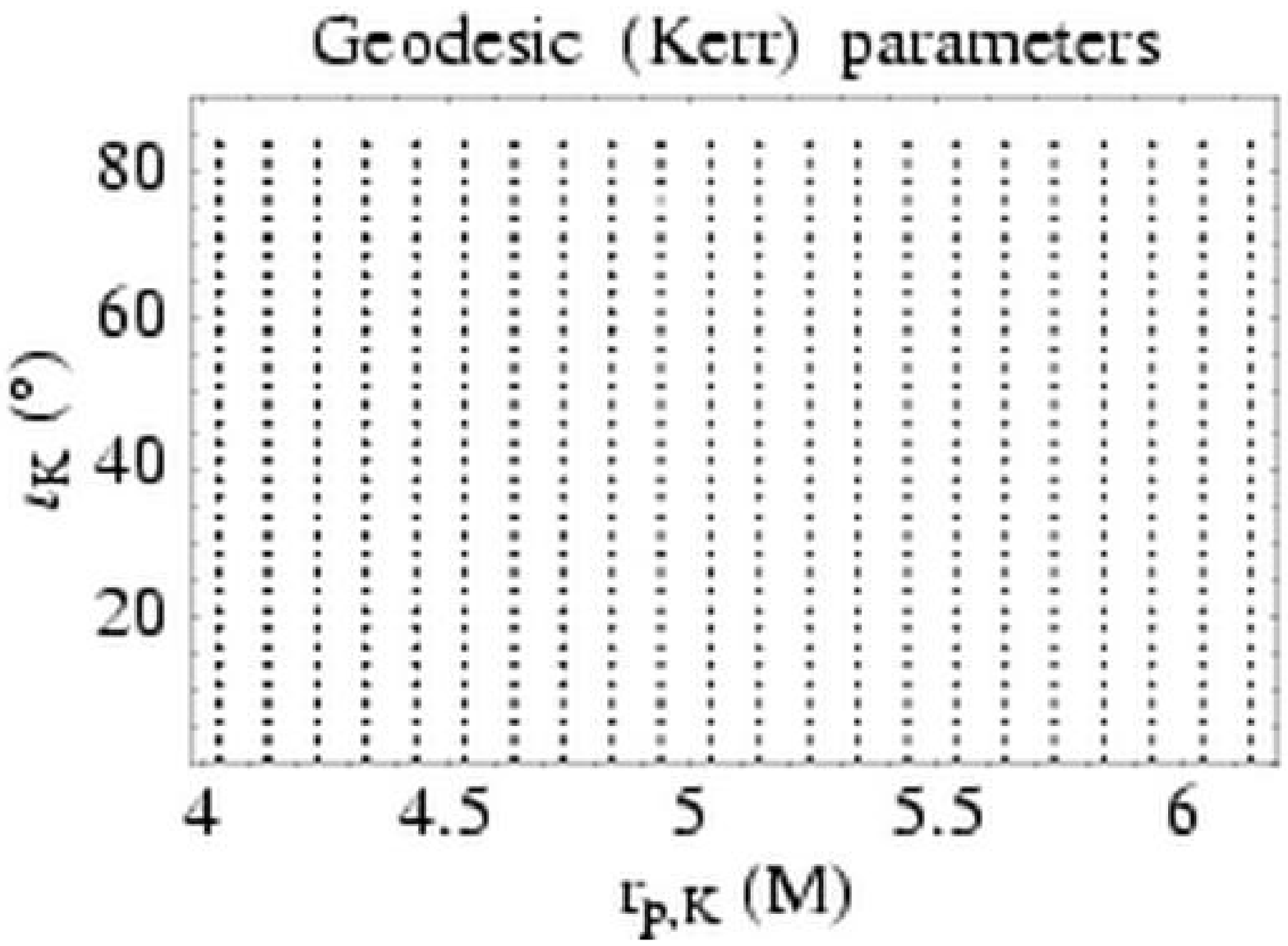}
	& \includegraphics[width=3in]{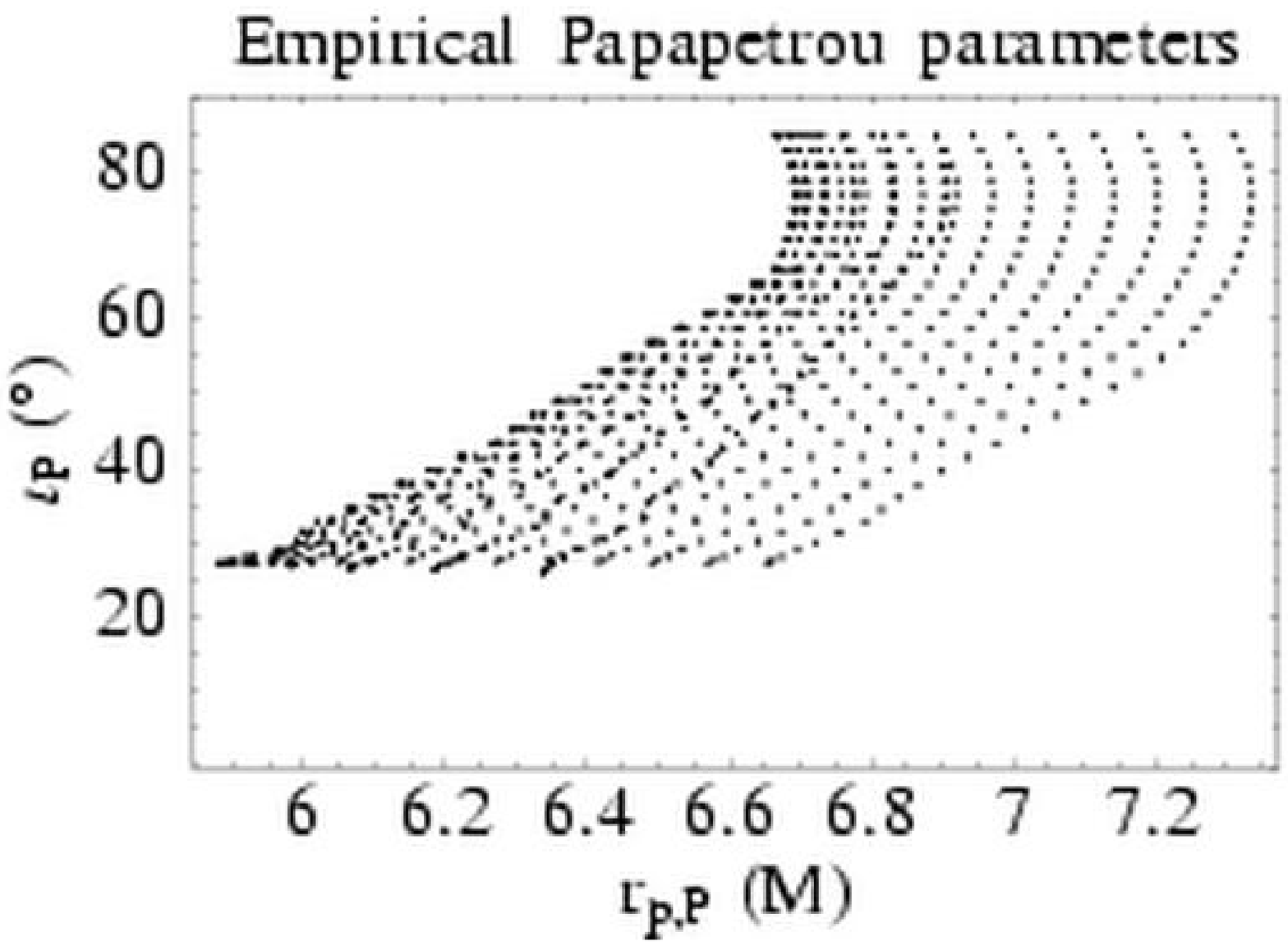}
    & \includegraphics[height=2in]{r_iota_1_0.5_1_0.2_-0.2_0_0.dat_sb.eps}\\
(a) & (b) & \medskip\\
\end{tabular}
\caption{\label{fig:a=0} $r_p$-$\iota$ map for $S=1$, $a=0$, 
and $e=0.5$. (a)~Requested parameters; (b)~empirical parameters.  The shading is
scaled to the same maximum Lyapunov exponent as Fig.~\ref{fig:e=0.5_S=1}.
No initial conditions are chaotic. Note that every column of~(a) is identical. 
This
is a result of the spherical symmetry of the $a=0$ (Schwarzschild) metric: all
inclination angles are equivalent.  As seen in~(b), this
symmetry is broken by the spin of the test particle.}
\end{figure*}

\end{document}